\documentclass[usenatbib]{mnras} %fleqn

\usepackage{newtxtext,newtxmath}
\usepackage{amsmath,bm} % amssymb clash with mras \usepackage{amssymb}	% Extra maths symbols
\usepackage{mathtools}
\usepackage{booktabs}
\usepackage[T1]{fontenc}

\usepackage{xcolor}

\DeclareRobustCommand{\VAN}[3]{#2}
\let\VANthebibliography\thebibliography
\def\thebibliography{\DeclareRobustCommand{\VAN}[3]{##3}\VANthebibliography}

\usepackage{enumitem}
\setlist[enumerate]{noitemsep,topsep=0em,labelwidth=*,leftmargin=0.6em}
\setlist[itemize]{noitemsep,topsep=0em,labelwidth=*,leftmargin=0.6em}

\usepackage{graphicx}	% Including figure files
\graphicspath{{Figures/figs/}}
\DeclareGraphicsExtensions{.pdf,.png,.jpeg}
\usepackage{amsmath}	% Advanced maths commands

\title[Extraction and Modelling in the Cosmic Web]{Swarm Intelligence-based Extraction and Manifold Crawling Along the Large-Scale Structure}

\author[P. Awad et al.]{
Petra Awad$^{1,2}$, Reynier Peletier$^{1}$, Marco Canducci$^{3}$, Rory Smith$^{4}$, Abolfazl Taghribi$^{2}$, 
\newauthor{Mohammad Mohammadi$^{2}$, Jihye Shin$^{5}$, Peter Ti{\v{n}}o$^{3}$, Kerstin Bunte$^{2}$}
\\
\\
$^{1}$University of Groningen, Kapteyn Astronomical Institute, 9747 AD Groningen, The Netherlands\\
$^{2}$University of Groningen, Bernoulli Institute for Mathematics, Computer Science and Artificial Intelligence\\
$^{3}$University of Birmingham, School of Computer Science, B15 1TT, Birmingham, United Kingdom\\
$^{4}$Universidad Technica Frederico de Santa Maria, Avenida Vicuña Mackenna 3939, San Joaquín, Santiago\\
$^{5}$Korea Astronomy and Space Science Institute (KASI)
776, Daedeok-daero, Yuseong-gu, Daejeon, 34055, South Korea 
}

\date{Accepted XXX. Received YYY; in original form ZZZ}

\pubyear{2022}

\def\P{{\mathbf P}}

\def\S{{\mathbf S}}

\def\cP{{\cal P}}

\newcommand{\RR}{{\mathbb{R}}}

\newcommand{\NN}{{\mathbb{N}}}

\renewcommand{\vector}[1]{\bm{#1}}

\renewcommand{\matrix}[1]{\bm{\MakeUppercase{#1}}}
\newcommand{\set}[1]{\mathcal{\MakeUppercase{#1}}}

\definecolor{RoyalBlue}{cmyk}{1, 0.50, 0, 0}
\begin{document}
%\tableofcontents

\label{firstpage}
\pagerange{\pageref{firstpage}--\pageref{lastpage}}
\maketitle

\begin{abstract}
The distribution of galaxies and clusters of galaxies on the mega-parsec scale of the Universe follows an intricate pattern now famously known as the Large-Scale Structure or the Cosmic Web. 
To study the environments of this network, several techniques have been developed that are able to describe its properties and the properties of groups of galaxies as a function of their environment. In this work we analyze the previously introduced framework: 1-Dimensional Recovery, Extraction, and Analysis of Manifolds (1-DREAM) on N-body cosmological simulation data of the Cosmic Web. 
The 1-DREAM toolbox consists of five Machine Learning methods, whose aim is the extraction and modelling of 1-dimensional structures in astronomical big data settings.
We show that 1-DREAM can be used to extract structures of different density ranges within the Cosmic Web and to create probabilistic models of them. For demonstration, we construct a probabilistic model of an extracted filament and move through the structure to measure properties such as local density and velocity.
%We then assess the performance of the algorithms and discuss the influence of the free parameters within the composite toolbox. 
We also compare our toolbox with a collection of methodologies which trace the Cosmic Web. We show that 1-DREAM is able to split the network into its various environments with results comparable to the state-of-the-art methodologies. A detailed comparison is then made with the public code DisPerSE, in which we find that 1-DREAM is robust against changes in sample size making it suitable for analyzing sparse observational data, and finding faint and diffuse manifolds in low density regions.

\end{abstract}


%%%%%%%%%%%%%%%%%%%%%%%%%%%%%%%%%%%%%%%%%%%%%%%%%%

%%%%%%%%%%%%%%%%% BODY OF PAPER %%%%%%%%%%%%%%%%%%

\section{Introduction}

%On the mega-parsec scale of the Universe, the spatial distribution of galaxies as well as clusters of galaxies is not uniform. In fact, looking at the outcomes of the Sloan Digital Sky Survey (SDSS) (\textcolor{blue}{Fukugita et al., 1998; Gunn et al., 1998}), one can see that there is a intricate inter-connected pattern that emerges at such a scale, and that this pattern forms a network that is termed the Cosmic Web (\textcolor{blue}{Bond et al., 1996}).    

%As described in Peebles (1980), the Cosmic Web emerges as the outcome of the anisotropic nature of gravitational collapse. The latter is the driving effect behind structure formation including the emergence of the Cosmic Web's different morphological components namely: clusters, filaments, and walls. The connection between these components can be summarized as follows: clusters are regions of intersection of filaments, and filaments are regions of intersection of walls (Cautun et al. 2012; Doroshkevich et al. 1980; Melott 1983; Pauls $\&$ Melott 1995; Shapiro et al. 1983; Sathyaprakash et al. 1996).

Large observational surveys such as the SDSS \citep{SDSS}, 6dFGS \citep{6DFGSOrg,6DFGS}, and 2MRS \citep{2MRSOrg, 2MRS} have repeatedly confirmed that galaxies and clusters of galaxies are distributed in the Universe in the form of an interconnected network known as the Cosmic Web \citep{BondEtal1996}. This network is the result of the anisotropic gravitational collapse which drives structure formation in the Universe and leads to the emergence of the main morphological components of the Cosmic Web namely: clusters, filaments, walls, and in relatively sparser regions to the emergence of cosmic voids \citep{Peebles1980}. In order to characterize the Cosmic Web physically as well as numerically, it is important to first define its main properties. One of the main properties of the Cosmic Web is the anisotropy arising from the presence of the different morphological structures which form it and from the shape asymmetry inherent to its different environments. Secondly, the mode of formation of the Cosmic Web has allowed for the emergence of interconnected structures whose densities vary across different scales \citep{DavisEtal1985, Jenkins1998, ColbergEtal2005, DolagEtal2006}. 
This leaves no space for clearly distinguishable structures at a given scale or density, and in turn increases the difficulty for differentiating between the regions belonging to the different environments. 
Similarly, the Cosmic Web spans six orders of magnitude in density with an overlap in the range of sizes and densities %of the different environments 
\citep{DoroshkevichEtal1980, KlypinEtal1983, PaulsEtal1995, SathyaprakashEtal1996}. 
This points to the fact that there is no optimal scale at which to identify the components of the Cosmic Web. In turn, this defines the hierarchical property and multi-scale nature in both mass and size of the Cosmic Web, with the velocity field surrounding the structures also being highly complex \citep{Sheth2004, ShethWeygaert2004, ShenEtal2006}.

It is therefore clear that developing (semi-)automated numerical algorithms which study the Cosmic Web in such a way that its different properties are taken into account is not an easy task.
Ultimately, the role of the developed tools is to locate and extract the lower-dimensional structures embedded in the potentially higher dimensional and massive % larger 
simulated data point clouds \citep{TaghribiEtal2022}. 
Therefore, a prominent problem that structure detection algorithms face is having to deal with a very large number of high-dimensional data points in addition to the presence of scatter or noise in the particle distributions along the structures, and outliers that affect the results of manifold %structure 
learning and dimensionality reduction techniques \citep{WuEtal2018, TaghribiEtal2022}. 
In some works, mathematical solutions were presented to face the problem of denoising given manifolds in a data set (i.e. extracting structures from a large scattered distribution of particles) such as resorting to the Longest Leg Path Distance (LLPD) in \citet{LLPD}. 
If the value of LLPD between a particle and its neighbors is larger than a predetermined threshold, then the particle is removed from the data set since it is considered as noise through this definition. 
Although this technique has its advantages such as successfully reducing the size of the point cloud, it has been shown to be problematic if the clustered data is highly curved and is of varying size \citep{LLPD, TaghribiEtal2022}. 
These limitations of LLPD makes it unreliable if applied to simulation point clouds of the Cosmic Web given their hierarchical nature.  

The different properties that the Cosmic Web possesses also complicate the ability to perform descriptive analysis by conventional astrophysical statistics used to quantitatively study the arrangement of mass in the Universe \citep{LibeskindEtal2018}. For example, the correlation function defined in \citet{Peebles1980} (the probability that another galaxy will be found within a certain distance from a given galaxy) is not sensitive enough to the complexity of patterns in mass and spatial distribution found in the Large-Scale Structure \citep{LibeskindEtal2018}. Therefore, it is necessary to look into newer approaches for tackling the task of tracing and analyzing structures of the Cosmic Web. In that pursuit, many novel methodologies have been developed, each employing different physical definitions for the structures in order to identify and classify the Cosmic Web environments within a given data set. Percolation techniques developed in \citet{BarrowEtal1985}, \citet{GrahamEtal1995}, and \citet{Colberg2007} provide a measure of filamentarity using a graph-theoretical construct termed the Minimum Spanning Tree (MST) of galaxy distributions. 
The branching of the MSTs are then used in works such as \citet{Colberg2007} and \citet{TREX} as a criterion to identify clusters and their branching filaments.
Stochastic techniques were also developed including the non-parametric formalism for two-dimensional distributions \citep{GenoveseEtal2010} which relies on the representation of filaments by their central axis, and the Bisous model \citep{TempelEtal2016} which represents cosmic structures as a series of connected and aligned cylinders.
Another type includes the phase-space methods such as the techniques developed in \citet{Shandarin2011}, and \citet{Abel&Kaehler2012}, and the ORIGAMI algorithm \citep{Origami1, Origami2}, all of which study the phase spaces of evolving mass distributions. Other methodologies include tessellation-based algorithms that strive to extract topological features from the underlying physical fields \citep{Weygaert&Schaap2009,Gonzalez&Padilla2010}. 
Studying the density field provides a link to morphology, while the tidal force field largely relates to the dynamical evolution of the Cosmic Web. The velocity field on the other hand provides information on the connection between the structures of the web and the velocity flow in and surrounding these structures. \citep{AragonCalvoEtal2007, HahnEtal2007, BondEtal2010, HoffmanEtal2012, CautunEtal2013, MetukiEtal2015}. 
\citet{AragonCalvoEtal2007} have followed this approach by creating the Multiscale Morphology Filter (MMF) that constructs a scale space by applying Gaussian smoothing at different scales to the density field. 
The Nexus formalism \citep{CautunEtal2013} then extends the MMF method by including appropriate filters for the tidal and velocity fields. 
Another known tessellation-based method is DisPerSE \citep{Disperse1, Disperse2}, a publicly available tool that relies on topological concepts such as Delaunay Field Estimation \citep[DTFE]{Schaap&Weygaert2000, Weygaert&Schaap2009} for the construction of a density field out of an input cosmological data set, and Discrete Morse Theory \citep{Forman1998, Gyulassy2008} for tracing and separating the environments of the Cosmic Web. Additionally, DisPerSE uses Persistence Homology \citep{EdelsbrunnerEtal2002} as a filtration technique for structures it classifies as insignificant. Given its public nature, DisPerSE has been frequently used in the literature such as in \citet{KleinerEtal2017}, \citet{KraljicEtal2018}, \citet{LaigleEtal2018}, and \citet{LuberEtal2019}.

%Introduce the algorithms here
In this work, we explore the toolbox 1-DREAM recently introduced in Astronomy \& Computing in \citet{CanducciEtal2022b}.
The toolbox consists of five main algorithms for the extraction and modelling of 1-dimensional astronomical structures. These algorithms can be used individually if desired, but are advised to be applied together and in the order presented in this work. The first methodology implements Ant Colony Optimization \citep{ACObook} for the highlighting of particles belonging to hidden manifolds (structures) within simulation data sets. The second methodology is also swarm intelligence-based and serves the identification of the mean curves (central axes) of the detected structures. The third algorithm attributes a dimensionality to the distributed points based on their local neighborhood, thus partitioning the data set into clusters (3-dimensional structures), walls (2-dimensional structures), and filaments (1-dimensional structures). The fourth technique further partitions the data containing 1-dimensional structures into a set of filaments represented by the ``skeletons" of the identified structures along with the set of particles surrounding each skeleton. Finally, the fifth algorithm provides for a given structure, a constrained Gaussian Mixture Model description centered on the structure's skeleton. 
When using the algorithms together, the 1-DREAM toolbox allows for the extraction of structures within the simulations and their subsequent modelling for further quantitative analysis. 

In \citet{CanducciEtal2022b} the five algorithms were presented as a coherent publicly available\footnote{%An implementation of t
%The toolbox is %publicly available at
Toolbox: \url{https://git.lwp.rug.nl/cs.projects/1DREAM}} 
framework, and the functionality of the toolbox was briefly demonstrated on three examples namely: a simulated jellyfish galaxy, a cosmic filament, and the tidal tail of Omega Centauri.
The aim of the current work is to explore the proposed toolbox more thoroughly when applied specifically to N-body cosmological data sets of the Cosmic Web. We explain how 1-DREAM extracts structures from a cosmological data cube, and as an example, we extract a cosmic filament and construct its probabilistic model in order to move along its central axis and measure local properties along and orthogonal to the filament. We also apply our toolbox on the data provided in \citet{LibeskindEtal2018} in which a systematic method of comparison is provided between many Cosmic Web tracing methodologies to test their ability to differentiate between its various environments. Using the provided standard analysis, we compare our results to the compilation of codes in \citet{LibeskindEtal2018}.
Finally, we perform a more detailed comparison with DisPerSE \citep{Disperse1} and find that 1-DREAM is more robust against changes in the sample size of the data which highlights its advantage at tracing filaments in low density regions of the Universe.

This paper is organized as follows: Section~\ref{sec:data} provides a description of the data sets used in this work consisting of Dark Matter particle distributions extracted from N-body cosmological simulations; Section~\ref{sec:GeneralFormalism} details the general formalism of the algorithms. % where their purpose and modes of action are defined.
Section~\ref{sec:Analysis} presents the results when applying our toolbox on the data provided in \citet{LibeskindEtal2018} and the discussion of these results based on the standard analysis defined in that same work. We perform a more detailed comparison with other state-of-the-art tools of the field and highlight some strengths of our method in Section~\ref{sec:Compare}.
Section~\ref{sec:Conc} then summarizes our work and suggests future developments.

%All papers should start with an Introduction section, which sets the work
%in context, cites relevant earlier studies in the field by \citet{Fournier1901},
%and describes the problem the authors aim to solve \citep[e.g.][]{vanDijk1902}.
%Multiple citations can be joined in a simple way like \citet{deLaguarde1903, %delaGuarde1904}.

%-----------------------------%
\section{Simulation Data}
\label{sec:data} % used for referring to this section from elsewhere
%-----------------------------%

%The main input on which 1\MC{-}DREAM runs is multi-dimensional distributions of particles or point clouds. 
%This in turn makes the results of N-body simulations ideal data sets to apply our methodologies on. 
To demonstrate the astronomical applicability of the introduced algorithms, we use two realistic cosmological data sets, both consisting of point-particles distributed in three-dimensional space.
The first data set is the output of a Dark Matter-only N-body cosmological simulation that is run using the GADGET-3 code \citep{Gadget2}. 
The initial conditions are generated at redshift $z=200$ using the Multi Scale Initial Condition software \citep[MUSIC]{Hahn&Abel2011}. 
The CAMB package \citep{LewisEtal2000} is then used to calculate the linear power spectrum. 
We produce a single cosmological volume with dimensions $120\times120\times120$~Mpc/h containing $\approx 7$~million particles in total.
The dark matter particles have a fixed mass of $1.072\times10^9 M_{\odot}/h$, and the cosmology assumed for the simulation is the following: $\Omega_m = 0.3$, $\Omega_{\Lambda} = 0.7$, $\Omega_b = 0.047$ and $h_0 = 0.684$. 
From the described simulation we use the output at redshift $z=0$ which consists of the masses and the three-dimensional components of the positions and velocities of all dark matter particles. 
This data set has been created and used in the following works: \citet{SmithEtal2021}, \citet{JheeEtal2022},
\citet{Smith2Etal2022, SmithEtal2022}, \citet{KimEtal2022}, \citet{ChunEtal2022}, and will be referred to as the \emph{N-cluster simulation} hereafter following the convention in \citet{JheeEtal2022}. This data set represents the typical data on which Cosmic Web-tracing algorithms are applied, and so will be used to explain the general formalism of the toolbox and to compare the properties of DisPerSE and 1-DREAM.

The second data set included in our investigation is the publicly available data introduced in \citet{LibeskindEtal2018}. 
It consists of Dark Matter particle distributions and a list of Dark Matter halos extracted from a GADGET-2 N-body simulation \citep{Gadget2}. 
The simulation box has dimensions of $200\times200\times200$~Mpc/h and contains $512^3$ Dark Matter particles. 
The bound particles are then grouped into halos by a Friend--of--Friend algorithm \citep{DavisEtal1985}. 
The cosmological parameters used for this simulation are the following : $\Omega_M = 0.31$, $\Omega_{\Lambda} = 0.69$, $n_s = 0.96$, $h = 0.68$, and $\sigma_8 = 0.82$.
This data was used by the authors in \citet{LibeskindEtal2018} to provide a unified comparison scheme between many existing Cosmic Web related algorithms. 
The work mainly relied on comparing the algorithms' classification of the particles or halos between belonging to clusters, walls, filaments, or voids. 
In Section~\ref{sec:Analysis}, we apply our toolbox on this data set thereby producing our own classification of the particles and halos, and thus provide grounds for comparing our results with a large set of other state-of-the-art methodologies.

%-------------------------------------------%
\section{General Formalism}
\label{sec:GeneralFormalism} % used for referring to this section from elsewhere
%-------------------------------------------%

In this section we provide a brief overview of the five algorithms introduced in this work and refer the reader to a detailed methodological explanation of the cumulative toolbox in \citet{CanducciEtal2022b} and to the individual papers where each algorithm was first introduced. 
As a reference to the different algorithms we explain the purpose of each one first and then move on to describing how each one operates. 
Finally, to better illustrate the functionality of the different parts of the toolbox, we demonstrate the pipeline of methodologies on a filament connecting two clusters, extracted from our cosmological simulation. 
%We should define what is meant by noise here also (before the description).

\begin{description}
    \item[LAAT \citep{TaghribiEtal2022}:] \emph{Locally Aligned Ant Technique} is developed for highlighting the contrast between high and low density regions in a given point cloud as well as detecting regions aligned with defined structures within the data. 
    The algorithm, inspired by Ant Colony Optimization, defines a pheromone quantity ``deposited" on the point cloud particles and is used to incentivize the choice of jumps in a random walk through neighborhoods within the particle distribution. 
    During the random walk, the pheromone accumulates on the particles that align with the directions of manifolds estimated within a neighborhood defined radius, 
    %aligned with manifolds defined here as structures distinguishable from random distributions of points or noise, 
    and evaporates on noise particles and background far from any structures. %the more randomly distributed particles. 
    The deposited pheromone amount can be interpreted as a measure of faintness of the structures, and thresholding is %can then be 
    used to extract the detected structures.%distinguish the structures from the noise. 

    \item[EM3A \citep{MohammadiEtal2022}:]%Standing for 
    \emph{Evolutionary  Manifold  Alignment Aware Agents} %EM3A 
    moves particles belonging to the manifolds towards their central axis, thus further enhancing the contrast between under-dense and over-dense regions in the data. This algorithm together with LAAT is said to ``denoise" the data, as in it uncovers the manifolds embedded within their scattered or noisy environments. Similar to multi-agent random walks, the motion of particles are enforced by biologically motivated ant-colony behaviour. Game theoretical principles are also applied to adapt parameters automatically.

    \item[DimIndex \citep{CanducciEtal2022a}:] This
    method makes use of the eigen-decomposition of local neighborhoods of particles for assigning a \emph{Dimensionality Index} to the structure those particles belong to. 
    The index is an indication of the most likely dimension of the structure to which a particle belongs. In other words, this algorithm assigns a number (either 1, 2, or 3) to each particle in the data set corresponding to the spatial dimensionality of the respective structure that the particles make up. 
    These indices can thus be used as labels to partition the data into its different dimensional portions by differentiating between points belonging to 1D structures (filaments), 2D structures (walls), and 3D structures (clusters).
   
    \item[Multi-Manifold Crawling \citep{CanducciEtal2022a}:] %Taking
    Based on the original data and the central axis of the manifolds recovered by EM3A as input, this %the following
    algorithm is applied on the recovered axes to construct their skeletal representations and partition the data into a set of skeletons and the respective group of particles surrounding them.
    Again, walking agents are utilized to ``crawl" along the detected structures and sample them in a discrete set of (roughly) equidistant points ordered along the direction of the structures. This allows to obtain a set of piece-wise linear curves each representing a structure in the data set.
    The recovered skeletons, refined using SGTM (explained next), can then serve as a central axis to move along and measure physical properties in longitudinal and orthogonal directions to this axis. 
    
    \item[Stream GTM \citep{CanducciEtal2022a}:] This algorithm is a varied formulation of \emph{Generative Topographic Mapping} which takes a given detected manifold, in our case restricted to be 1-dimensional, and constrains the points belonging to it to a Gaussian Mixture Model centered on the stream's skeleton. The constrained mixture is then trained using the Expectation and Maximization technique \citep{Bishopbook} to create a probabilistic model describing the unmodified particle distribution around the skeleton retrieved by Crawling as a collection of Gaussians. The model can then provide the likelihoods of given particles to belong to the studied manifold. In other words the probabilistic model relaxes the notion of a radius beyond which a structure ends, and substitutes it with a measure of probability for particles to belong to the modelled structure. 
    
\end{description}

%Normally the next section describes the techniques the authors used.
%It is frequently split into subsections, such as Section~\ref{sec:maths} below.

\begin{figure*}
\centering
\includegraphics[width=\textwidth]{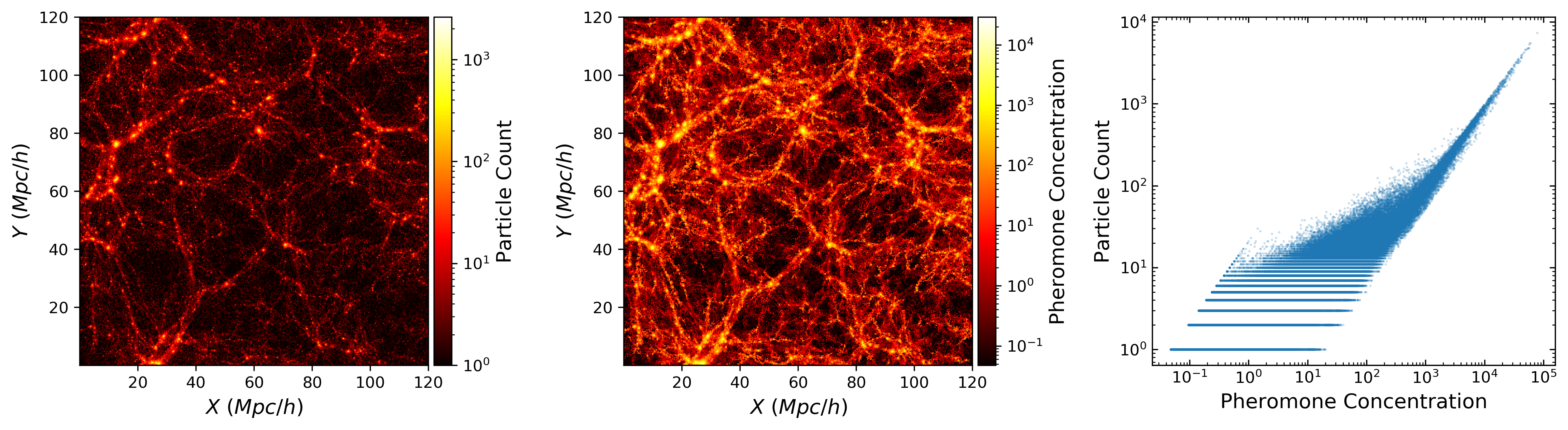} % Figures/figs/ .png
\caption{The left panel shows a slice of thickness $20$~Mpc/h from our simulation data. 
Brighter regions correspond to structures of higher density than fainter regions. 
The middle panel shows the same slice after applying LAAT to the entire data cube. 
Brighter regions now correspond to structures where the ``deposited" pheromone is more concentrated. 
After two runs of LAAT, the pheromone concentrates more on densely populated regions and less on sparser areas.
%The colorbar of the left panel also shows how much the density contrast has been enhanced between the different regions with a difference of $6$~orders of magnitude as compared to the left panel with a difference of $3$~orders of magnitude.
In the right panel, we plot in logarithmic scale the particle count versus pheromone concentration in each cell of of a $120 \times 120 \times 120$ binning of the simulation cube. At high number densities we see a linear relation between the pheromone and local density, but observe a wide scatter between the two plotted quantities at lower densities. This suggests that the relation between the pheromone concentration and the local density is not a simple power-law scaling. 
}
\label{fig:PheromoneDistribution}
\end{figure*}

\subsection{LAAT}
\label{sec:LAAT} % used for referring to this section from elsewhere

To begin with the description of our methodologies, first consider a data set $\set{Q} = \{\vec{x}_1, \vec{x}_2, \dots, \vec{x}_n \}$ consisting of the position vectors of $N$ particles such that $\vec{x}_i \in \mathbb{R}^D$. Then there exists $D$ principal components in a spherical neighborhood $\set{N}^i_r \coloneqq \set{B}(\vec{x}_i,r)$ of radius $r$ around a point at $\vec{x}_i$. 
We call $\vec{v}_d$ and $\lambda_d$ the local eigenvectors and corresponding ordered eigenvalues with $d=1,2,...,D$, respectively. 
LAAT then consists of a random walk in which agents jump from a particle belonging to the data set to the next particle. The high preference jumps are chosen according to the following two properties: 
jumps along the dominant eigenvectors are favored, and 
paths accumulating higher %higher priority paths are those where an 
amounts of artificially deposited pheromone get higher priority \citep{ACObook}.
Given a path $(\vec{x}_j-\vec{x}_i)$ between particles $i$ and $j$, the relative normalized weighting of the alignment of this path with a local eigenvector $\vec{v}_d$ is given as follows:
\begin{equation}
    w_{d}^{(i,j)} = 
    \frac{|\cos \alpha_d^{(i,j)}|}{\sum\limits_{d^\prime=1}^D |\cos \alpha_{d^\prime}^{(i,j)}|} \enspace .
%     |\cos \alpha_d^{(i,j)}| \left(\sum\limits_{d^\prime=1}^D |\cos \alpha_{d^\prime}^{(i,j)}|\right)^{(-1)} \enspace .
%     \frac{|\cos \alpha_d^{(i,j)}|}{\sum\limits_{d^\prime=1}^D |\cos \alpha_{d^\prime}^{(i,j)}|} \enspace .
    \label{cosineweight}
\end{equation}
%Where 
Here $\alpha_d^{(i,j)}$ is the angle between $(\vec{x}_j-\vec{x}_i)$ and $\vec{v}_d$.
Considering the normalized eigenvalues $\tilde{\lambda}^{(i)}_d$ (s.t. $\sum_{d=1}^D \tilde{\lambda}^{(i)}_d = 1$), we define the preference of the jump from $\vec{x}_i$ to $\vec{x}_j$ that is aligned with the local eigenvectors. 
This preference and its normalized version are given by the following: 
\begin{align}
%\begin{equation}
E^{(i,j)} &= \sum\limits_{d=1}^D w_{d}^{(i,j)} \cdot \tilde{\lambda}_{d}^{(i)} \enspace, % tODO notation change?
\label{Eformula}\\
%\end{equation} 
%\begin{equation}
\tilde{E}^{(i,j)} &=\frac{E^{(i,j)}}{\sum\limits_{j^\prime\in \set{N}^i_r} E^{(i,j^\prime)} } \enspace .
\label{Enormalizeformula}
%\end{equation}
\end{align}
Furthermore, we define an amount of pheromone $F^j(x)$ for a particle at $\vec{x}_j$ at a time $t$ (iteration in the random walk). 
Thus, the above preference for jumps will allow for the accumulation of the pheromone on the particles aligned with the manifolds. 
Inspired by nature, we incorporate an evaporation rate $0<\zeta<1$ in the definition of the pheromone which serves to decrease its amount on the particles less visited by the agents. 
Given the above, the pheromone quantity and its normalization within the neighborhood of $\vec{x}_i$ are written as:
\begin{align}
%\begin{equation}
F^{j}(t+1) = (1-\zeta)\cdot F^{j}(t) \enspace ,
\label{evaporationformula}\\
%\end{equation}
%\begin{equation}
\tilde{F}^j(t) = \frac{F^j(t)}{\sum\limits_{j^\prime\in \set{N}^i_r} F^{j^\prime}(t) } \enspace.
\label{pheromonenormalized}
%\end{equation}
\end{align}
Combining equations (\ref{Enormalizeformula}) and (\ref{pheromonenormalized}) allows us to define the total preference of the jump from $\vec{x}_i$ to $\vec{x}_j$ and based on that the corresponding jump probabilities. 
We provide these two quantities respectively: 
\begin{align}
%\begin{equation}
    V^{(i,j)}(t) &= (1-\kappa)\tilde{F}^j(t)+\kappa \tilde{E}^{(i,j)} \enspace ,
\label{vformula} \\
%\end{equation}
%\begin{equation}
    P(j|i,t) &= \frac{\exp(\omega V^{(i,j)}(t))}{\sum\limits_{j^\prime\in \set{N}^{(i)}_r} \exp(\omega V^{(i,j^\prime)}(t))} \enspace .
\label{pformula}
%\end{equation}
\end{align}
Here, $\kappa \in [0,1]$ is a parameter which adjusts the relative importance of the pheromone and manifold alignment terms, and  $\omega>0$ is the inverse temperature  \citep{TaghribiEtal2022}. 
The remaining hyper-parameters for this random walk are the number of agents $N_\mathrm{ants}$, the number of epochs or times the random walk is re-initiated $N_\mathrm{epochs}$, and the number of steps $N_\mathrm{steps}$ that each agent takes within an epoch. An in-depth explanation of the influence of $\kappa$ and $\omega$ on the results is provided in Appendix~\ref{sec:ParameterChange} including the recommended values for all parameters of the algorithm.

To initiate the random walk, a random starting particle is chosen such that its neighborhood is dense enough, and so within a given epoch, the agents will perform the random walk on the particles in $\mathcal{Q}$ for $N_\mathrm{steps}$ and with the jump probabilities given in equation~\ref{pformula}. 
At the end of each epoch, the multiplicity of visits to each particle is counted and the value of the pheromone quantity for these particles is updated using: %in the following manner:

\begin{equation}
    F^{j}(t) = F^{j}(t-1)+\nu (j) \gamma \enspace .
\label{pheromoneupdate}
\end{equation}
Here, $\gamma$ is the constant amount of pheromone deposited, and $\nu (j)$ is the multiplicity of visits for particle $j$. Given the defined jump probabilities, and the enforced pheromone evaporation rate, running the random walk for several epochs will allow the pheromone to accumulate along the particles aligned with the manifolds in the data set, and will also lead to the pheromone's dissipation in more scattered regions. 
It is then possible to choose a threshold for the final pheromone value that would filter out points belonging to less prominent structures.  
A conceptually similar approach to LAAT is the Monte-Carlo Physarum Machine (MCPM) algorithm \citep{BurchettEtal2020} where instead of the ant behavior, MCPC mimics the mode of growth of ``slime mold" for revealing the network of structures within the Cosmic Web. Whereas MCPC finds optimal connections between galaxies or Dark Matter halos however, LAAT highlights the particles that are aligned with structures of matter.
%MCPM has been used in \citep{BurchettEtal2020} to explore the relation between the Inter-Galactic Medium and the Cosmic Web.}

In Figure~\ref{fig:PheromoneDistribution} we demonstrate the result of running LAAT on the N-cluster simulation data of the Cosmic Web. In the left panel we present a slice with a thickness of $20$~Mpc/h from the data cube containing different cosmic structures with varying densities. 
The brighter regions correspond to places of higher density while fainter regions correspond to those of lower density and almost-empty regions such as voids. 
In the middle panel we see the same slice after running LAAT on the entire data cube. 
It is evident how the pheromone amount will, after running for several epochs, accumulate on the structures identified. 
We can also see how places of higher density such as clusters and thick filaments will accumulate more pheromone concentrations than regions of lower density.
We then bin the simulation particles within a $120 \times 120 \times 120$ grid, and plot, in logarithmic scale, the particle count versus the pheromone concentration within each grid cell. We observe that for large particle counts, there exists a linear relation between the local density and pheromone amount. However, for smaller densities, we observe a wide scatter between the two quantities. This scatter confirms that LAAT is performing more than a simple power-law transformation of the local density in order to enhance the contrast between high and low density regions. It also shows that some structures though faint, are found and highlighted by LAAT.
%This therefore demonstrates LAAT's ability to %distinguish and 
%highlight regions of varying densities within the Cosmic Web. 
As mentioned before, %previously, 
the pheromone concentration can then be used to threshold and select %as a thresholding value to select 
the particles belonging to the %se
different regions.

\subsection{EM3A}
\label{sec:EM3A} % used for referring to this section from elsewhere

We now explain %the steps with which 
how EM3A moves particles of the data set towards the central axes of the detected manifolds. 
The first step consists of defining a strategy for recognizing the manifold structure. 
Given a manifold $M$ in the data, this strategy uses the eigen-decomposition of the covariance matrix of the local neighborhoods for each particle to define the local tangent space of $M$. % to $M$ at each of these particles. 
In other words for a given point $\vec{x}_i$ of $M$, the tangent space to the manifold at that particle is given by the set of eigenvectors of the covariance of the neighborhood centered at $\vec{x}_i$. 
A random walk is then started such that the walking agents are reinforced to move data points closer to the detected manifolds by moving the particles orthogonal to their corresponding tangent space. 
Using the eigenvectors of the neighborhood centered at $\vec{x}_i$, we construct the matrix $U$ whose columns are the calculated eigenvectors and we let $\vec{\mu}$ be the %kernel 
average of $\vec{x}_i$'s neighbors. 
Using these two quantities we define the distance from the point $\vec{x}_i$ to the manifold $M$:
\begin{equation}
        \delta^{\set{M}}(i) = \Vert(I - UU^T)(\vec{\mu} - \vec{x}_i)\Vert \enspace ,
    \label{DistToManifold}
\end{equation}
where $\Vert.\Vert$ is the Euclidean norm. 
%As for t
The weights and probabilities to jump to another particle $\vec{x}_j$ in the neighborhood %, they can be 
are defined by: %correspondingly: 

\begin{align}%\begin{equation}
    w(\vec{x}_i, \vec{x}_j) &= 
      \begin{cases} 
      1-\frac{\delta^{M}(i)}{b} & \delta^{M}(i) \leq b \\
      0 & \delta^{M}(i) > b \enspace ,
      \end{cases}
    \label{EM3Aweights} \\
%\end{equation}
%\begin{equation}
    P(\vec{x}_i, \vec{x}_j) &= \frac{w(\vec{x}_i, \vec{x}_j)}{\sum\limits_{m \in N^i_r}w(\vec{x}_i, \vec{x}_m)} \enspace .
    \label{EM3Aprobs}
%\end{equation}
\end{align}
This definition for the weights encourages the agents to remain close to the manifold with the parameter $b$ chosen in such a way that $50\%$ of neighbors have non-zero weights.

In addition to the walk, the agents also move the data points: pick them and drop them. Therefore, we define the pick-up probability for the particles visited by the agents. 
Given the particle at $\vec{x}_j$, the probability for moving it closer to the manifold is defined by:
\begin{equation}
    P_\mathrm{pick}(\vec{x}_j) = \frac{1 - w(\vec{x}_j, \vec{x}_j)}{\sum\limits_{m \in N^i_r}\big(1 - w(\vec{x}_i, \vec{x}_m)\big)} \enspace.
    \label{pickprobs}
\end{equation}
This implies that the probability to be picked up increases if the particle in question is farther away from the tangent space. 
If the point is picked up, it is then moved along the complement of the tangent space with the following displacement update formula modulated by the amount of displacement $\eta>0$:
\begin{equation}
        \vec{x}^\mathrm{new}_j= \vec{x}^\mathrm{old}_j +  \eta(I - UU^T)(\vec{\mu} - \vec{x}^\mathrm{old}_i) \enspace .
    \label{EM3Aupdate}
\end{equation}
In this work, we use a specific version of EM3A termed EM3A+, in which the number of agents employed is equal to the number of particles in the data set. 
Therefore, an agent is initialized at every point in the data set, and the same steps of finding the nearest manifolds and moving the particles closer to them proceeds. 
This choice, though is more computationally expensive, eliminates any stochastic property of this algorithm and allows for more consistent results for each run. 
Furthermore, denoising of data sets using the above steps, similar to the Manifold Blurring Mean Shift (MBMS) algorithm \citep{MBMS}, is dependent upon the choice of radius for the neighborhoods of the particles. 
To limit this dependence on the radius, EM3A implements methods from evolutionary game theory by representing a range of calculated radii as evolutionary strategies and afterwards trying to compute the ``fittest" strategy between them. 
We refer the reader to \citet{CanducciEtal2022b} and \citet{MohammadiEtal2022} for a more detailed description of %this last property of 
the algorithm. 

%Refer back to them as e.g. equation~(\ref{eq:quadratic}).

\begin{figure*}
\centering
\includegraphics[width= \textwidth]{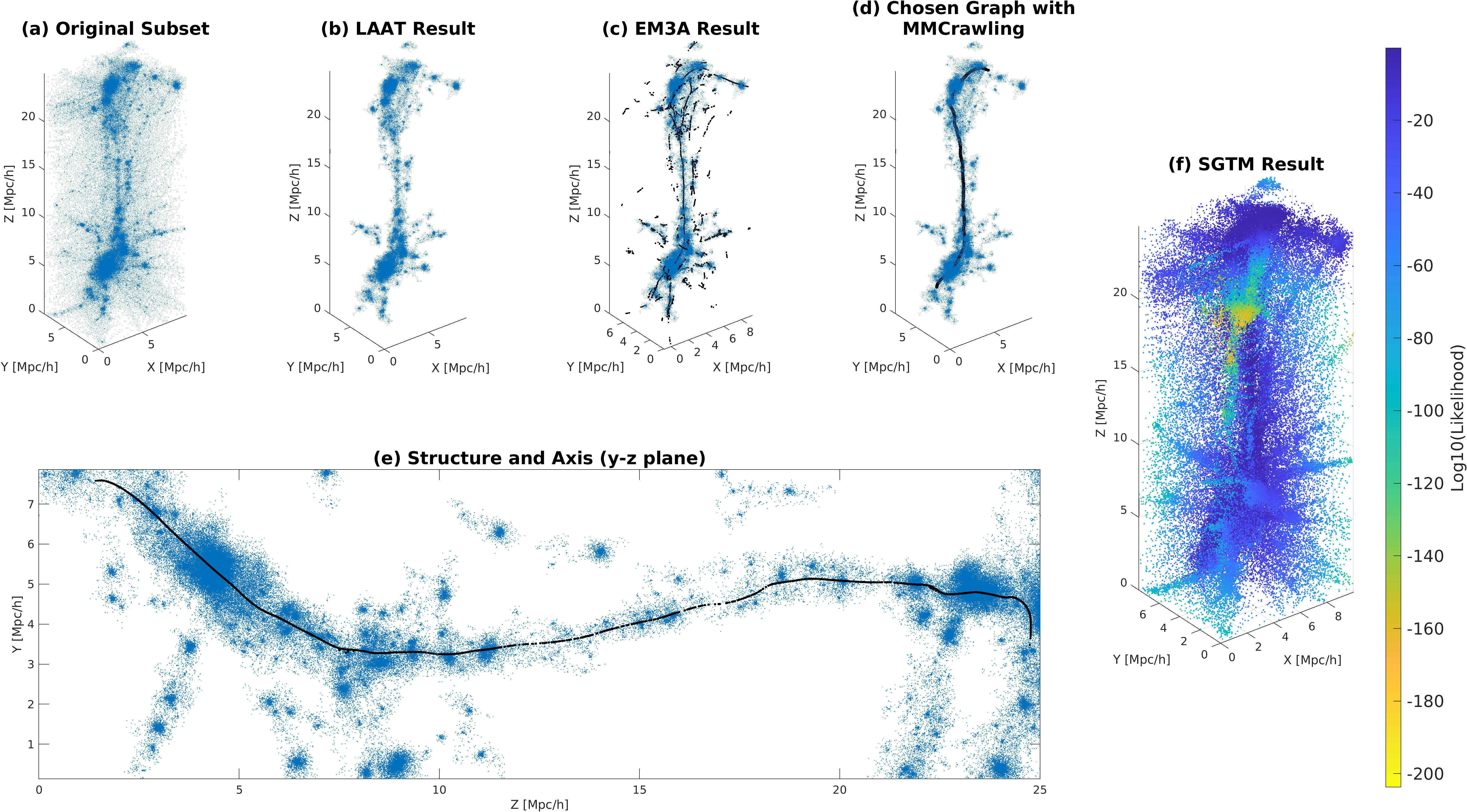}
\caption{Each panel corresponds to a step in extracting and modelling a filament within our simulation data. Panel (a) shows the original subset of the data containing the filament connected by two clusters. Panel (b) shows the main structures extracted using LAAT. Panel (c) shows the result of denoising the different detected structures using EM3A+ where we recover the central axes of all detected structures. In this step, we restrict the two clusters to be 1-dimensional instead of considering their 3-dimensional nature. Panel (d) highlights MMCrawling's graph representation of the longest recovered axis of panel (c). The isolation of this axis is also achieved using the MMCrawling algorithm. Panel (e) shows a projection of the extracted filament on the $y-z$ plane with its central axis plotted in black. Panel (f) shows the result of modeling this structure using SGTM. The color-scheme corresponds to the calculated likelihood of each particle to belong to the structure given the constructed model where regions in blue correspond to high likelihoods.}
\label{fig:Pipeline}
\end{figure*}

\subsection{Dimensionality Index}
\label{sec:DimIdx} % used for referring to this section from elsewhere

The following algorithm attributes a number to each particle in the data set specifying the dimension of the structure that the particle belongs to. 
Given a particle at $\vec{x}_i$ the dimensionality index of this particle is calculated by first evaluating the set of normalized eigenvalues of its neighborhood given by  $\tilde{\Lambda}_i = \Lambda_i/\sum_{k=1}^D \Lambda_i^k$ where $D$ we remind is the dimension of the space containing the particle ($D=3$ in the current work). 
A large eigenvalue indicates a significant eigenvector in a given direction and therefore a prominent contribution to the local dimension of the manifold. 
If we then consider the simplex with vertices centered at $\vec{e}_1 = (1, 0 ,0), \vec{e}_2 = (1/2, 1/2, 0), \vec{e}_3 = (1/3, 1/3 ,1/3)$, the normalized eigenvalues in descending order of each particle's neighborhood could be thought of as a position on the simplex. 
If the particle's corresponding simplex location falls directly on $\vec{e}_1$ then the particle's neighborhood is 1-dimensional with 100\% certainty, and similarly for $\vec{e}_2$, and $\vec{e}_3$ if the neighborhoods are 2 or 3-dimensional respectively. 

In a realistic setting however, a neighborhood's eigenvalue-spectrum will lie somewhere on the simplex between these vertices. 
Computing the dimensionality index of a particle belonging to a neighborhood then consists of measuring the geodesic distance under the Fisher metric from its corresponding position on the simplex to each of the simplex's vertices. 
The vertex closest to the position thus determines the corresponding particle's index. 
More formally, the dimensionality index $\Delta_i$ for a particle at $\vec{x}_j$ and the geodesic distance between any two points with positions $\tilde{\Lambda}_i$ and $\tilde{\Lambda}_m$ on the simplex are given respectively by: % the following:
\begin{align}
    \Delta_i &= \arg\min_j d_J(\tilde{\Lambda}_i,\vec{e}_j) \enspace , \\ d_J(\tilde{\Lambda}_\ell,\tilde{\Lambda}_m) &= 2\arccos{\left(\sum_{k=1}^{D} \sqrt{(\tilde{\Lambda}_\ell^k \cdot \tilde{\Lambda}_m^k)}\right)} \enspace.
\end{align}
Since there is no defining edge between one structure of the Cosmic Web and another, we spatially smooth the dimensionality index by adding a smoothing functionality to mimic the smoothness of the structures. In this way, we attribute similar dimensionality indices to particles in close neighborhoods.
We therefore define a smoothing Gaussian Kernel $\mathcal{K}$ between every eigen-spectrum computed $\tilde{\Lambda}_i$ and every vertex $\vec{e}_j$ of the simplex. 
This Gaussian Kernel and its normalization are given respectively:
\begin{align}
\label{eq:Kernel}
    \mathcal{K}(\tilde{\Lambda}_i;\vec{e}_j) &= \exp\left[- \frac{d_J(\tilde{\Lambda}_i,\vec{e}_j)^2}{2s^2} \right] \enspace, \\
%\end{equation}
%\begin{equation}
    \overline{\mathcal{K}}_i(j) &= \frac{\mathcal{K}(\tilde{\Lambda}_i;\vec{e}_j)}{\sum_{k=1}^{D}\mathcal{K}(\tilde{\Lambda}_i;\vec{e}_k)} \enspace.
%\end{equation}
\end{align}
In equation~\ref{eq:Kernel}, $s$ is the geodesic distance between any vertex and the circumcenter of the simplex (i.e. the point equidistant to all vertices of the simplex).
A second kernel $c$ %that 
smooths between the individual particle positions $i$, and $j$ in the data set and is given by $c(i,j) = \exp\left[-\lVert \vec{x}_i - \vec{x}_j \rVert^2 / (2 r^2)\right]$. 
This kernel %will 
defines the weights for calculating the probability to attribute a given index (represented by a vertex $j$) to a particle $i$.  
The normalized probability is then provided here: 
\begin{equation}
\label{eq:SmoothIdx_P}
    P_i(j) = \frac{\sum\limits_{\vec{x}_l \in \set{N}(\vec{x}_i,r)} c(i,l) \cdot  \overline{\mathcal{K}}_l(j)}{\sum\limits_{\vec{x}_l \in \set{N}(\vec{x}_i,r)} c(i,l)}    
    \enspace.
\end{equation}

Finally, the smoothed dimensionality index $\Delta^S_i$ attributed to a particle $i$ is the index $j$ of the simplex vertex for which particle $i$ has maximum probability $P_i(j)$:
\begin{equation}\label{eq:Final_D}
\Delta^S_i = \arg\max_j P_i(j) \enspace.
\end{equation}
In Section~\ref{sec:Analysis}, we propose to use this algorithm to partition the data between points belonging to the different structures of the Cosmic Web. 
By calculating the index of each particle in the data set, we are able to separate the data between particles belonging to clusters ($\Delta_i^S = 3$), %points belonging to 
walls ($\Delta_i^S = 2$), and %points belonging to 
filaments ($\Delta_i^S = 1$).

\subsection{MMCrawling}
\label{sec:Crawling} % used for referring to this section from elsewhere

Continuing to the remaining algorithms, MMCrawling partitions the data into separate filaments each represented by a graph (a set of vertices or nodes and edges connecting them), and a set of particles surrounding the graph. In other words, MMCrawling divides the data into a set of "skeletons" and sparse representations of the structures respectively. The method builds these sets by initiating an agent moving recursively through the data and following the steps recounted here:

%\textbf
\paragraph*{Initialization:}
We denote by $\tilde{\set{Q}}$ the resulting denoised point distribution after applying EM3A on the data set $\set{Q}$, and $\set{R}$ the set of points that have not been visited yet by MMCrawling. 
An initial random position $\vec{x}_0$ is chosen in the data to commence the walk or crawling. 
Similar to the previous algorithms, the eigenvalues and eigenvectors of the neighborhood centered at $\vec{x}_0$ are computed, and the normalized eigenvector $\hat{v}_0$ with largest corresponding eigenvalue is assumed to span the tangent space to a given manifold $M$ at $\vec{x}_0$, and is taken to be the initial direction for crawling on the manifold. 
The initial position and main eigen-direction thus allow us to create two other positions on either side of $\vec{x}_0$ and along the direction of $\hat{v}_0$. 
The two new candidate positions are the following:
\begin{equation}\label{eq:Cr_Estimates}
\vec{c}_n^{\pm} = \vec{x}_0 \pm \beta \cdot r \cdot \hat{v}_0 \enspace.
\end{equation}
Here, $r$ is the radius of the neighborhood around $\vec{x}_0$, and $\beta$ is referred to as the jump tolerance, i.e. the parameter controlling the distance between the previous positions of the graph nodes and the new ones. The parameter $\beta$ therefore, helps regulate the effect of outlier points on the eigenvalue decomposition. 
Additionally, $n$ is the index of the iteration of the algorithm. 
The steps defined so-far are not sufficient to maintain the crawling close to the manifold, and so instead of using the candidates $\vec{c}_n^{\pm}$ as the new positions visited, we select their closest neighbor in $\set{R}$ as an alternative under the condition that these new positions are still within the considered neighborhood. 
The two selected positions for the manifold representation %and the condition they should satisfy 
are therefore: 
\begin{align}
%\begin{equation}
\label{eq:ClosestN}
\vec{x}_n^\pm &= \arg\min_{\vec{x} \in \set{R}}( \|\vec{x} -  \vec{c}_n^{\pm} \|) \enspace ,\\
%\end{equation}
\intertext{and satisfy the following condition:}
%\begin{equation}
\label{eq:crawlingCondition}
&\|\vec{x}_n^\pm - \vec{x}_0\| \leq r \enspace.
%\end{equation}
\end{align}
The initial points representing the manifold so far are grouped in the set $\overline{\set{P}} = \{\vec{x}_0, \vec{x}_1^+, \vec{x}_1^-\}$, and the set containing their lower-dimensional counterparts can also be defined accordingly as $\set{P} = \{0, 1, -1\}$.
Given the particles belonging to those two sets, the remaining particles in the neighborhood are covered and therefore unnecessary, thus they are removed from  $\set{R}$ and considered as part of the sparse representation of the structure.
The points in the set $\set{P}$ are taken to be the first three nodes of the graph representation with the \emph{projected node} being connected to its \emph{projecting node} by an edge.

%\textbf
\paragraph*{MMCrawling Update}: 
After initializing the first three nodes, in every next iteration $n$, the following steps will be applied on each node identified in the preceding iteration $n - 1$. 
The manifold is first explored using the first detected direction i.e. starting with $ \vec{x}_1^+$. 
Subsequently these three steps are performed: finding the neighborhood of the particle at that position, performing eigen-value decomposition and selecting the largest eigen-direction, followed by projecting a node in that direction using~\eqref{eq:Cr_Estimates} and~\eqref{eq:ClosestN}, and finally depleting the un-selected particles from the neighborhood and adding them to the set of sparse representations.
These steps are repeated until no suitable candidates are found within the neighborhood of the last projected node. 
In that case, the end of the manifold is found, and crawling is halted in that direction. 
The same is then repeated using $\vec{x}_1^-$ until the other end of the manifold is encountered. 
Each node projected is then saved in the set $\overline{\set{P}}$, the low-dimensional set  $\set{P}$ is updated, and each parent node (projecting node) is connected by an edge to its child node (the projected node). 
Once manifold $M$'s representation is recovered, and so long as $\set{R}$ has not yet been depleted, the steps of initialization and crawling update are then repeated until all manifolds identified in the data set have been recovered and the neighborhoods of all the points have been subtracted from $\set{R}$. %The running of 
Hence the algorithm %then 
terminates once $\set{R}$ is completely depleted, or if specified, once its size reaches a given lower threshold.   

\subsection{SGTM}
\label{sec:AGTM}

Finally, the last algorithm in our toolbox is Stream-GTM or SGTM standing for \emph{Stream Generative Topographic Mapping} derived from GTM \citep{Bishopbook}, which is used for density modeling of high-dimensional noisy data sets. 
The method follows a probabilistic approach to model structures in a given data set as constrained Gaussian mixtures.
In other words, the distribution of particles forming a given noisy manifold will be attributed a likelihood to belong to the manifold since the structure will be modelled as a mixture of Gaussian probability distributions. 
In this work, the manifolds are considered to be one-dimensional, however previous work demonstrated the efficiency of this modelling technique for higher dimensional manifolds with unknown topology \citep{CanducciEtal2022a} or given spherical topology \citep{Canducci2021,TAGHRIBI2022376}.

To begin, the low-dimensional representation of a manifold previously stored in set $\set{P}$ is re-scaled so that it lies within the interval [-1, 1]. The particles corresponding to this representation are stored in set $\set{X}$ and the re-scaling is defined as follows:
\begin{equation}
    x_\ell = -1 + \frac{p_\ell - \min(\set{P})}{\max(\set{P}) - \min(\set{P})} \qquad \forall p_\ell \in \set{P}.
\end{equation}
The mapping between these centers and the centers in the original data ($\overline{\set{P}}$) can be achieved by using radial basis functions (RBFs). 
The latter are highly useful tools for approximating the basis of a given vector space of interest and the interaction terms between the basis vectors. 
They can therefore be used to map the rescaled low-dimensional centers back to the data space. 
We denote by $\phi$ the symbol for the RBFs and $y(x_l)$ the function mapping a point $x_l \in \set{X}$ to its counterpart in the data. 
Moreover, taking $\sigma$ as the %to be the 
mean distance between all adjacent centers, the mapping function $y$ and the RBF function $\phi$ between two centers $x_l, x_s \in \set{X}$ take the following form: % of the following:
\begin{align}
%\begin{equation}
\label{eq:RBF}
y_j(x_\ell) &= \sum_{i=1}^{L} w_{j\ell} \phi(x_i, x_\ell) \qquad \forall x_i \in \set{X} \enspace , \\
%\end{equation}
%\begin{equation}\label{eq:RBFs}
\phi(x_s, x_\ell) &= \exp\left[-\frac{(x_s - x_{\ell})^2}{2\sigma^2} \right] \enspace .
%\end{equation}
\end{align}
Where $L$ is the number of RBFs, in this case the size of $\set{X}$, and $w_{j\ell}$ is the weight associated to the $j-$th coordinate of the map of $x_\ell$ onto $\overline{\set{\cP}}$. 
More concretely, we define $\tilde{x}$ to be the column vector containing the points in $\set{X}$, and $\Phi(\tilde{x})$ to be the matrix with entries $\Phi_{s \ell} = \phi(x_s, x_\ell)$, and similarly for defining the weight matrix $W$. 
Thus,~\eqref{eq:RBF} can be written in its matrix form: 
\begin{equation}
    \vector{y}(\tilde{x};W) = \Phi(\tilde{x})W \enspace .
\end{equation}
The probabilistic model of a given manifold should be aligned with the structure and as explained previously can be given as a mixture model of multivariate Gaussians. 
Consequently, a probabilistic model forced to be aligned with the manifold is taken to be a flat mixture model of multivariate Gaussian distributions centered on the nodes belonging to $\overline{\set{P}}$. 
The Gaussians and the resulting mixture model are defined respectively as:
\begin{align}
%\begin{equation}
\label{eq:NoiseGTM}
p(\vec{x}|x_\ell, \matrix{\Sigma}_\ell,W)  &=  
    \frac{1}{[(2\pi )^D |\matrix{\Sigma}_\ell|]^\frac{1}{2}} \exp{\left( -\frac{\Delta\vec{x_\ell}^T \matrix{\Sigma}_\ell^{-1} \Delta\vec{x_\ell}}{2} 
 \right)}, \\
%\end{equation}
%\begin{equation}\label{eq:sum_int}
p(\vec{x}|W,\matrix{\Sigma}) &= \frac{1}{L}\sum_{\ell = 1}^{L} 
p(\vec{x}|x_\ell,\matrix{\Sigma}_\ell,W) \enspace.
%\end{equation}
\end{align}
%Where 
Here, $\Delta\vec{x_\ell} = y(x_\ell;W) - \tilde{x}$, and $\matrix{\Sigma}_\ell$ is the manifold aligned covariance matrix of the $\ell$-th Gaussian, while $\matrix{\Sigma}$ is the collection covariance matrix of all Gaussians.
A final quantity to define is the log-likelihood of the weight matrix. 
Given a configuration of the Gaussian mixture model, the log-likelihood of the weights of the mixture is defined as: % the following:
\begin{equation}\label{eq:CompLogL}
 \set{L} (W) = \sum_{n = 1}^{N} \ln \left\{\frac{1}{L}\sum_{\ell 
= 1}^{L} p(\vec{x}_n|x_\ell,\matrix{\Sigma}_\ell, W) \right\} \enspace.
\end{equation}
After this initial configuration for the multivariate Gaussians is set, the model needs to be trained in order to predict the optimal Gaussian mixture (defined by its centers, covariance matrices and the weights) that fits the manifold. 
This is therefore an optimization task that is performed by maximizing the log-likelihood defined in~\eqref{eq:CompLogL} to compute the best fitting $y_\ell$, $\matrix{\Sigma}_\ell$, and $W$. 
The training is performed using the Expectation Maximization (EM) method whose details are thoroughly outlined in \citet{Bishopbook}.

A final note is made on the number of Gaussian distributions used to model a given manifold. 
%Since the Gaussian probability distributions are initialized on the low-dimensional representation of the manifold as generated by MMCrawling, the number of Gaussians in the Gaussian Mixture of SGTM is therefore determined by this initialization. 
The centers of the Gaussians, before the EM-training, are assumed to be the positions of the graph nodes generated by MMCrawling, and the choice of initial covariance matrices which determine the size of each Gaussian is explained on page 6 of \citet{CanducciEtal2022a}. This initialization serves as a prior to the training performed by SGTM that then determines the optimal sizes and positions to model the data as a Gaussian mixture, having centers constrained to lie on a one-dimensional manifold. For MMCrawling (being an iterative procedure) it is not possible to set the number of centers a priori. Their number depends on the value of hyperparameters $r$ (radius of the particle neighborhood) and $\beta$ (jump tolerance). 
%Given the low-dimensional constraint and the local information provided by $r$, the underlying manifold is sampled, along its elongation, by a set of points linked in a chain. 
%The distance between adjacent points on the chain is always within the interval $[\beta \cdot r; ~r]$. 
The optimal number of Gaussian distributions is thus set by MMCrawling, and this number is not modified by SGTM.
%As opposed to kernel density estimation, SGTM (initialized with the points sampled via MMCrawling) is robust against overfitting and allows for an efficient generalization of the model to unobserved data.

\subsection{Method Discussion}
\label{sec:proscons}

We provide here a brief discussion of 1-DREAM's algorithms %in the hope of further clarifying 
to further clarify their intended usage. 
The methodologies explained in Sections~\ref{sec:LAAT}~through~\ref{sec:AGTM} have first been presented and their individual function analyzed extensively %published separately 
in the corresponding papers %works 
mentioned at the beginning of Section~\ref{sec:GeneralFormalism}. 
%In these works, the authors have shown how the algorithms could function individually. 
In \citet{CanducciEtal2022b}, we have shown %demonstrated 
how the methods could be coherently combined such that the output of one serves as input for another, to detect and subsequently model astronomical structures. 
The application versatility of different combinations of the algorithms have %also 
been demonstrated %in the development of 
in that contribution based on three astrophysical examples: 
the tails of a jellyfish galaxy, a cosmic filament, and the stellar streams of Omega Centauri. 
In this work, we %focus our attention on the Cosmic Web to show 
analyze in detail how 1-DREAM can be utilized in N-body simulations of the Cosmic Web. %that setting. 
LAAT highlights detected structures %the density contrast between regions 
of varying density in the simulation, and %by applying a thresholding criterion, separating 
separates these from noise using a threshold. %various regions. 
EM3A on the other hand, determines %locates 
the central axes of cosmic structures. 
DimIndex separates the data based on the dimensionality of the structures, to define which %between 
particles belong %ing 
to clusters, filaments, walls, and voids. 
MMCrawling partitions the data into a set of filament axes and a set of the particles surrounding each axis. SGTM then models the distribution of particles in each filament as a Gaussian mixture model. 
In this order, the output of each algorithm fits directly with the next. 
We thus recommend using the algorithms of 1-DREAM as a combination in the given order. %, yet 
However, we kept them as modules so one can still utilize a given algorithm or several of them individually depending on the intended usage.

%what are the pros and cons of each algorithm? 
% LAAT + Robustness - Dependence on scale / Needs prior knowledge of the data
% EM3A + Robustness - Dependence on scale / Large Complexity
% DimIndex + Smoothing - Dependence on scale
% MMCrawling + probabilistic treatment (Sectioning the data) - Dependence on scale
% SGTM + probabilistic treatment (GGM modelling) - Dependence on scale
The remaining discussion of the toolbox will briefly cover the advantages and shortcomings of each algorithm. 
The main advantage to note for the detection algorithms %of the  ``extraction" part 
of 1-DREAM, which are %consists of 
LAAT and EM3A, is the consistency of their output despite their stochastic nature. 
In other words, though the distribution of agents is initialized randomly at the start of every run, LAAT and EM3A retrieve the located structures with minimal variability in their count or nature. 
We develop and empirically substantiate %prove 
these claims further in Section~\ref{sec:Compare} and Appendix~\ref{sec:Stochasticity} respectively. %Secondly, a
An advantage of DimIndex is its ability to distinguish not one but all environments of the cosmic web. 
%This is also performed in a way that allows for 
Note also, that a smooth transitioning between the structures can be achieved by applying the local smoothing kernels. 
%Thirdly, 
Lastly, the modelling part of the 1-DREAM toolbox, embodied by MMCrawling and SGTM, allows for a statistical approach in the modelling of the detected structures. % defining probabilistic models of them. 
%has the advantange of following a statistical approach in defining models of cosmic structures. 
Since the structures of the Cosmic Web span a wide range of sizes (lengths and cross-sections), outlining particles %regions 
belonging to structures and the regions around them is not trivial. 
Thus, instead of defining a strict separation %outline 
between these regions, the sectioning of the data provided by MMCrawling followed by the probabilistic modelling of SGTM provides a probability estimate for each point to belong to a given modelled structure.
%Other advantages concerning EM3A+ is its ability to recover axes that are well aligned with the centers of cosmic filaments, and that are robust under subsampling of the data (we develop and prove these claims further in Section~\ref{sec:Compare}).

Among the 
shortcomings to keep in mind, is that LAAT and EM3A rely on a random walk which is typically applied to
%are computationally expensive algorithms, since they rely on a random walk generated within 
a large distribution of particles. 
While the implementation of LAAT allows the user to trade-off a higher memory usage to gain speed, EM3A is computationally more expensive.
Secondly, the parametrization of the algorithms requires some prior knowledge on the properties of the data (although mainly about the characteristic scale of the manifolds in the data). 
For various astrophysical settings, the values of the algorithms' parameters may need to be adjusted to fit the nature of the current particle distribution. 
However, in the case of our N-body simulations of the Cosmic Web, we have suggested the most fitting parameter settings in Table~\ref{table:parameters} of Appendix~\ref{sec:MethodAnalysis}, which showed good results for this type of data. %work best with this type of data. 
Thirdly and most importantly, all algorithms of 1-DREAM rely on a single scale approach, meaning that one set of results is produced for each choice of neighborhood radius $r$, and so the results produced by these algorithms vary with the change of that choice.
This shortcoming can be overcome with the development of calibration techniques for the neighborhood radius or modifying the algorithms to include a multi-scale implementation. We leave such explorations for future work however, and demonstrate the usage of the algorithms in their current implementation. 
%On the other hand, we provide recommended values for the neighborhood radius parameter throughout the paper, and in Table~\ref{table:parameters}.

\subsection{Demonstration on a Cosmic Subset}
\label{sec:Application}

%Figure 2
To provide a better understanding of the use of our toolbox 
%when applied on datasets of the Cosmic Web, we describe
we demonstrate and analyze here the results we obtain when running LAAT, EM3A+, MMCrawling, and SGTM on a cosmic filament with two clusters on either end. 
The applicability of the remaining Dimensionality Index algorithm to Cosmic Web data is then %demonstrated 
shown in the subsequent Section~\ref{sec:Analysis}. 
The subset of the N-cluster simulation data which we use in this section contains $\approx$ 500,000 particles and is shown in panel (a) of Figure~\ref{fig:Pipeline}. 
Many structures can be observed in this subsection including two clusters connected by a filament with an approximate length of $\approx 20$~Mpc/h as well as several other smaller filaments surrounding the larger filament and clusters. 
One can also observe that these structures are embedded within a region of lower density where the particles are more randomly and sparsely distributed.
We aim to isolate the longer filament in this subset and to measure along it the local density and particle velocities using our proposed methodologies. 

%LAAT result + parameters
To isolate the regions of higher density, we run LAAT on this small subset of the Cosmic Web and remove all particles that have accumulated the least amount of pheromone. 
To obtain this result, we set $18$ agents to run for $100$ epochs taking $12000$ steps in each epoch. The values of these parameters are chosen such that their product is 10 to 100 times the number of data set points. This ensures that every particle in the data set is at least visited once by an agent.
The radius of neighborhoods is fixed at the recommended value $r=0.5$~Mpc/h which provides a large enough scope of the structures in each neighborhood while keeping the running-time feasible.
%\footnote{
A more detailed discussion of the running-time will be 
provided in the gitlab repository of 1-DREAM. %}  
After thresholding using the minimum amount of pheromone, the particles which satisfy this condition are shown in panel (b) of Figure~\ref{fig:Pipeline}. 
We can observe that the filtered-out particles are the sparsely distributed particles surrounding the denser regions in the subset. 
One can also run LAAT on the filtered-out particles to study any remaining fainter structures that were not identified in the first run due to the presence of more dominant structures (refer to Appendix~\ref{sec:MethodAnalysis} for further details).

%EM3A result + parameters
We then attempt to find the central axes of the identified structures using EM3A+. 
We fix the neighborhood radius to be $1$~Mpc/h and allow the agents to run for $10$ epochs.
We then observe the result of moving the particles belonging to each structure orthogonal to the local tangent space of the structures at the particles' positions. 
The results we obtain are demonstrated in panel (c) of Figure~\ref{fig:Pipeline}. 
The points in blue are the initial positions of the particles which is also the output of the LAAT filtration, while the points in black show the new positions occupied by those same particles. 
One can see how the new positions trace the central axes of the many structures identified within this subset.

%Crawling result + parameters
Since our aim is to look at the main filament connecting the two clusters, we use MMCrawling to generate a set of graph representations of all the axes produced, and choose the longest for subsequent modeling. %the graph corresponding to the structure of our choice. 
In this part we use 1~Mpc/h for the size of the neighborhood radius as smaller sizes misrepresent this structure, and $\beta = 0.6$ such that $\beta \cdot r = 0.6$~Mpc/h is the projecting distance used for adding a node. 
Note that for filaments that show higher curvature, smaller values of $\beta$ are needed to trace their correct shape. 
The chosen graph resulting from this procedure is shown in panel (d) of Figure~\ref{fig:Pipeline}. 
A different viewing profile of the filament and chosen axis from EM3A+ are shown in panel (e) of the same figure. 
%SGTM result
Using the recovered axis, we now use SGTM to create a multivariate Gaussian distribution on each of the axis nodes and commence the training to compute the centers, covariance matrices, and weights of the mixture of Gaussians that best model the distribution of particles forming the studied structure. 
This probabilistic model provides for each of these particles a likelihood for belonging to the given structure. 
The particles shown in panel (f) of Figure~\ref{fig:Pipeline} are color-coded according to their likelihood to belong to the model of the filament. 
We observe how particles closer to the detected filament have a much higher likelihood than particles farther away.

\begin{figure*}
\centering
%\begin{subfigure}[b]{0.81\textwidth}
\parbox{0.03\textwidth}{\rotatebox{90}{Slice 29}}\parbox{0.8\textwidth}{
   \includegraphics[width=\linewidth]{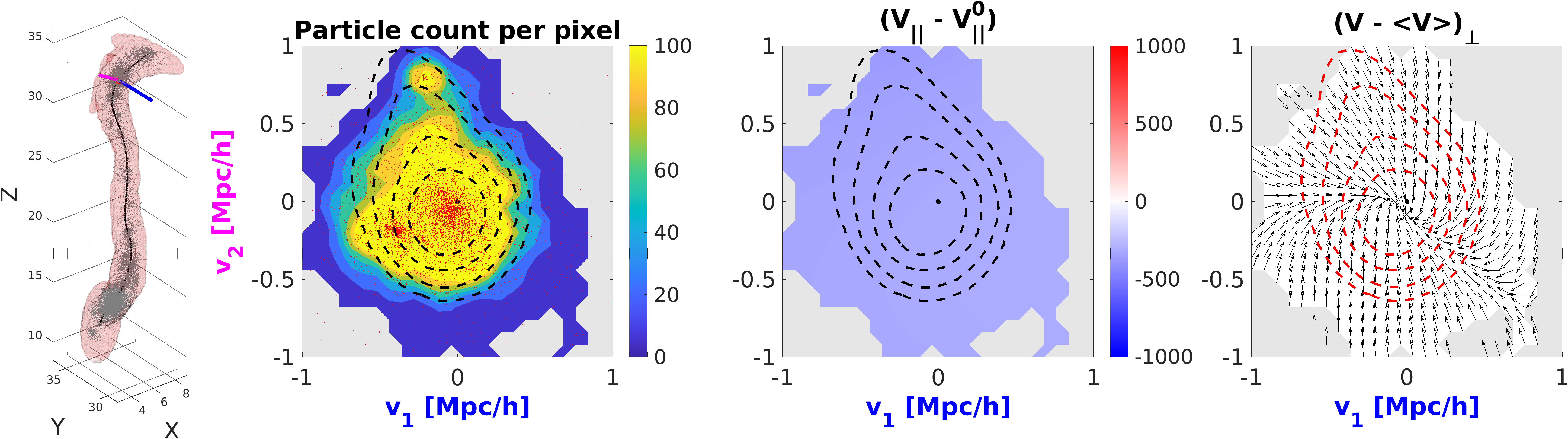}}\\%   \label{fig:Slice0029} 
%\end{subfigure}
%\begin{subfigure}[b]{0.81\textwidth}
\parbox{0.03\textwidth}{\rotatebox{90}{Slice 46}}\parbox{0.8\textwidth}{
   \includegraphics[width=1\linewidth]{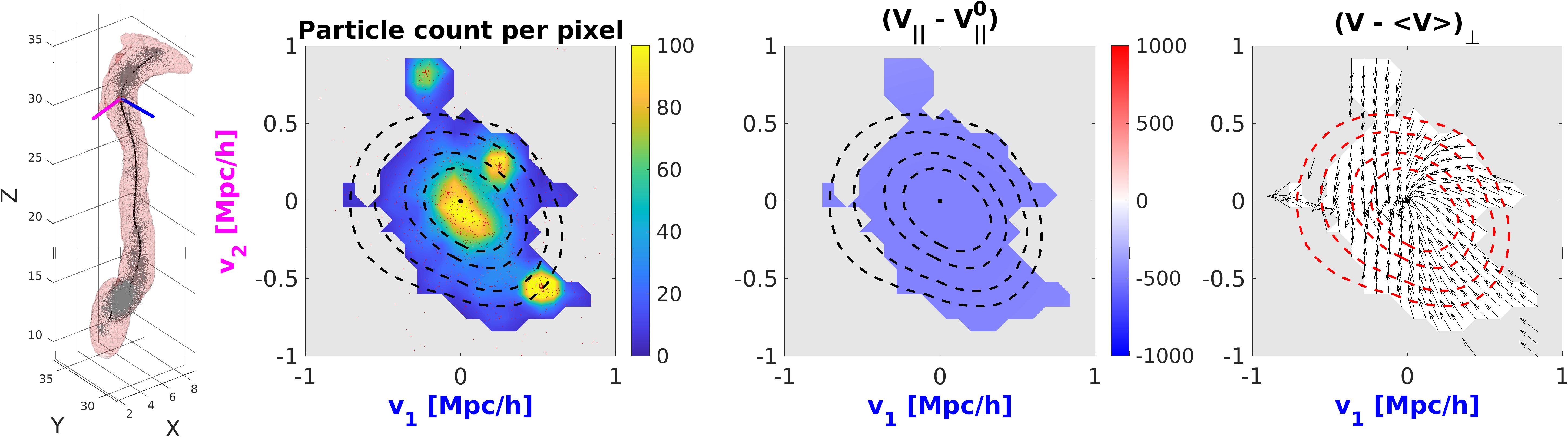}}\\%   \label{fig:Slice0046}
%\end{subfigure}
%\begin{subfigure}[b]{0.81\textwidth}
\parbox{0.03\textwidth}{\rotatebox{90}{Slice 90}}\parbox{0.8\textwidth}{
   \includegraphics[width=1\linewidth]{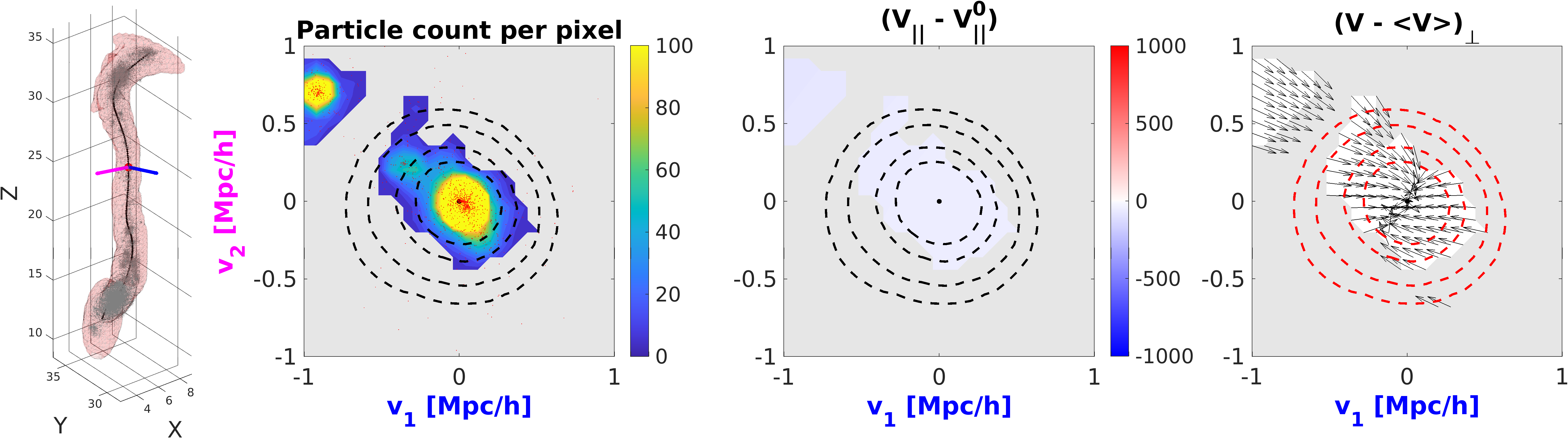}}\\%   \label{fig:Slice0090}
%\end{subfigure}
%\begin{subfigure}[b]{0.81\textwidth}
\parbox{0.03\textwidth}{\rotatebox{90}{Slice 136}}\parbox{0.8\textwidth}{
   \includegraphics[width=1\linewidth]{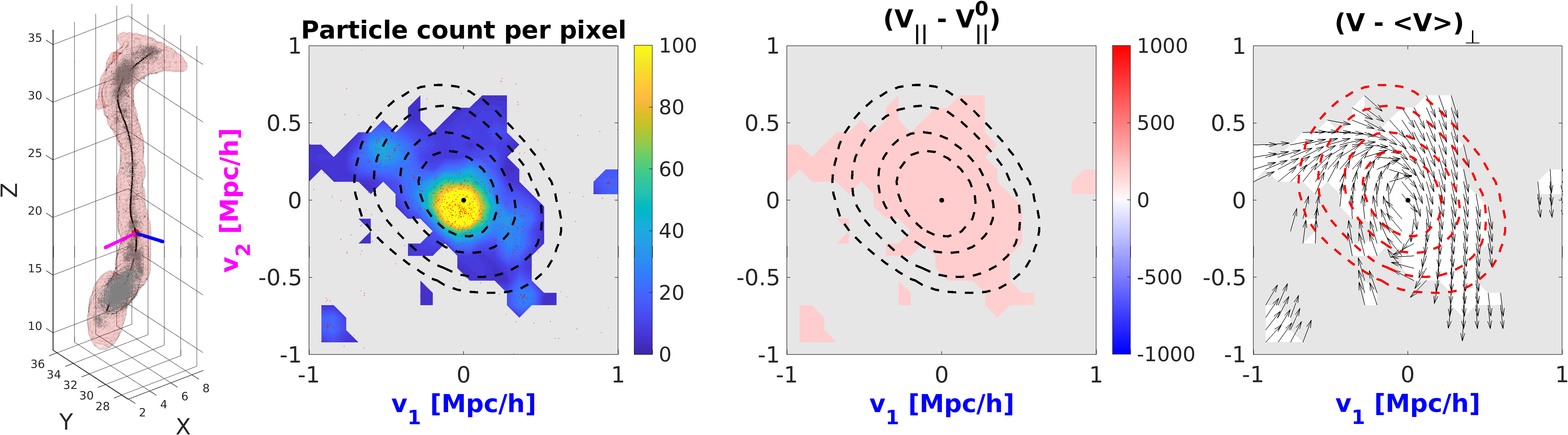}}\\%   \label{fig:Slice0140}
%\end{subfigure}
%\begin{subfigure}[b]{0.81\textwidth}
\parbox{0.03\textwidth}{\rotatebox{90}{Slice 180}}\parbox{0.8\textwidth}{
   \includegraphics[width=1\linewidth]{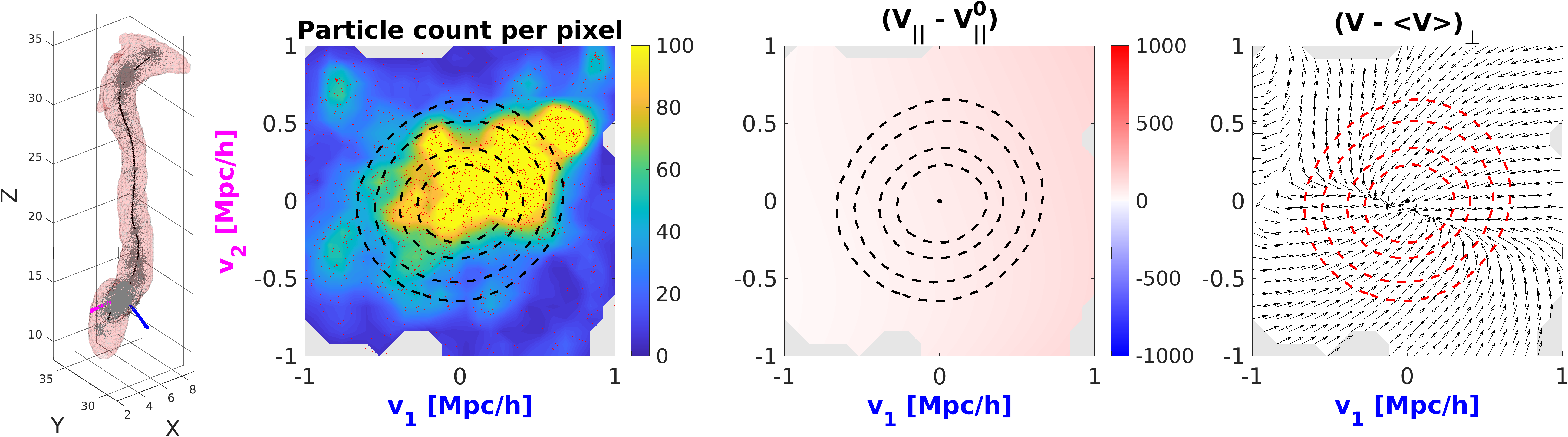}}\\%   \label{fig:Slice0180}
%\end{subfigure}
\caption{Each row corresponds to a snapshot of moving on the structure's axis and the measurements made in cross-sectional planes or slices perpendicular to it. The calculation of the thickness of the slice is further described in the text. The left-most column shows the structure with the maximum-likelihood iso-surface surrounding it in pink. The remaining columns from left to right are the measurements made in each cross-section of: the particle count per pixel, velocity parallel to the axis, and velocity perpendicular to the axis. The concentric dotted curves correspond to iso-contours of the probabilistic model of the structure where particles within the inner contours have a higher probability to belong to the structure than particles within the outer contours. The gray regions are masked areas where the number of particles per pixel is too low to provide precise statistics.}  
%The remaining columns from left to right are the measurements made in each cross-section of: 
%The left-most column shows the filament with the maximum-likelihood iso-surface surrounding it.
%The remaining columns from left to right are the measurements made in each cross-section of:

%The left panel shows a filament and its maximum-likelihood iso-surface followed by three cross-sectional  visualizations or slices perpendicular to its axis showing:
%the particle count per pixel, velocity parallel to the axis, and velocity perpendicular to the axis.
%Each row depicts a snapshot moving along the filament.

%Each row corresponds to a snapshot of moving on the structure's axis and the measurements made in cross-sectional planes perpendicular to it. 
%The left-most column shows the filament with the maximum-likelihood iso-surface surrounding it.
%The remaining columns from left to right are the measurements made in each cross-section of:

\label{fig:orthogonalSlices}
\end{figure*}
%Cross-section plot
Making use of the results obtained so far, we now move along the axis connecting the centers determined by SGTM, and attempt to measure the local density and velocity of the particles forming the structure. 
To increase the sampling of this axis, we apply a simple cubic spline interpolation. 
We also define two ortho-normal directions $v_1$ and $v_2$ at each location of the axis which span the orthogonal plane at that particular location. 
The two vectors are plotted in blue and magenta in the first column of Figure~\ref{fig:orthogonalSlices}. 
This plane along with a thickness scaled according to the distance between the individual nodes of the axis defines a cross-section, and moving along the axis allows us to access all cross-sections of the structure. 
Each row in Figure~\ref{fig:orthogonalSlices} corresponds to the measurements of the local particle density (second column), velocity parallel to the axis (third column), and velocity perpendicular to the axis (fourth column) for each location shown on the structure in the first column. 
The dashed circles correspond to the likelihood iso-contours computed by the probabilistic model. Regions inside the inner iso-contours have a larger likelihood to belong to the structure than regions within the outer iso-contours. 
Additionally, the grey areas correspond to masked regions where the number of particles in each grid element is less that 5 and hence, not enough to draw meaningful statistics. 
The masking is performed so that we focus on regions that are better populated and so have more reliable measurements.

The cross-sections displaying the local density confirm that the distribution of matter in a filament is not completely uniform. 
The concentration of matter increases as we approach each cluster connected by the filament as demonstrated by the larger number of particles within the iso-contours. 
We observe that the cross-sectional shape of the filament can vary along the filament as well. 
Part of this could be explained by the existing nearby filaments that are not considered in this particular application. 
Regarding the velocities, the colors in the second column correspond to the average parallel velocity in each grid cell of the cross-section. 
One can observe how starting from the top of the structure when moving downward, the color switches from blue to red indicating the switch in the direction of motion of the particles.
This is in accord with the theory that matter is continuously pulled from filaments towards the clusters by passing the saddle point where the flow reverses direction when one of the clusters becomes the greater attractor \citep{KraljicEtal2019}. 
We also point out that the largest average velocity signified by the darker colors are in the second and fourth row of the figure. 
This shows that not only is the motion of particles directed towards the clusters, but also this motion is accelerating as the particles get closer. 
This velocity decreases again within the clusters as is expected given that now we moved from the region where material falls into the cluster into the region where material is falling both in and out of the cluster at the same time. 
Finally, we inspect the motion of particles perpendicular to the structure and observe that this motion tends to be primarily directed towards the axis showing that not only is there a flow of matter towards the clusters, but also a flow of matter from around the filament, towards it \citep{CodisEtal2012, WangEtal2014,LaigleEtal2015, KraljicEtal2019}.

\begin{figure*}\centering
\includegraphics[width=0.9\linewidth]{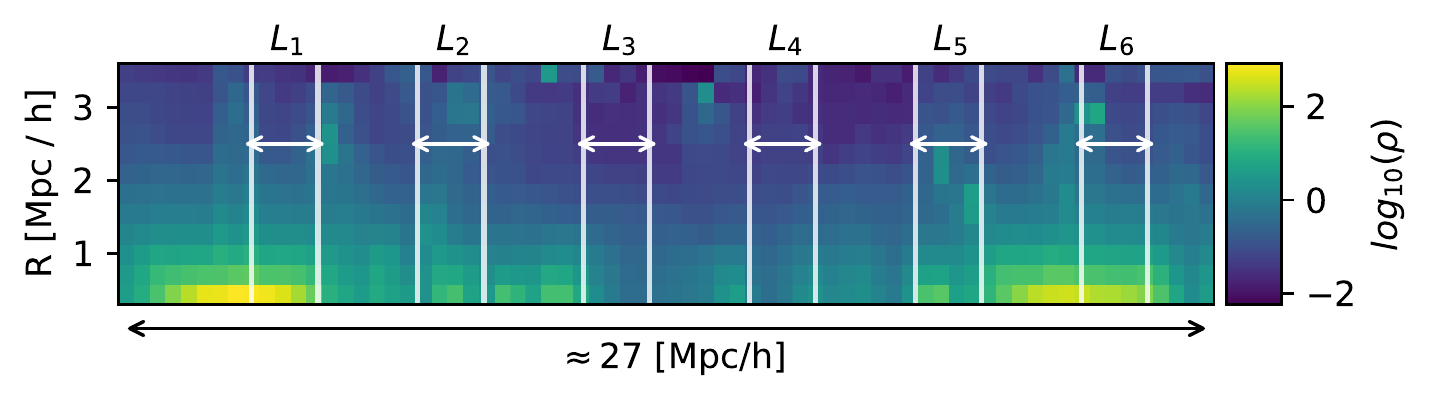}\\
%\caption{%We show a d
%Density biplot of the same structure presented in Figures~\ref{fig:Pipeline} and~\ref{fig:orthogonalSlices}. 
%The construction of biplots is detailed in \citet{CanducciEtalSubmitted}. 
%The $x$-axis corresponds to the graph axis of the structure with an estimated length of $\approx 27.4$~Mpc/h. 
%The $y$-axis corresponds to the radial direction away from the structure's center. 
%The regions labeled $L_1$ through $L_6$ correspond to windows from which we evaluate radial density profiles in Figure~\ref{fig:DensityProfiles}.}
%\end{figure}
%\begin{figure*}
%\centering
\includegraphics[width=0.8\linewidth]{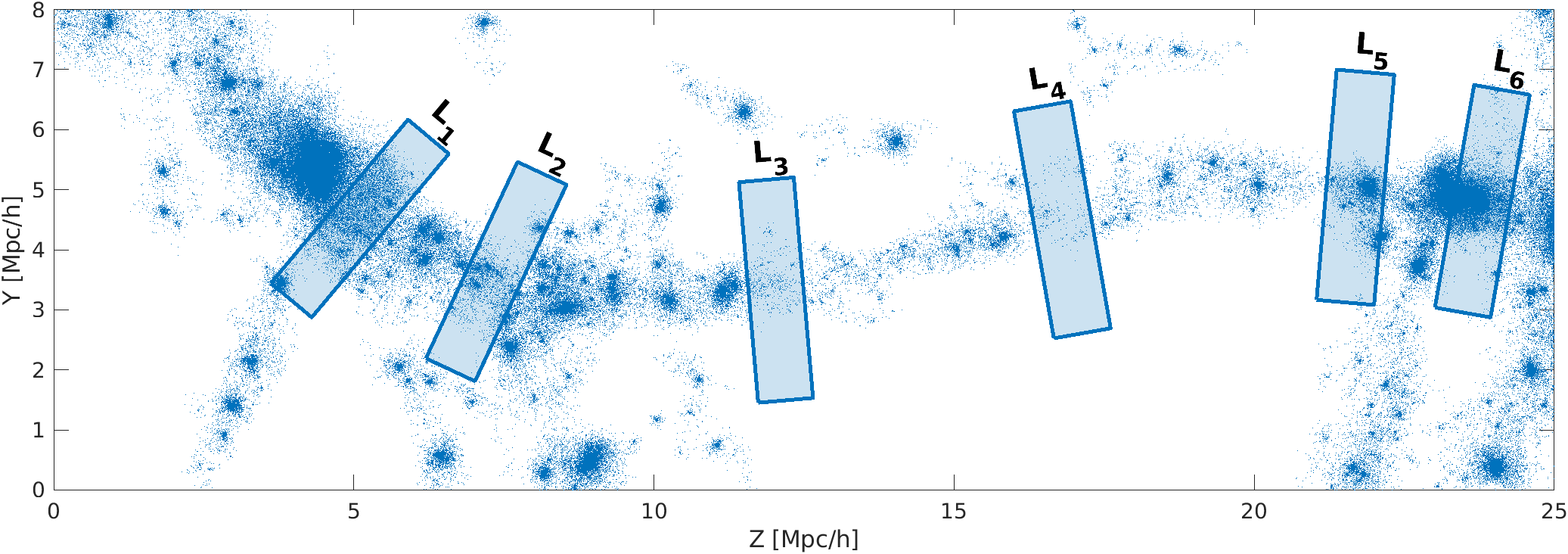}
\caption{
Top: Density biplot of the same structure presented in Figures~\ref{fig:Pipeline} and~\ref{fig:orthogonalSlices}. 
%The construction of biplots is detailed in \citet{CanducciEtalSubmitted}. 
The $x$-axis corresponds to the graph axis of the structure with an estimated length of $\approx 27$~Mpc/h. 
The $y$-axis corresponds to the radial direction away from the structure's center. The regions labeled $L_1$ through $L_6$ correspond to the windows from which we evaluate radial density profiles in Figure~\ref{fig:DensityProfiles}. Bottom: we show the location of the windows on the physical length of the structure. 
%We demonstrate the regions of interest $L_1$ through $L6$ shown previously in Figure~\ref{fig:Biplot}. 
The regions considered within the biplot span both sides of the structure symmetrically with respect to the constructed graph axis, and are orthogonal to it. 
%We show this figure to clarify where on the structure are the windows of interest taken.
}
\label{fig:Biplot}
%\label{fig:Lpositions}
\end{figure*}
%Biplot and profiles
Another method for studying the properties of the structure at hand using the results provided by our procedures is to look at the structure's radial density profiles. 
Again, we start with the trained axis recovered by SGTM, but in this case, we define equidistantly-spaced concentric cylinders centered on the nodes, with a length equal to the distance separating adjacent nodes. 
We then look at the particle counts within each cylinder so that the properties of the modelled structure, such as its local density, can be studied radially from the centers of the cylinders, and longitudinally along the lengths of the cylinders.
%\petra{such that moving radially away from the cylinder's center allows for studying properties of the modelled structure in the radial direction. }
%such that moving concentrically outwards allows for radial calculations, and moving along the structure allows for longitudinal calculations. 
We present in the top panel of Figure~\ref{fig:Biplot} the resulting 2D density profile where the $x$-axis is the central trained axis and the $y$-axis is the radial distance away from the axis. 
%One can see 
The large densities $\rho$ are contained within clusters and in the regions close to them, and %how sparse 
the filament becomes sparse close to the center. 
The length of the entire structure shown on the $x$-axis is the result of summing the lengths of the segments connecting the nodes of the axis.

\begin{figure}
\centering
%\begin{subfigure}[t]{\columnwidth}
%\centering
\includegraphics[width=1\linewidth]{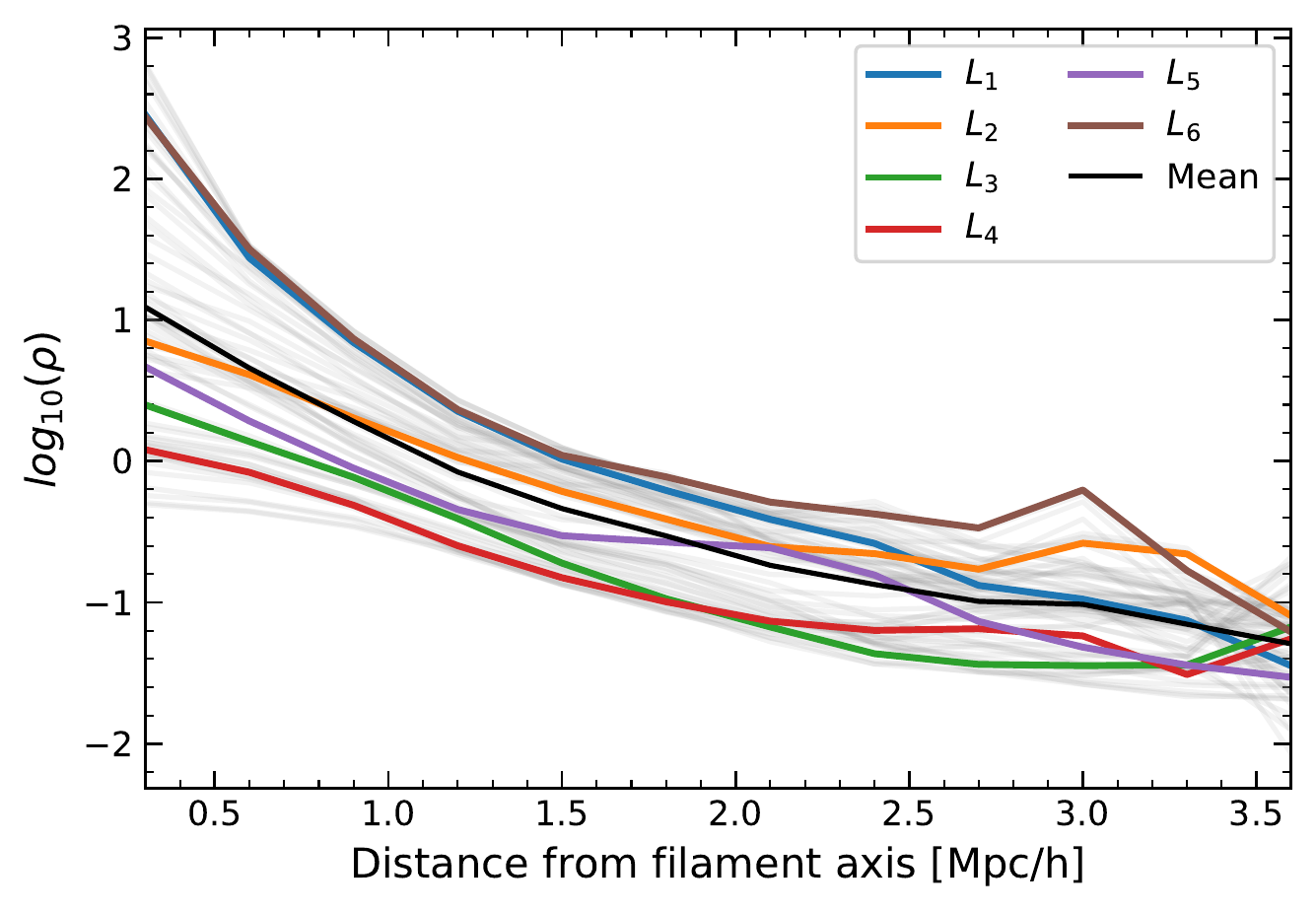} 
%\caption{}
%\end{subfigure}%
\\%~ 
%\begin{subfigure}[t]{\columnwidth}
%\centering
\includegraphics[width=1\linewidth]{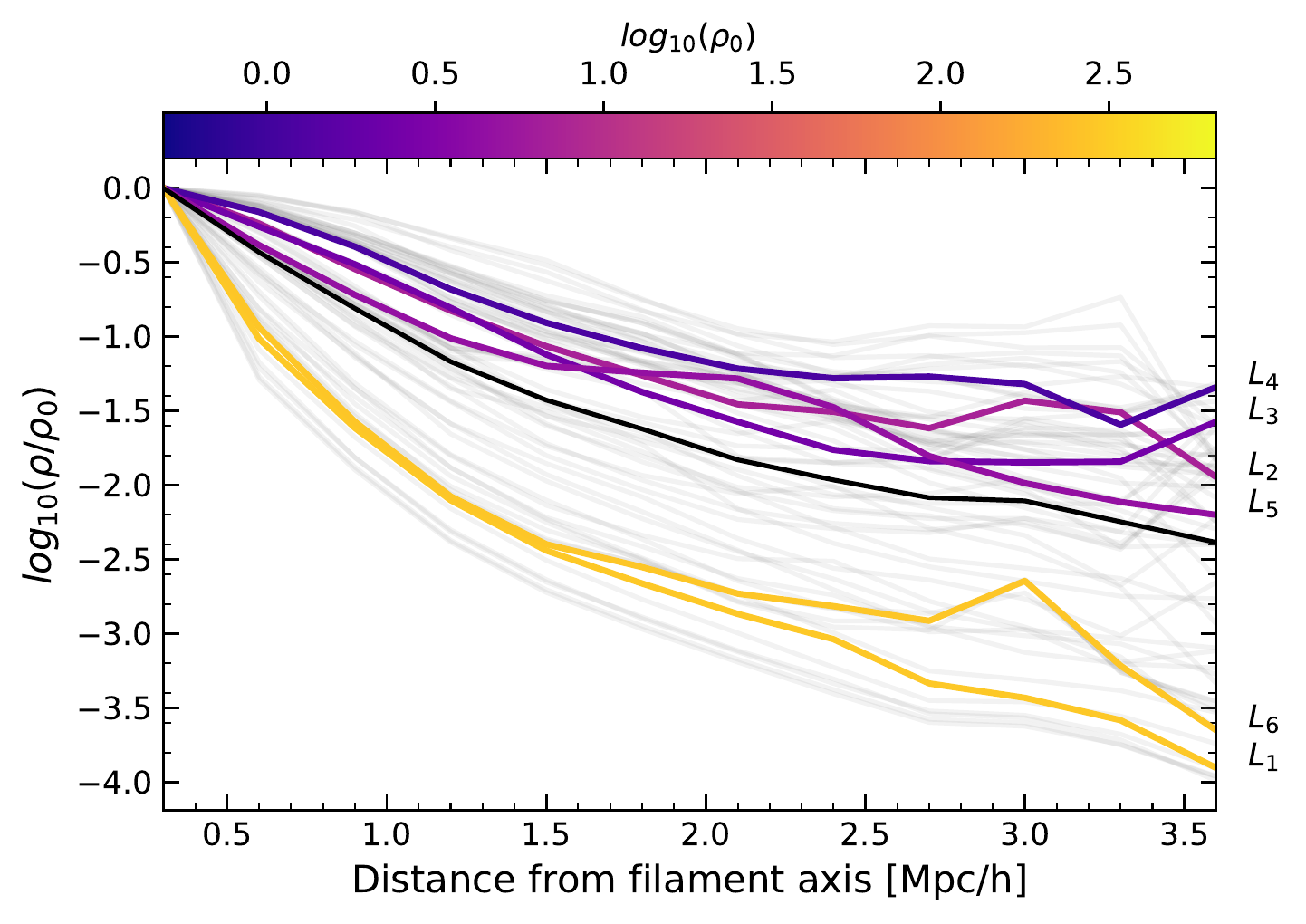}
%\caption{}
%\end{subfigure}
\caption{%Panel (a) 
The top panel shows the average density in the windows specified in Figure~\ref{fig:Biplot}, starting from the structure's axis and moving radially outward. 
The grey lines represent the radial density profiles along the entire filament in windows of size 5 pixels ($\approx2$~Mpc/h). 
The average of all the profiles calculated is shown in black. 
The bottom panel %Panel (b) 
shows the same profiles after normalizing by the density of the region closest to the middle axis for each specified window. %The same applies to the profiles shown in grey. 
The average of all profiles is again shown in black.}
\label{fig:DensityProfiles}
\end{figure}

To study the radial density profiles in more detail, we plot at each location on the axis, the variation of particle density in units of particles$/$Mpc$^3$ as we move radially outwards. 
The results are shown by the grey lines in Figure~\ref{fig:DensityProfiles}, top %left 
panel. 
We choose six windows labeled $L_1$ through $L_6$ and compute the average density in the direction of the radial axis within these specified windows.
For better illustration of where these windows are approximately located on the original length of the structure, we show the windows $L_1$ through $L_6$ as the rectangular regions (cylindrical in 3D) in Figure~\ref{fig:Biplot}. %\ref{fig:Lpositions}
The corresponding profiles are shown with the colored lines in Figure~\ref{fig:DensityProfiles}. 
$L_1$ and $L_6$ show the density profile within a part of the upper and lower clusters respectively, while the rest focuses on places on the filament in between. 
The black line is the mean of all radial density profiles for this particular structure. 
One can observe the monotonically decreasing nature of the density as we move radially outward from the filament from $R=0$ as far as $R=2.5$~Mpc/h, after-which given the increase in the scattering of the particles, the profiles may reach the mean density of the universe or even enter an under-dense region such as a void.

One way to separate the profiles that start from comparatively high densities such as the profiles within $L_1$ and $L_6$ is to normalize each radial density profile by the central density $\rho_0$ measured closest to the axis at each given location on the structure. 
We display the result of this procedure in Figure~\ref{fig:DensityProfiles}, bottom %right 
panel, where the color in each profile in the selected windows is now representative of the normalizing factor $\rho_0$. 
The black line is the average of all radial profiles shown in grey. 
We can see how the more concentrated regions $L_1$ and $L_6$ contain a higher central density, and their profiles are steeper compared to the profiles on different parts of the filament. 
We can also see how within the same filament, one can obtain regions of varying densities which reiterates the results of Figure~\ref{fig:orthogonalSlices}.

%Figures and tables should be placed at logical positions in the text. Don't
%worry about the exact layout, which will be handled by the publishers.
%Figures are referred to as e.g. Fig.~\ref{fig:example_figure}, and tables as
%e.g. Table~\ref{tab:example_table}.
%\begin{figure}
%	% To include a figure from a file named example.*
%	% Allowable file formats are eps or ps if compiling using latex
%	% or pdf, png, jpg if compiling using pdflatex
%	\includegraphics[width=\columnwidth]{example}
 %   \caption{This is an example figure. Captions appear below each figure.
%	Give enough detail for the reader to understand what they're looking at,
%	but leave detailed discussion to the main body of the text.}
%   \label{fig:example_figure}
%\end{figure}
%\begin{table}
%	\centering
%	\caption{This is an example table. Captions appear above each table.
%	Remember to define the quantities, symbols and units used.}
%	\label{tab:example_table}
%	\begin{tabular}{lccr} % four columns, alignment for each
%		\hline
%		A & B & C & D\\
%		\hline
%		1 & 2 & 3 & 4\\
%		2 & 4 & 6 & 8\\
%		3 & 5 & 7 & 9\\
%		\hline
%	\end{tabular}
%\end{table}

%------------------------------%

%---------------------------%
\section{Comparison Between DimIndex and  Other Cosmic Web Tracing Algorithms}
\label{sec:Analysis} % used for referring to this section from elsewhere
%---------------------------%

The work of \citet{LibeskindEtal2018} has provided data sets containing Dark Matter particles and corresponding halo distributions generated as described in Section~\ref{sec:data}. 
The effort was conducted in an attempt to provide a quantitative basis for the comparison of results generated by several Cosmic Web tracing methods. 
The mentioned results study the ability of classifying the particles/halos in the data set between the different structures of the Cosmic Web. 
Since our proposal for the Dimensionality Index algorithm (hereafter DimIndex) separates data points between those belonging to 1, 2, and 3 dimensional structures, we explore here the possibility of using this methodology to classify the data given by \citet{LibeskindEtal2018} between filaments, walls, and clusters, respectively. We apply our analysis on $5\%$ of the particle data set provided (amounting to $\approx 7$ million particles) to maintain a feasible time and memory usage, and follow the analysis steps suggested in \citet{LibeskindEtal2018}. 
This will allow for the comparison of our cosmic structure classification method with other current methods in the literature.

Since DimIndex is able to distinguish between particles belonging to the three possible dimensional structures, we still need to perform a step that picks out the particles belonging to voids which cannot be assigned a dimension with our current formalism.
The filtration method we apply in pursuit of that goal consists of first denoising the data set using EM3A+, i.e. we first require that the points move closer to the central axes of the structures they respectively belong to. 
Since regions inhabited by any of the three structures are denser than regions enclosed by voids, we expect EM3A+ to enhance that density contrast, and so to provide a better outline between regions belonging to clusters and those belonging to voids. 
We refer to the unaltered data as the \emph{original} data set, %We call the \emph{original} data set that which is unaltered, 
and the one resulting from applying EM3A+ as the \emph{denoised} data set. 
Therefore, to filter out the points belonging to the voids, we fix a radius $r$ and consider the neighborhoods with that radius centered around the points belonging to both the original and the denoised data sets. 
If a given particle lies far from any structure, then we expect that the neighborhoods of that point in both data sets to be sparsely populated. 
Therefore, aside from the definition given by DimIndex to particles belonging to the filaments, walls, and clusters of the Cosmic Web, our definition for particles belonging to voids is: the particles whose neighborhoods in \emph{both} the original and denoised data sets have a smaller number of points than a chosen threshold $\tau > 0$ \citep{CanducciEtal2022a}. 
In this work, we fix $\tau = 5$ particles. 
With this definition in mind, we first filter out the particles belonging to voids, and then run DimIndex on the remaining particles to partition the data set between the other environments. 
We therefore note that EM3A+ has only been used in this case as a step to construct a filtration technique that distinguishes particles belonging to very sparse regions such as voids. 
The main comparison with the algorithms stated in \citet{LibeskindEtal2018} however, is performed with the results provided by the DimIndex algorithm.

Our results depend critically on the choice of neighborhood radius set by DimIndex around the particles, since choosing a larger radius than is fitting for the current data will include undesirable particles from neighboring structures. Meanwhile, choosing a smaller radius than needed would leave out particles that could increase/decrease the influence of a certain eigen-direction of particle distribution; in both cases therefore, the local dimensionality of the structure will be inaccurately calculated. 
We thus perform our analysis using different neighborhood radii namely $0.25, 0.5, 0.75$, and $1$~Mpc/h and present the results for each case.    

\begin{figure*}
\centering
\includegraphics[width=0.9\textwidth]{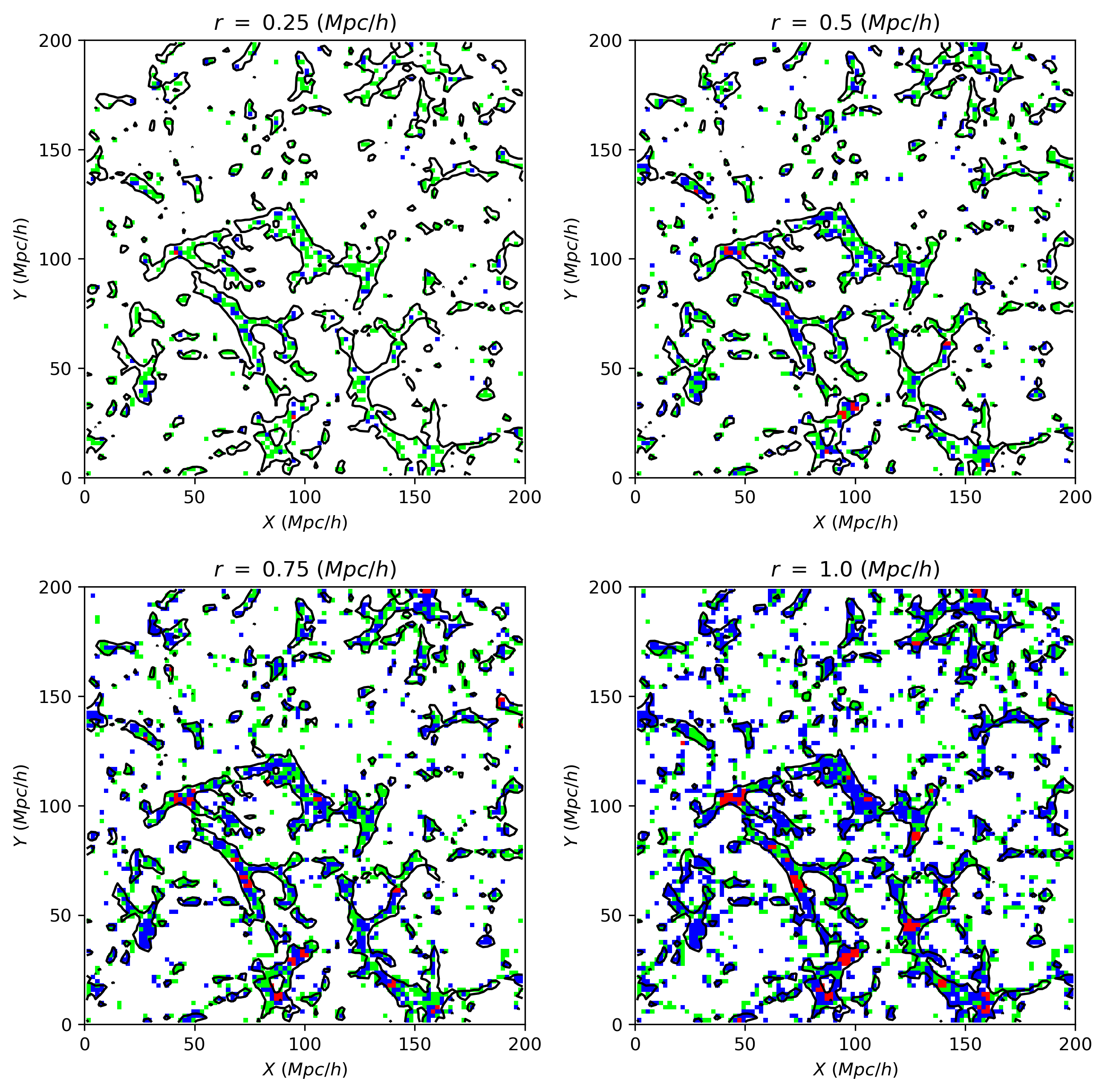}
\caption{Results of using DimIndex to classify the Dark Matter particles in the public data set of \citet{LibeskindEtal2018} between structures of the Cosmic Web. 
We present our results on the same $2$~Mpc/h thick slice of the data cube considered in the mentioned work. 
The black lines represent the $\delta=0$ contours and the colors red, blue, and green correspond to squares classified as belonging to clusters, filaments, and walls respectively. Each panel shows the results obtained for the %considered 
neighborhood radius values of % for DimIndex (
$0.25$, $0.5$, $0.75$, and $1$~Mpc/h. %). 
The results of this classification on the entire data cube is further analyzed in Figures~\ref{fig:AllPDF}, \ref{fig:AllMassFunctions}, and~\ref{fig:AllFractions}.}
\label{fig:VisualCompare}
\end{figure*}

The particles in the data cube are binned within a $100 \times 100 \times 100$ sized box therefore giving a $2$~Mpc/h length for the side of each grid cell of the box, and we attempt to assign an index to each cell. 
The dimensionality indices of all particles within a cell are averaged so that the index attributed to each cell in the cube is the calculated average rounded to the nearest integer value. 
In Figure~\ref{fig:VisualCompare} we present the result of our classification when using the mentioned radii by visualizing a slice of $2$~Mpc/h thickness from the entire data cube.
Squares in blue, green, and red represent the regions belonging to filaments, walls, and clusters respectively. 
We observe that, as expected, the algorithm classifies the particles within the over-density contours between a mixture of the three different structures, and the major amount of space remaining is attributed to voids. 
As for the effect of the choice of radius, we observe that for the smallest chosen value of $r$, namely $r=0.25$~Mpc/h, the majority of the cells are attributed to walls, then filaments, and almost no cells in the slice are classified as belonging to clusters. This is because the radius is too small to capture the 3-dimensional nature of the distribution of particles in the neighborhoods especially that the typical scales of clusters is of the order of $1$~Mpc/h.
With increasing radius, we observe that less points are classified as walls and more as filaments and clusters. 
With the largest radius considered $r=1$~Mpc/h, we observe that the regions classified as clusters can be seen more easily. 
We explain these changes with the fact that for small radii, the neighborhoods will be of smaller size, and so will be occupied by a fewer number of particles. 
This in turn will show up as an increase in the number of particles filtered out i.e. classified as belonging to voids. 
Additionally, taking a very small radius acts as a zoomed in perspective of the structures, and so the results will be less telling of the properties of the local manifold, and more of the total distribution of particles within individual neighborhoods. 
As we increase the radius, which acts as a larger scale perspective, the particles in each neighborhood will be better representative of the local dimensionality of the structure, and so we observe that filaments and walls will be detected more adeptly. 
We note that for much larger radii, it is possible to run again into the problem of falsely estimating the local dimensionality since in this scenario, multiple manifolds can fall within the same neighborhood and bias the eigen-direction estimation. 

%the larger the radius of the neighborhoods is, the more points and possibly small structures could be included withing that neighborhood. Therefore, when performing the local eigenvalue decomposition, the result will be less telling of the properties of the local manifold, and more of the total distribution of points within the region that are isotropically distributed and so will lead to comparable weights for all eigendirections. In other words, the points in that region will be more likely to be classified as a 3-dimensional object. 

\begin{figure*}
\centering
\includegraphics[width=\textwidth]{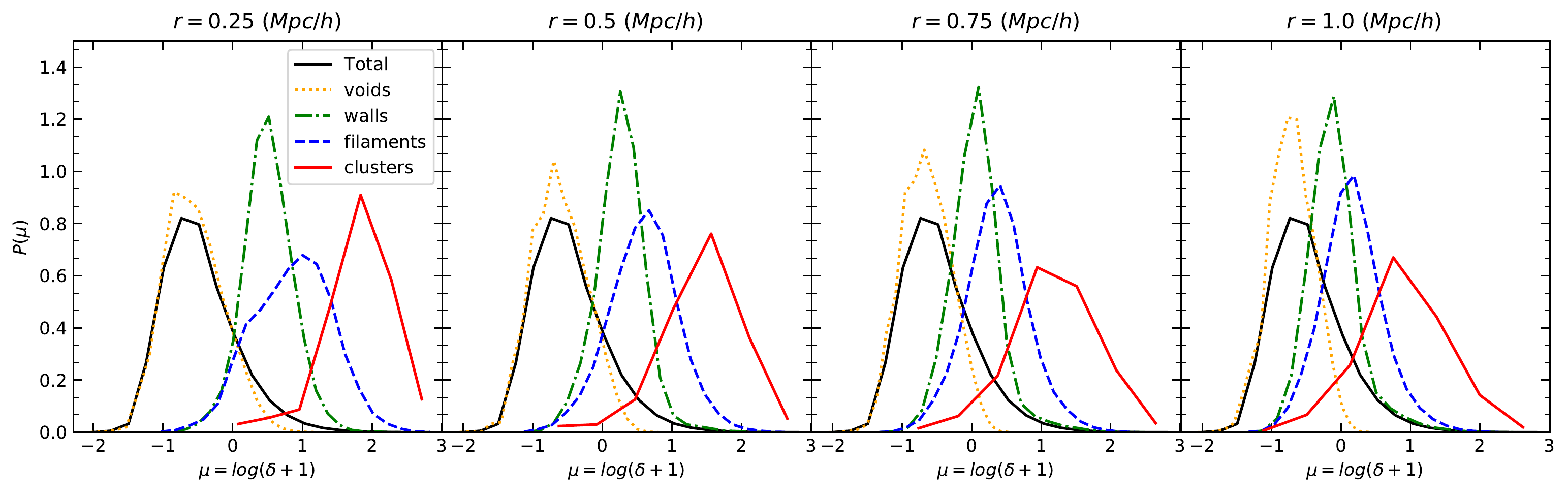}
\caption{We show the Probability Distribution Function (PDF) of the density contrast $1 + \delta$ as a function of the Cosmic Web environment (clusters, filaments, walls, and voids) obtained from the classification performed with DimIndex. 
In black is the PDF of $1 + \delta$ for the entire simulation cube. 
Each panel portrays the results for the specified neighborhood radius $r$. 
Note that all PDFs are normalized to unit area.}
\label{fig:AllPDF}
\end{figure*}

\begin{figure*}
\centering
\includegraphics[width=2\columnwidth]{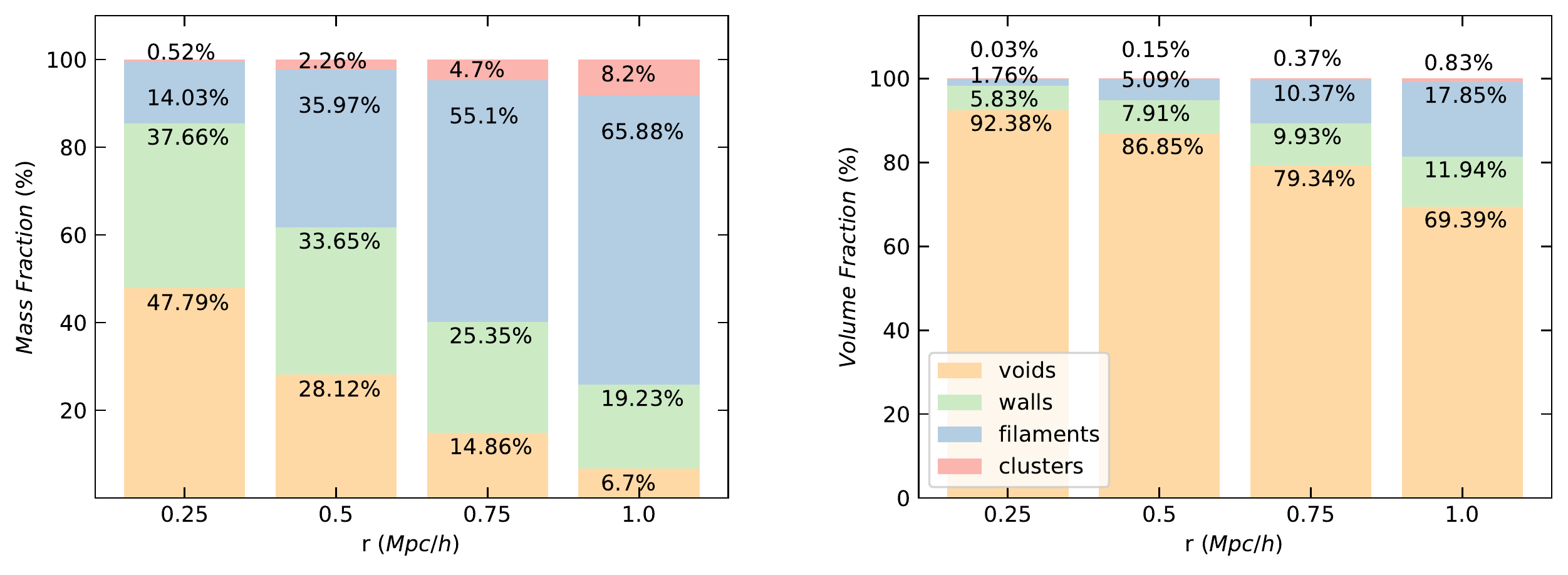}
\caption{We show the mass filling fraction (left panel) and volume filling fractions (right panel) for the environments classified with DimIndex. 
Each bar represents our results for an assumed neighborhood radius, and the size of the colored regions corresponds to the fraction of the entire bar. 
The percentage fractions for each environment is provided as well. 
The method of computing these quantities is detailed in the text.}
\label{fig:AllFractions}
\end{figure*}

\begin{figure*}
\centering
\includegraphics[width=\textwidth]{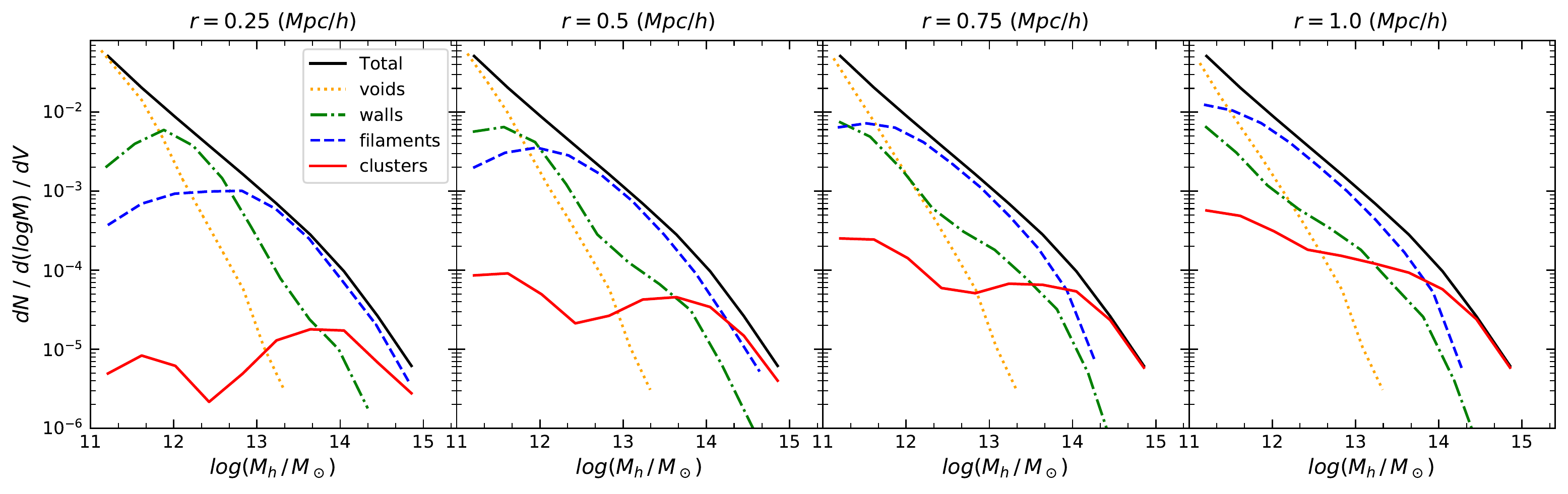}
\caption{We show the cumulative halo mass function of all halos within the data set of \citet{LibeskindEtal2018} as a function of the Cosmic Web environment (clusters, filaments, walls, and voids) as obtained from the classification performed with DimIndex. 
In black is the total halo mass function, and each panel portrays the results for the considered neighborhood radius $r$.}
\label{fig:AllMassFunctions}
\end{figure*}

We study our results more quantitatively %studied 
by first plotting the density probability distribution functions (PDFs) of the classified particles. 
The result of this analysis is shown in Figure~\ref{fig:AllPDF} for the different neighborhood radii where $\delta = \rho/\overline{\rho}$ is the number density or particle count in each cell of the gridded data divided by the mean density of the Universe ($\overline{\rho} = 512^3 / 100^3$). 
We observe that in the cases of all considered radii, clusters lie in overdense regions which hold a wide range of environmental densities (3 orders of magnitude) while the most under-dense regions are attributed to voids. 
Filaments and walls on the other hand, occupy the regions between $\mu = -1$ and $\mu = 2$ where $\mu = \log_{10}(\delta + 1)$. 
While the PDF of walls is located almost equally between the over-dense and under-dense regions, the PDF of filaments occupies a larger portion in the over-dense side. 
Such results are to be expected as discussed in \citet{CautunEtal2014}. 
With respect to the changes we see as a function of the increase of neighborhood radius $r$ (each result demonstrated by the individual panels), we observe similar trends as what we described in Section~\ref{sec:Analysis} for Figure~\ref{fig:VisualCompare}. 
We therefore, discuss a calibration method to adapt the neighborhood radius parameter at the end of this section.

%However, we see that our algorithm recovers very similar distributions for filaments and walls whereas one would expect that filaments would occupy ranges of higher density than walls. From this we can conclude that for the choice of parameters used, Dimensionality Index has a hard time differentiating between filaments and walls. Additionally with increase of radius, we observe that the distributions of filaments, walls, and clusters become very similar. This again indicates that high values for $r$ will lead to unrealistic results and that more physically motivated outcomes could be achieved with a suitable smaller radius. 
Other interesting quantities to calculate and that serve to analyze the results of our classification are the mass and volume filling fractions attributed to the different Cosmic Web structures. 
Similar to what has been done in \citet{LibeskindEtal2018}, the mass fraction is calculated by summing up the number of particles in all the cells of the cube with the same index. 
This quantity is then normalized by dividing by the total number of particles in the simulation. 
On the other hand, the volume fraction is calculated by counting all the volume elements with the same index and dividing by the total number of volume elements.
Figure~\ref{fig:AllFractions} demonstrates the results of these calculations. 
For small radii we observe that the mass is concentrated in voids ($47.8\%$), which is a very large fraction compared to measurements provided by previous studies and methods.
This concentration is greatly improved as we increase the radius until we see that only $6.7\%$ of the Universe's mass is concentrated in voids when employing $r=1$~Mpc/h. 
The mass fraction in walls similarly decreases from $37.7\%$ to $19.2\%$ while the mass in filaments and clusters increases from $14\%$ to $65.9\%$ and $0.5\%$ to $8.2\%$ respectively.
As for the volume filling fraction, we observe that in all cases of considered radii, voids take up the largest volume fraction of the Universe ($92.4 \%$ for $r=0.25$~Mpc/h to $69.4\%$ for $r=1$~Mpc/h) and the remaining space is distributed between the rest of the environments.
We observe that the volume occupied by clusters remains very small (between $0.03\%$ and $0.8\%$). 
For the more realistic results provided using the largest chosen radius, we note a $17.9\%$ volume filling fraction in filaments, and $11.9\%$ in walls. Using Table~2 in \citet{LibeskindEtal2018}, we can conclude that the results of our best-case scenario ($r=1$~Mpc/h) are comparable to the mass and volume fractions calculated by the following algorithms: 
V-web \citep{HoffmanEtala2012}, CLASSIC \citep{Kitaura&Angulo2012}, NEXUS+ \citep{CautunEtal2013}, MMF-2 \citep{AragonCalvoEtal2007}, ORIGAMI \citep{Origami1} and MSWA \citep{Ramachandra&Shandarin2015}.

%Radius $0.25$~Mpc/h shows that the mass fraction of the Universe is concentrated in voids which of course is illogical. We attribute this result to the filtering scheme we have used since the smaller the radius, the sparser the neighborhoods would be compared to the threshold $\tau$, and so the more likely that the points will be considered to be far from any nearby manifold, i.e. more likely to belong to voids. 

Finally, we take a look at the halo distribution by plotting the halo mass function to see how the mass of halos is distributed between the different structures according to our classifications. We note that no separation between central and satellite halos is attempted and so we expect the halos to occupy a wide range of masses.
We first provide each halo an index corresponding to the dimensionality of the structure it belongs to. 
This is performed by binning the provided halo positions within the grid previously defined, and attributing to each halo, the index given to the respective cell it is found in.
The masses of the halos are also provided in the \citet{LibeskindEtal2018} data set and so these masses of halos classified as belonging to either voids, walls, filaments or clusters are used to plot the cumulative halo mass function for each of these environments. 
This procedure is repeated for all chosen values of the neighborhood radius. 
We illustrate our results in the different panels of Figure~\ref{fig:AllMassFunctions}. 
%What can be observed in all radius cases is that the most massive halos are located in clusters and less massive halos can be found in the halos and walls followed by the voids being host to the least massive halos. If we look at the curve for filaments and walls, we observe that these structures also contain a large fraction of the halos and that they are mostly dominated by those with low mass as opposed to the clusters which are dominated by halos of larger masses ($>5.10^{14} M_{\odot}$). 
For voids, we observe that they are dominated by the least massive of halos with a cut-off at halos with masses larger than  $10^{13.5} M_{\odot}$. 
With respect to the rest of the environments, looking at the changes in the different panels i.e. changes with larger neighborhood radius, we observe similar trends as was apparent in the previous figures discussed in this section. 
For our best case scenario demonstrated by the right-most panel of Figure~\ref{fig:AllMassFunctions}, we observe mass functions similar to what is documented in the literature: we see that the most massive halos ($ \geq 10^{14.5} M_{\odot}$) are found solely in clusters, and the least massive halos ($ \leq 10^{11.5} M_{\odot}$) are located predominantly in voids. 
The mass range in between is occupied by halos that are classified as belonging to filaments mostly and to walls and clusters secondly.

%As the radius is increased from $0.25$ to $1$~Mpc/h, we see that less halos are getting classified in the voids and subsequently more of them will be distributed between the rest of the structures. A bigger radius will tend to envelope a larger number of points and so the points in the neighborhood will most likely be attributed to 3-dimensional distributions, followed by two dimensional distributions, and finally the least likely would be to find only a filamentary distribution of points in such a large radius. The logic can clearly be seen in the lower right panel of Figure~\ref{fig:AllMassFunctions} where the largest fraction of halos and the most massive ones are located in clusters, then walls, and a smaller number of halos is found to be contained within filaments. 

In comparison to the methods discussed in \citet{LibeskindEtal2018}, 1-DREAM's DimIndex is able to separate the environments of the Cosmic Web and provide results within the ranges predicted by most of those methods namely %particularly 
V-web, NEXUS+, MMF-2, ORIGAMI, and MSWA. 
The analysis we provide in Figures~\ref{fig:AllPDF} to~\ref{fig:AllMassFunctions} can be easily juxtaposed with the figures portrayed in \citet{LibeskindEtal2018} to compare the different classifications. %An advantage DimIndex provides over methods such as 
In contrast to DisPerSE \citep{Disperse1} and Spineweb \citep{AragonCalvoEtal2010} 
%is that unlike those methods, 
DimIndex can identify clusters and not just walls and filaments. 
The case is similar for algorithms that can only identify filaments such as Bisous \citep{TempelEtal2016}, FINE \citep{Gonzalez&Padilla2010}, and MST \citep{AlpaslanEtal2014}.
The second point to discuss is that it is necessary to choose a reasonable value for the neighborhood radius in order to obtain physically realistic outcomes of our implementation.
Regarding the choice of this parameter, it is possible to implement calibration techniques to find its preferred value. 
One suggestion for such a calibration is to refer to observational surveys, and look at measurable quantities performed on Cosmic Web data. 
One example is the work of \citet{TempelEtal2014} who produced a catalogue of filaments from the SDSS along with their distribution of lengths. 
It is therefore possible to take the same selection of data, and calibrate our radius parameter to give a similar distribution of filament lengths. 
This would be possible given the capabilities of 1-DREAM's MMCrawling algorithm to construct graph representations of detected filaments in a data set and to calculate their individual lengths. 
This calibration is left for future developments of the toolbox.
%We leave this calibration proposition however for future developments of this toolbox. 

%----------------------------------------%
\section{More Detailed Comparisons with the Disperse Code}
\label{sec:Compare} % used for referring to this section from elsewhere
%----------------------------------------%

\begin{figure*}
\centering
%\begin{subfigure}[b]{\textwidth}
 \includegraphics[width=1\linewidth]{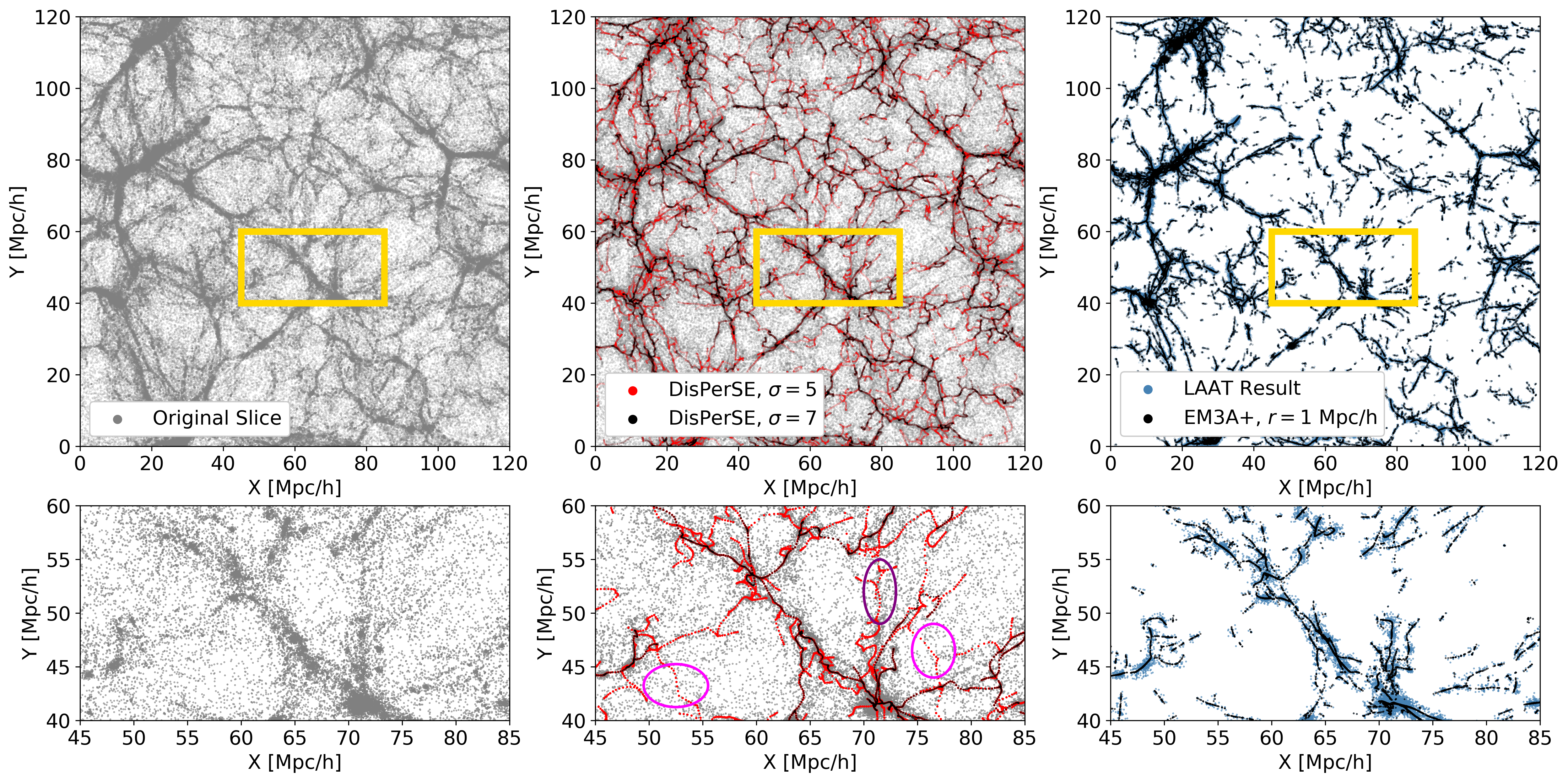}
%   \caption{}
%   \label{fig:em3aSlice}
%\end{subfigure}
%\begin{subfigure}[b]{\textwidth}
%   \caption{}
%   \label{fig:disperseSlice}
%\end{subfigure}
\caption{Comparison between EM3A+ results and DisPerSE results when applied on a slice of thickness $10$~Mpc/h of the simulation data. The top row shows the whole slice in the $x-y$ plane, while the bottom row is a zoom-in region indicated by the yellow rectangle in the top row. The left column displays the original data in gray
%, and the points selected by a run of LAAT in light-blue, where each represents the input dataset for DisPerSE and EM3A+ respectively.
The middle column displays the results of DisPerSE for two persistence ratios $\sigma=7$ (black) and $\sigma=5$ (red) when applied on the original data. The selected regions in purple and magenta are further discussed in the text. The right column displays the results of EM3A+ (black) when applied on the particles filtered out by LAAT indicated by the blue particles.}

\label{fig:SliceComparison}
\end{figure*}

%introduced methodologies. and furthermore compare them to previously developed techniques for the detection and tracing of structures in Cosmic Web data sets. 
Since our toolbox's EM3A+ and the publicly available code DisPerSE\footnote{\url{http://www2.iap.fr/users/sousbie/web/html/indexd41d.html}} \citep{Disperse1} have a similar function of tracing cosmic web filaments, we compare the two algorithms in this section. 
DisPerSE is a widely used method particularly helpful in analyzing the filamentary network of the Cosmic Web in reliance on several topological concepts such as Delaunay Field Tesselation Estimation \citep[DTFE]{Schaap&Weygaert2000, Weygaert&Schaap2009,Cautun&Weygaert2011}, 
and Discrete Morse Theory \citep{Forman1998, Gyulassy2008}.
%\textcolor{red}{One advantage that 1-DREAM has is its optimization in terms of computational complexity. 
%Unlike DisPerSE or Nexus+ \citep{CautunEtal2013}, which run on the underlying fields of the data (density fields for DisPerSE and density/velocity/shear fields for Nexus+), our algorithms run directly on the particle distributions which significantly reduces the complexity of the computations thus making our tools computationally cheap in comparison.} 
%A shortcoming of this property however is the large number of parameters needed to initialize the algorithms.
%As discussed in Section~\ref{sec:Analysis}, the results are robust against changes of a large number of these parameters and the rest could be fine-tuned by keeping in mind their respective effects on the results (also explained in Section~\ref{sec:Analysis}). 
%looking into a multi-scale implementation of our toolbox would be a worthwhile attempt. 
%Description of Disperse and what it does
%Talk about what other works find out as well. 
%Furthermore, since DisPerSE is a well-known and publicly available method for analyzing data sets of the Cosmic Web, we highlight \petra{the advantages of EM3A+} by providing an example comparison with it in addition to the general comparison with the \citet{LibeskindEtal2018} data set. 
It runs on the continuous density field created by applying DTFE on the set of point cloud data. 
It then evaluates the positions in the field where the gradient vanishes, i.e. the critical points, and then classifies those points as local maxima, minima, or saddle points in reliance on the Hessian matrix evaluated over the field. 
Following the flow of the density gradient, DisPerSE creates connections between the identified critical points which induces the tesselation of the field into regions belonging to manifolds of varying properties. 
From these manifolds, DisPerSE is able to identify the regions belonging to walls and to filaments. 
Finally Persistence Homology \citep{EdelsbrunnerEtal2002} is used to filter out insignificant structures.
%as a filter for structures that are found to be not significant. 

In works such as \citet{TREX} and \citet{TaghribiEtal2022}, a comparison between the algorithms presented in either works and DisPerSE is performed by running each methodology on the whole data cube, and inspecting their results in tracing filamentary structures on a chosen slice from the entire data set. 
We perform a similar analysis in this work when comparing EM3A+ and DisPerSE. 
For a fair comparison, we use the recommended procedure to run both algorithms, which is to use the original data as input for DisPerSE, and the result of LAAT filtration as input for EM3A+. 
We take a slice of thickness $10$~Mpc/h from the N-cluster simulation data and run LAAT to extract the prominent structures contained in it.  
The original slice is shown in gray in the top left panel of Figure~\ref{fig:SliceComparison} and the particles extracted by LAAT are shown in blue in the top right panel of the same figure.
%The original slice and the particles extracted by LAAT are shown in grey and blue in the top left panel of Figure~\ref{fig:SliceComparison} respectively.
We then apply DisPerSE on the original slice using two values for the persistence ratio $\sigma$ and display the results in the top middle panel of Figure~\ref{fig:SliceComparison}: $\sigma=7$ in black and $\sigma=5$ in red. In the top right panel of Figure~\ref{fig:SliceComparison} we apply EM3A+ using a neighborhood radius $r=1$~Mpc/h on the particles extracted by LAAT. The result of EM3A+ is shown in black. The bottom panels represent a zoom-in plot of the area encompassed by the yellow rectangle in the corresponding panels above them.

In the results provided by DisPerSE, we see that high $\sigma$ values trace the largest and densest structures in the slice but miss out on smaller structures such as the filament outlined by the purple ellipse. 
When using smaller values for $\sigma$, we observe that the fainter structures can be recovered at the cost of detecting many unclear structures that are not visibly present in the data such as the bridges encircled by the magenta ellipses. 
When performing statistical studies of the properties of galaxies as a function of their distance to the Cosmic Web structures, such possibly fake detections may create a bias in the results. On the other hand, we observe that EM3A+ has a much lower chance of producing false positive tracings as its purpose is to move particles towards the center of the closest detected structure. This makes it more reliable to use in studies with statistical natures.

\begin{figure*}
\centering
\includegraphics[width=\textwidth]{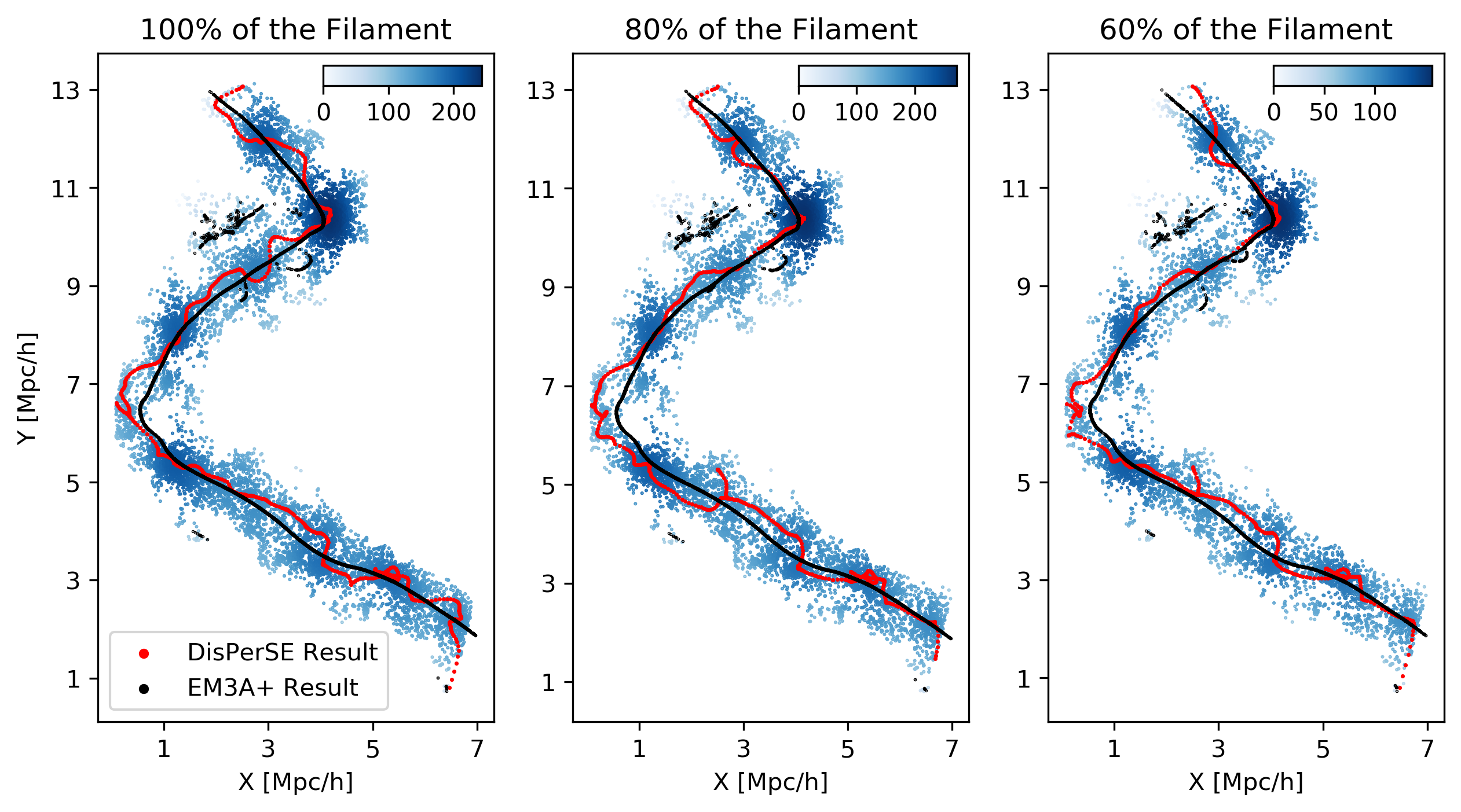}
\caption{Comparison between the axis retrieved by DisPerSE (red) and that retrieved by EM3A+ (black) on a filament extracted from the N-cluster simulation data. 
This experiment is performed on the full unsampled filament (taking $100\%$ of its particles), and after randomly sampling $80\%$ and $60\%$ of the filament's particles. 
The filament itself is shown as a number density plot in blue projected onto the $x-y$ plane. 
Darker blue regions correspond to areas of high density, in contrast with light blue regions. 
The three panels show the axes obtained for these three cases respectively.}
\label{fig:SimpleResults}
\end{figure*}

In addition to this comparison, we apply both EM3A+ and DisPerSE on a random filament extracted from the N-cluster simulation data, and assess the abilities of both algorithms to trace the middle axis of the detected filament. 
The extracted filament is shown in the left panel of Figure~\ref{fig:SimpleResults} where we present the projection of the position of its particles along the $x-y$ plane. 
Each particle is color-coded according to its local density where darker blue areas are denser than lighter regions of the filament. 
The same set of points making up the filament is provided for both EM3A+ and DisPerSE, and optimal parameters were chosen for either algorithm. 
For EM3A+, we use a neighborhood radius of $1$~Mpc/h and run for $10$~epochs, while for DisPerSE, we choose a high persistence ratio of $\sigma = 7$ and the smoothing parameter set to $10$. We select this value of $\sigma$ since any higher value leads to tracing a portion of the filament only while missing the rest.
The smoothing parameter allows for averaging the position of the Delaunay vertices 10 times to smooth the retrieved axis. 
The results from both EM3A+ and DisPerSE are shown as the black and red lines respectively in Figure~\ref{fig:SimpleResults}. 
The immediate result we see is that both algorithms are able to detect the general shape of the structure well. 
However, the axis resulting from DisPerSE shows several twists and turns that follow the areas of higher density rather than remain close to the middle of the filament. 
This behaviour is expected given the reliance of DisPerSE on density field estimation for the creation of the axis vertices. 
On the other hand, since EM3A+ relies on the estimated distance to the detected manifold to move the particles closer towards it, this acts as a density-independent approach for tracing the central axis of the structure. 
We observe as a result that the axis represented in black tends to stay in the middle of the filament and move along it more directly, without winding as much as the red axis does.

Furthermore, we discuss the reliability of the results when reducing the number of particles in the simulation by randomly sampling $80\%$ and $60\%$ of the filament's particles. 
Similar to how the left panel of Figure~\ref{fig:SimpleResults} presents the results when considering the initial un-sampled filament, the middle panel and right panels demonstrate the results after considering $80\%$ and $60\%$ of the filament's particles respectively. 
Similar outcomes are observed as discussed for the $100\%$ case with respect to the evenness of the created axis. 
We can see that with different samplings of the filament, the twists and bends in the axis traced by DisPerSE do not stay in the same place. 
This behaviour is not observed for the case of EM3A+ for the same reasons  previously discussed in this section. 

\begin{figure*}
\centering
\includegraphics[width=0.83\textwidth]{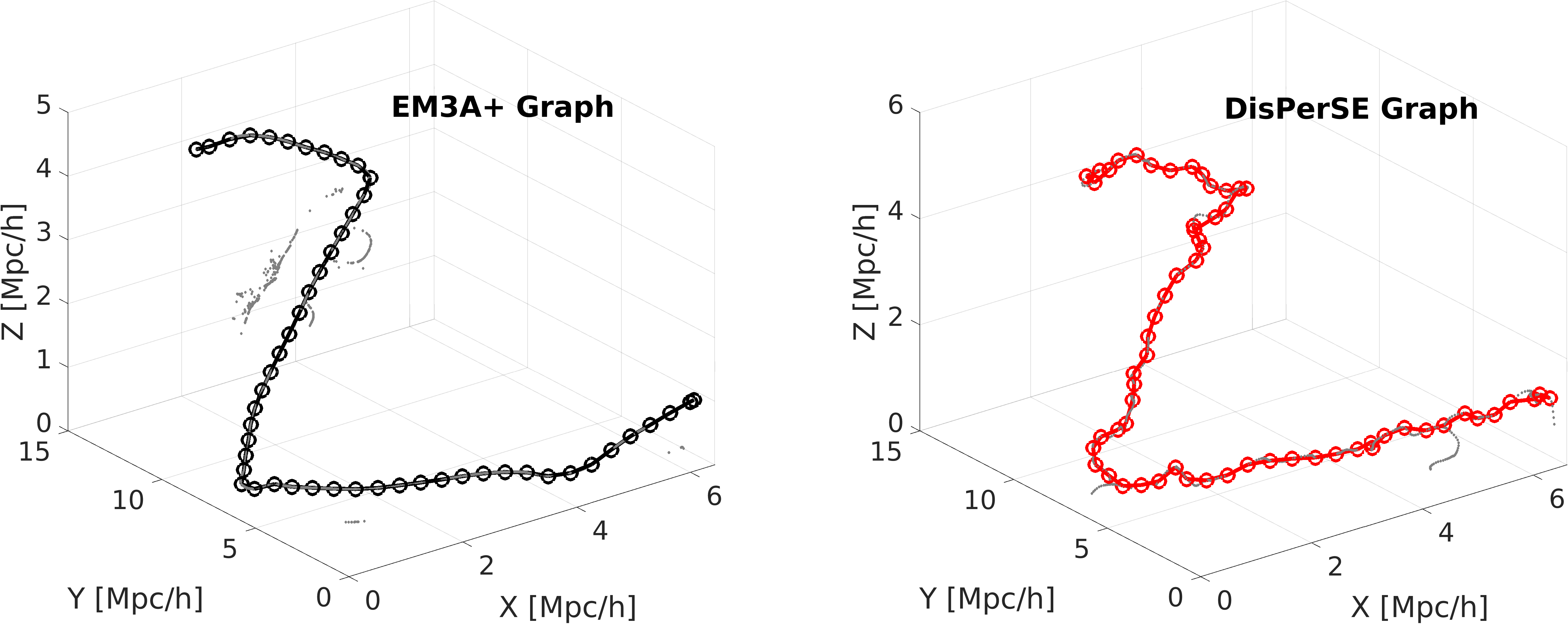}
\caption{Graph representations of the central axis found by EM3A+ (left column) and DisPerSE (right column).  
In grey are the axes initially recovered by the two algorithms. 
In black and red are the graphs resulting from MMCrawling with the circles representing the node positions of the graphs. 
Refer to the text for the parameters used in this construction and their justification.}
\label{fig:100percent}
\end{figure*}

%Describe Crawling methodology
% + results in Figures plane Intersect and graphDistances
Working with the results of either algorithms separately, we attempt to study the differences seen as a function of sampling in more detail. 
For this analysis, we employ the same crawling mechanism constructed for the cross-sectional visualization in Figure~\ref{fig:orthogonalSlices}. 
We first construct the graph representation of all retrieved axes. 
%We present in Figure~\ref{fig:100percent}
%,\ref{fig:80percent}, and \ref{fig:60percent} 
%the constructed graphs for the axes created in the $100\%$ case for EM3A+ (black) and DisPerSE (red)
%The graphs are superimposed upon the axes shown by the grey points in order to show the faithfulness of the graphs to the shape of the axes retrieved by either algorithms. 
To run MMCrawling on all axes of EM3A+, we use the neighborhood size $r = 0.5$~Mpc/h and jump tolerance $\beta = 0.4$. 
The jump tolerance $\beta$ controls the distance between the projecting and projected nodes of the MMCrawling algorithm. Therefore, if a modelled structure shows several bends, it is recommended to choose a smaller value for $\beta$. By choosing a smaller value, the crawling is performed along smaller steps, thus capturing local curvature that will be missed if the steps had been larger (i.e. larger $\beta$). Accordingly, to run MMCrawling on all axes of DisPerSE, we use $r = 0.5$~Mpc/h and $\beta = 0.3$. This slight difference in parametrization is therefore chosen so we can fairly represent the axes created by both EM3A+ and DisPerSE. We present in Figure~\ref{fig:100percent} the constructed graphs for the axes created in the $100\%$ case for EM3A+ (black) and DisPerSE (red).
The graphs are superimposed upon the axes shown by the grey points in order to show the faithfulness of the graphs to the shape of the axes retrieved by either algorithm.

\begin{figure*}
\centering
%\begin{subfigure}[t]{\columnwidth}
%    \centering
    \includegraphics[width=1\columnwidth]{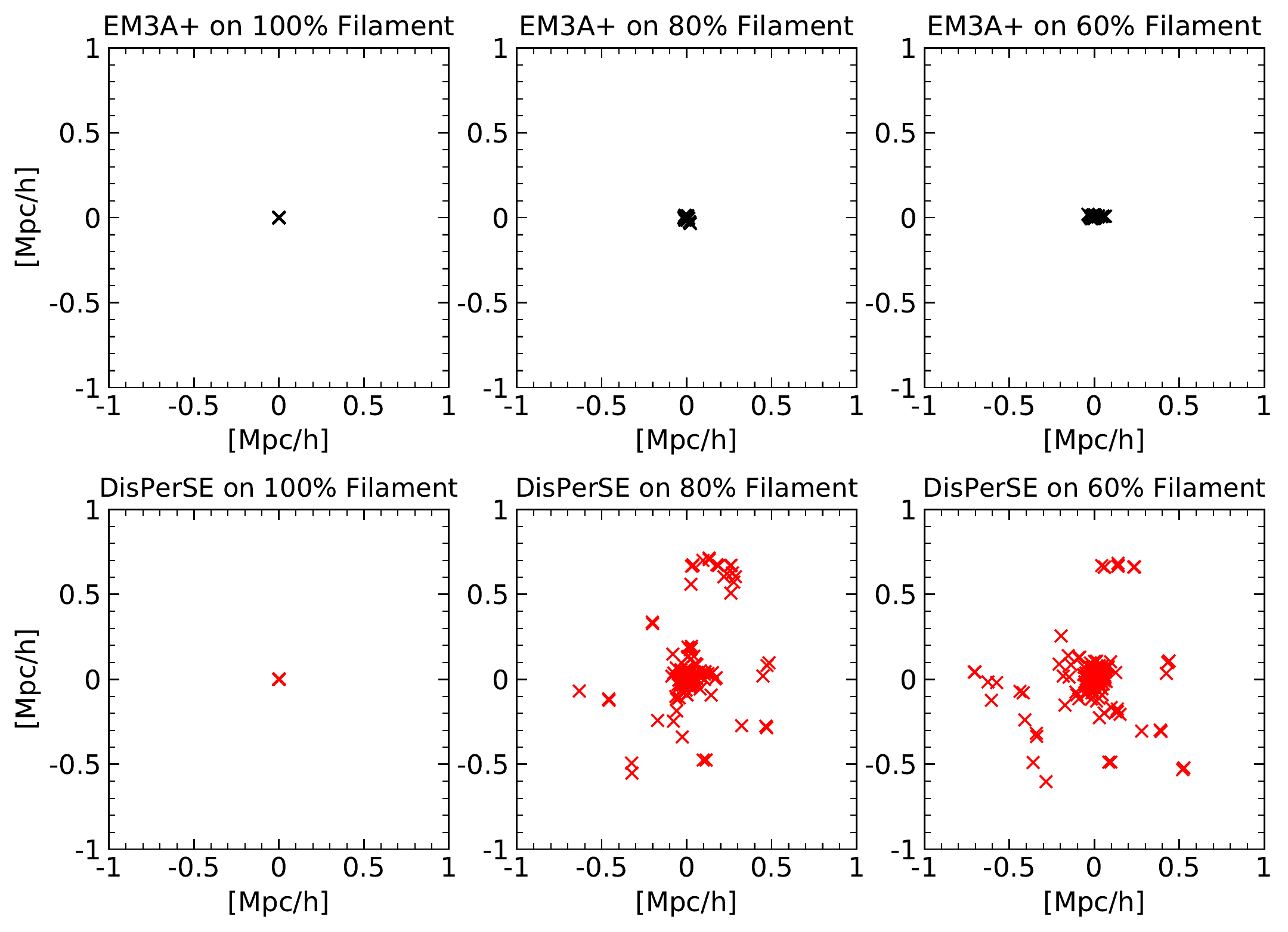}
%    \caption{}
%\label{fig:PlaneIntersect}
%\end{subfigure}%
~ 
%\begin{subfigure}[t]{\columnwidth}
%    \centering
    \includegraphics[width=1\columnwidth]{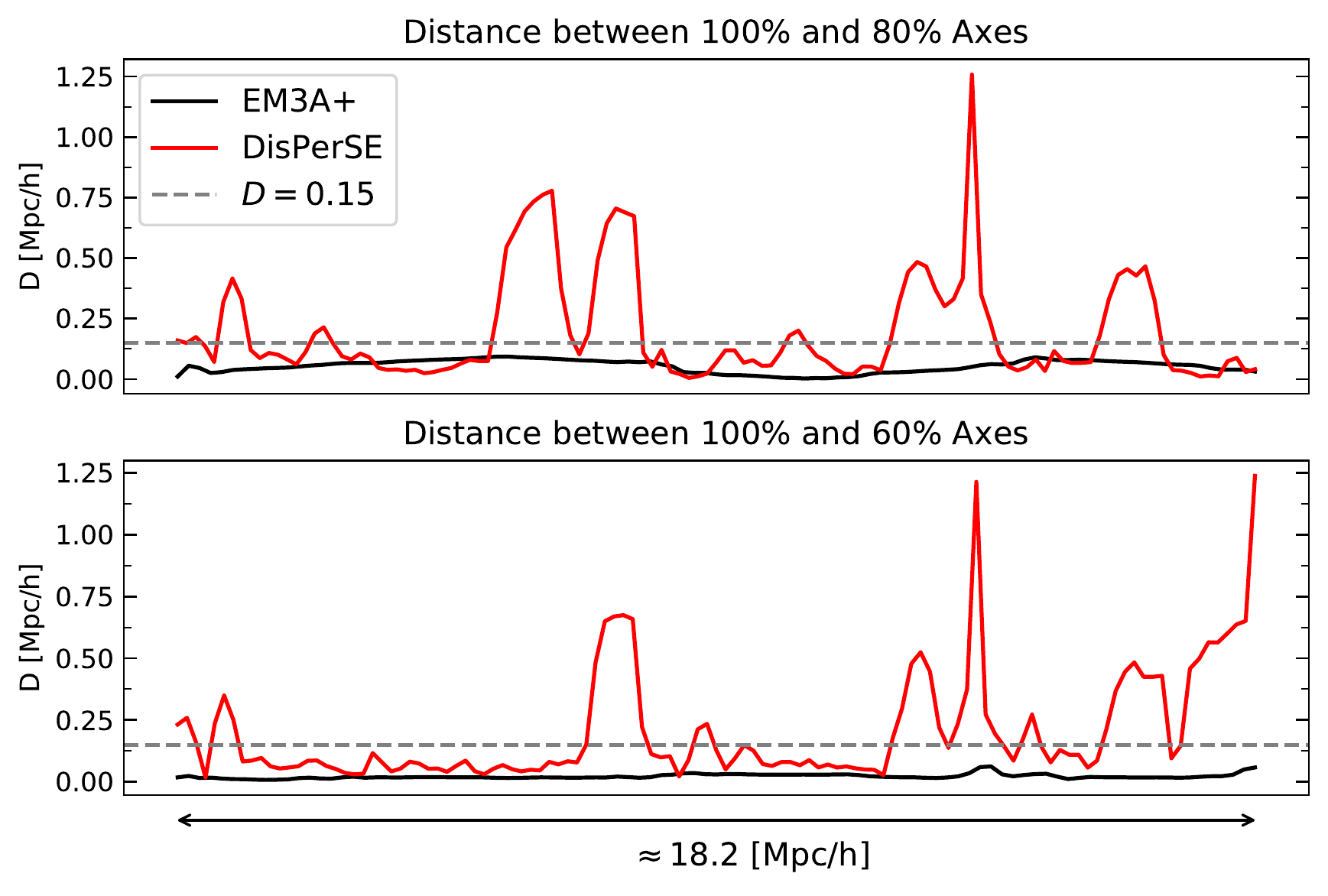}
%    \caption{}
%\end{subfigure}
\caption{Both panels attempt to measure the stability of EM3A+ (black) and DisPerSE (red) when the same filament used as input is sub-sampled by $80\%$ and $60\%$. 
The first column %In panel (a) 
shows the result of crawling on the reference ($100\%$) and measuring the intersection position between the orthogonal plane centered at this axis and the axes resulting from the sub-sampled cases. 
Column 2 shows the stacked intersections with the $80\%$ sampled case while crawling along the entire reference axis. 
Column 3 presents the similarly computed results for the $60\%$ case. 
%An ideal case is presented in the first column where the $100\%$ axes are compared with themselves. Panel (b) 
The right panel shows the result of the same analysis but when measuring the distances $D$ between the subsampled cases and the $100\%$ case for each of EM3A+ and DisPerSE. The dashed grey line at $D = 0.15$~Mpc/h contains $100\%$ of the intersection points in the case of EM3A+ but only $64\%$ for DisPerSE.}
\label{fig:GraphDistance}
\end{figure*}

%The value of $\beta$ in the latter is smaller in order to capture all details of the windings observed.

We then take the axis which results from using the un-sampled ($100\%$ case) filament as a reference (Figure~\ref{fig:100percent}) on which we move or ``crawl" on. 
As we crawl, a plane orthogonal to and centered on the axis at each visited position is considered. 
We then evaluate the intersection points between this plane and the axes resulting from the $80\%$ and $60\%$ sampled filaments. 
The intersection points therefore represent a measure of difference between the axes compared. 
To visualize this comparison, at each crawling position, we project the intersection point on the orthogonal plane and repeat this step for all crawling positions. 
We then observe the stack of the intersections on the plane and compare them to the origin.
The results for EM3A+ and DisPerSE are presented in the left panel of Figure~\ref{fig:GraphDistance} %\ref{fig:PlaneIntersect} 
In the ideal case where the compared axes are exactly the same, one should observe the same result as we see in the upper and lower left squares where we compare the axis of the $100\%$ case to itself. 
We see that all the interactions are exactly at the origin which means that there are no deviations from the reference. 
Looking at the $80\%$ and $60\%$ cases, we can see the deviations discussed. 
We observe that for EM3A+, the deviations from the reference are negligible as all intersections lie extremely close to the center. 
This indicates %shows 
that down-sampling the data produces little effect on the results under the optimal choice of parameters. 
On the other hand, although the results with DisPerSE show that the majority of the intersections lie close to the origin, we see large scatter as we down-sample the filament. 

As a final way to quantify the effect of changing the sampling of the data, we measure the distance $D$ between the intersection points with the plane and the reference axis. 
The results are presented in the right panel of Figure~\ref{fig:GraphDistance}, where we plot, across all crawling positions along the filament, the distance between the reference and the $80\%$ case (top panel), and between the reference and the $60\%$ case (lower panel). 
On the x-axis we present the estimated length of the filament calculated by summing the individual edges of the EM3A+ graphs. 
The results reiterate what we have seen in Figure~\ref{fig:SimpleResults} and the left pane of  Figure~\ref{fig:GraphDistance}. %\ref{fig:PlaneIntersect}. 
We see how the distance between the references and the sampled axes varies to within $\approx 1.25$~Mpc/h in the case of DisPerSE while it stays less than $0.1$~Mpc/h for EM3A+. This can also be quantified using the horizontal dashed line passing through $D = 0.15$~Mpc/h. We can see that all intersections with the EM3A+ axis lie within this distance to the central axis as opposed to only $64\%$ of the intersections in the case of DisPerSE.

We note that in terms of running time, EM3A+ is in the process of becoming optimized and so for the time being is slower than DisPerSE. For more details, a description of the typical running time required by each algorithm in 1-DREAM will be provided in the gitlab description of the toolbox. On the other hand, the findings provided in this section demonstrate the stability of EM3A+ as a function of downsizing the data on which it is run.
The upcoming observational surveys of galaxies within the Cosmic Web will provide catalogues of galaxies belonging to the studied structures. Since observational data tends to be sparse, tracing these structures would require methods that are robust against density variations. 
We therefore suggest that EM3A+ would be a useful algorithm for such research. 
If, on the other hand, %what is required is 
a method that traces the local density variations and rapidly time-evolving structures in the individual filaments is required, then DisPerSE would be a more adept methodology to employ.

%---------------------------------%
\section{Summary and Conclusion}
\label{sec:Conc} % used for referring to this section from elsewhere
%---------------------------------%

In this work we apply the toolbox 1-DREAM (1-Dimensional Recovery, Extraction, and Analysis of Manifolds) on N-body cosmological simulation data of the Cosmic Web. 
1-DREAM has been introduced previously in the work of \citet{CanducciEtal2022b}, where it has been briefly demonstrated on a filament of the Cosmic Web, as well as on the tidal tails of a Jellyfish galaxy, and the stream of Omega Centauri in the halo of the Milky Way. 
In this work we explain in depth how this toolbox is appropriate for use in the analysis of Cosmic Web environments, cosmic filaments in particular, while leaving in-depth statistical studies and explorations using 1-DREAM for future applications and research:

\begin{itemize}
    \item We present the publicly available algorithms comprising the toolbox, the first two of which are swarm-intelligence based methods for the extraction of structures from the simulation data, while the rest are methods for characterizing and modeling these structures.
    
    \item We provide an example filament taken from a subset of the cosmological simulation, and explain the steps we follow in extracting this filament from the entire subset and constructing a probabilistic model of it using our proposed toolbox.
    
    \item We demonstrate the possibility of moving along the constructed axis of the filament to measure local properties in lengthwise and orthogonal directions. The properties we focus on are local density and the velocity field perpendicular and parallel to the filament. As a further demonstration of the usefulness of the toolbox, we show how the velocity flow orthogonal to the axis varies, and how the particle velocity tends to accelerate towards the clusters at either end of the filament. We also provide radial density profiles of the filament and show how these profiles vary at different positions along the filament.
    
    \item The public simulation data provided by \citet{LibeskindEtal2018} and the analysis defined therein is introduced as a common method of comparing the ability of various algorithms in identifying the environments of the Cosmic Web. We show that 1-DREAM is successful in differentiating the various environments to a degree that is comparable with the state-of-the-art tools compiled in \citet{LibeskindEtal2018}. 
    
    \item We compare between 1-DREAM and the publicly available algorithm DisPerSE and show that 1-DREAM produces axes that are more aligned with the centers of cosmic filaments, and that our technique is very stable against down-sampling of the simulation data. %\petra{We also show that our toolbox is less likely to produce tracings of cosmic structures that are possibly not truly present in the data.}

\end{itemize}

Since 1-DREAM operates on a single-scale approach, the results are dependent on the scale parameter defined by the size of neighborhoods considered around each simulation particle. 
We thus suggest two possible future developments for this toolbox to improve on the discussed issue. 
On one hand, it is possible to isolate a large-enough subset of the simulation data and calibrate the neighborhood radius to give a desired physical outcome such as specific mass/volume fractions of the Cosmic Web environments. 
Instead of relying on simulation data, one could also calibrate the parameters using measurements of some properties of cosmic filaments in observational surveys. 
The second suggestion is to allow for a multi-scale solution in which the neighborhood radius parameter becomes adaptive to the local density. 
These two suggestions are possible directions in which the future versions of the toolbox could be taken but are not necessary for the functioning of the different algorithms.

Given this overview of 1-DREAM, future work may also include an in-depth exploration of the properties of cosmic filaments. In the case of both observational and simulation data, it is important to have algorithms that produce robust results under smaller data set sizes. We therefore suggest 1-DREAM as a tool to analyze data from both observational surveys and cosmological simulations.

%Document ends here
%%%%%%%%%%%%%%%%%%%%%%%%%%%%%%%%%%%%%%%%%%%%%%%%%%
%Keep these sections
\section*{Acknowledgements}
This work is supported by the DSSC Doctoral Training Programme of the University of Groningen. We thank the students Joel During and Jeroen Klooster for making the algorithms suggested in this work more user-friendly.

\section*{Data Availability}
Access to the N-cluster simulation data used in this paper is provided through the public gitlab repository link to the 1-DREAM toolbox: \url{https://git.lwp.rug.nl/cs.projects/1DREAM}. The remaining subsets used in this work may be provided upon request.


\bibliographystyle{mymnras}
\bibliography{2022Awad} % if your bibtex file is called example.bib

\begin{thebibliography}{}
\makeatletter
\relax
\def\mn@urlcharsother{\let\do\@makeother \do\$\do\&\do\#\do\^\do\_\do\%\do\~}
\def\mn@doi{\begingroup\mn@urlcharsother \@ifnextchar [ {\mn@doi@}
  {\mn@doi@[]}}
\def\mn@doi@[#1]#2{\def\@tempa{#1}\ifx\@tempa\@empty \href
  {http://dx.doi.org/#2} {doi:#2}\else \href {http://dx.doi.org/#2} {#1}\fi
  \endgroup}
\def\mn@eprint#1#2{\mn@eprint@#1:#2::\@nil}
\def\mn@eprint@arXiv#1{\href {http://arxiv.org/abs/#1} {{\tt arXiv:#1}}}
\def\mn@eprint@dblp#1{\href {http://dblp.uni-trier.de/rec/bibtex/#1.xml}
  {dblp:#1}}
\def\mn@eprint@#1:#2:#3:#4\@nil{\def\@tempa {#1}\def\@tempb {#2}\def\@tempc
  {#3}\ifx \@tempc \@empty \let \@tempc \@tempb \let \@tempb \@tempa \fi \ifx
  \@tempb \@empty \def\@tempb {arXiv}\fi \@ifundefined
  {mn@eprint@\@tempb}{\@tempb:\@tempc}{\expandafter \expandafter \csname
  mn@eprint@\@tempb\endcsname \expandafter{\@tempc}}}

\bibitem[\protect\citeauthoryear{{Abel}, {Hahn}  \& {Kaehler}}{{Abel}
  et~al.}{2012}]{Abel&Kaehler2012}
{Abel} T.,  {Hahn} O.,   {Kaehler} R.,  2012, \mn@doi [\mnras]
  {10.1111/j.1365-2966.2012.21754.x}, \href
  {https://ui.adsabs.harvard.edu/abs/2012MNRAS.427...61A} {427, 61}

\bibitem[\protect\citeauthoryear{{Alpaslan} et~al.,}{{Alpaslan}
  et~al.}{2014}]{AlpaslanEtal2014}
{Alpaslan} M.,  et~al., 2014, \mn@doi [\mnras] {10.1093/mnras/stt2136}, \href
  {https://ui.adsabs.harvard.edu/abs/2014MNRAS.438..177A} {438, 177}

\bibitem[\protect\citeauthoryear{{Arag{\'o}n-Calvo}, {Jones}, {van de Weygaert}
   \& {van der Hulst}}{{Arag{\'o}n-Calvo} et~al.}{2007}]{AragonCalvoEtal2007}
{Arag{\'o}n-Calvo} M.~A.,  {Jones} B.~J.~T.,  {van de Weygaert} R.,   {van der
  Hulst} J.~M.,  2007, \mn@doi [\aap] {10.1051/0004-6361:20077880}, \href
  {https://ui.adsabs.harvard.edu/abs/2007A&A...474..315A} {474, 315}

\bibitem[\protect\citeauthoryear{{Arag{\'o}n-Calvo}, {Platen}, {van de
  Weygaert}  \& {Szalay}}{{Arag{\'o}n-Calvo}
  et~al.}{2010}]{AragonCalvoEtal2010}
{Arag{\'o}n-Calvo} M.~A.,  {Platen} E.,  {van de Weygaert} R.,   {Szalay}
  A.~S.,  2010, \mn@doi [\apj] {10.1088/0004-637X/723/1/364}, \href
  {https://ui.adsabs.harvard.edu/abs/2010ApJ...723..364A} {723, 364}

\bibitem[\protect\citeauthoryear{{Barrow}, {Bhavsar}  \& {Sonoda}}{{Barrow}
  et~al.}{1985}]{BarrowEtal1985}
{Barrow} J.~D.,  {Bhavsar} S.~P.,   {Sonoda} D.~H.,  1985, \mn@doi [\mnras]
  {10.1093/mnras/216.1.17}, \href
  {https://ui.adsabs.harvard.edu/abs/1985MNRAS.216...17B} {216, 17}

\bibitem[\protect\citeauthoryear{Bishop}{Bishop}{2006}]{Bishopbook}
Bishop C.~M.,  2006, Pattern Recognition and Machine Learning (Information
  Science and Statistics).
Springer-Verlag, Berlin, Heidelberg

\bibitem[\protect\citeauthoryear{{Bond}, {Kofman}  \& {Pogosyan}}{{Bond}
  et~al.}{1996}]{BondEtal1996}
{Bond} J.~R.,  {Kofman} L.,   {Pogosyan} D.,  1996, \mn@doi [\nat]
  {10.1038/380603a0}, \href
  {https://ui.adsabs.harvard.edu/abs/1996Natur.380..603B} {380, 603}

\bibitem[\protect\citeauthoryear{{Bond}, {Strauss}  \& {Cen}}{{Bond}
  et~al.}{2010}]{BondEtal2010}
{Bond} N.~A.,  {Strauss} M.~A.,   {Cen} R.,  2010, \mn@doi [\mnras]
  {10.1111/j.1365-2966.2010.16823.x}, \href
  {https://ui.adsabs.harvard.edu/abs/2010MNRAS.406.1609B} {406, 1609}

\bibitem[\protect\citeauthoryear{{Bonnaire}, {Aghanim}, {Decelle}  \&
  {Douspis}}{{Bonnaire} et~al.}{2020}]{TREX}
{Bonnaire} T.,  {Aghanim} N.,  {Decelle} A.,   {Douspis} M.,  2020, \mn@doi
  [\aap] {10.1051/0004-6361/201936859}, \href
  {https://ui.adsabs.harvard.edu/abs/2020A&A...637A..18B} {637, A18}

\bibitem[\protect\citeauthoryear{{Burchett}, {Elek}, {Tejos}, {Prochaska},
  {Tripp}, {Bordoloi}  \& {Forbes}}{{Burchett} et~al.}{2020}]{BurchettEtal2020}
{Burchett} J.~N.,  {Elek} O.,  {Tejos} N.,  {Prochaska} J.~X.,  {Tripp} T.~M.,
  {Bordoloi} R.,   {Forbes} A.~G.,  2020, \mn@doi [\apjl]
  {10.3847/2041-8213/ab700c}, \href
  {https://ui.adsabs.harvard.edu/abs/2020ApJ...891L..35B} {891, L35}

\bibitem[\protect\citeauthoryear{Canducci, Taghribi, Mastropietro, de Rijcke,
  Peletier, Bunte  \& Tino}{Canducci et~al.}{2021}]{Canducci2021}
Canducci M.,  Taghribi A.,  Mastropietro M.,  de Rijcke S.,  Peletier R.,
  Bunte K.,   Tino P.,  2021, in Yin H.,  et~al., eds, Intelligent Data
  Engineering and Automated Learning -- IDEAL 2021. Springer International
  Publishing, Cham, pp 493--501

\bibitem[\protect\citeauthoryear{{Canducci} et~al.,}{{Canducci}
  et~al.}{2022a}]{CanducciEtal2022b}
{Canducci} M.,  et~al., 2022a, \mn@doi [Astronomy and Computing]
  {10.1016/j.ascom.2022.100658}, \href
  {https://ui.adsabs.harvard.edu/abs/2022A&C....4100658C} {41, 100658}

\bibitem[\protect\citeauthoryear{{Canducci}, {Tiño}  \&
  {Mastropietro}}{{Canducci} et~al.}{2022b}]{CanducciEtal2022a}
{Canducci} M.,  {Tiño} P.,   {Mastropietro} M.,  2022b, \mn@doi [Artificial
  Intelligence] {10.1016/j.artint.2021.103579}, 302, 103579

\bibitem[\protect\citeauthoryear{{Cautun} \& {van de Weygaert}}{{Cautun} \&
  {van de Weygaert}}{2011}]{Cautun&Weygaert2011}
{Cautun} M.~C.,  {van de Weygaert} R.,  2011, arXiv e-prints, \href
  {https://ui.adsabs.harvard.edu/abs/2011arXiv1105.0370C} {p. arXiv:1105.0370}

\bibitem[\protect\citeauthoryear{{Cautun}, {van de Weygaert}  \&
  {Jones}}{{Cautun} et~al.}{2013}]{CautunEtal2013}
{Cautun} M.,  {van de Weygaert} R.,   {Jones} B. J.~T.,  2013, \mn@doi [\mnras]
  {10.1093/mnras/sts416}, \href
  {https://ui.adsabs.harvard.edu/abs/2013MNRAS.429.1286C} {429, 1286}

\bibitem[\protect\citeauthoryear{{Cautun}, {van de Weygaert}, {Jones}  \&
  {Frenk}}{{Cautun} et~al.}{2014}]{CautunEtal2014}
{Cautun} M.,  {van de Weygaert} R.,  {Jones} B. J.~T.,   {Frenk} C.~S.,  2014,
  \mn@doi [\mnras] {10.1093/mnras/stu768}, \href
  {https://ui.adsabs.harvard.edu/abs/2014MNRAS.441.2923C} {441, 2923}

\bibitem[\protect\citeauthoryear{{Chun}, {Shin}, {Smith}, {Ko}  \&
  {Yoo}}{{Chun} et~al.}{2022}]{ChunEtal2022}
{Chun} K.,  {Shin} J.,  {Smith} R.,  {Ko} J.,   {Yoo} J.,  2022, \mn@doi [\apj]
  {10.3847/1538-4357/ac2cbe}, \href
  {https://ui.adsabs.harvard.edu/abs/2022ApJ...925..103C} {925, 103}

\bibitem[\protect\citeauthoryear{{Codis}, {Pichon}, {Devriendt}, {Slyz},
  {Pogosyan}, {Dubois}  \& {Sousbie}}{{Codis} et~al.}{2012}]{CodisEtal2012}
{Codis} S.,  {Pichon} C.,  {Devriendt} J.,  {Slyz} A.,  {Pogosyan} D.,
  {Dubois} Y.,   {Sousbie} T.,  2012, \mn@doi [\mnras]
  {10.1111/j.1365-2966.2012.21636.x}, \href
  {https://ui.adsabs.harvard.edu/abs/2012MNRAS.427.3320C} {427, 3320}

\bibitem[\protect\citeauthoryear{{Colberg}}{{Colberg}}{2007}]{Colberg2007}
{Colberg} J.~M.,  2007, \mn@doi [\mnras] {10.1111/j.1365-2966.2006.11312.x},
  \href {https://ui.adsabs.harvard.edu/abs/2007MNRAS.375..337C} {375, 337}

\bibitem[\protect\citeauthoryear{{Colberg}, {Krughoff}  \&
  {Connolly}}{{Colberg} et~al.}{2005}]{ColbergEtal2005}
{Colberg} J.~M.,  {Krughoff} K.~S.,   {Connolly} A.~J.,  2005, \mn@doi [\mnras]
  {10.1111/j.1365-2966.2005.08897.x}, \href
  {https://ui.adsabs.harvard.edu/abs/2005MNRAS.359..272C} {359, 272}

\bibitem[\protect\citeauthoryear{{Davis}, {Efstathiou}, {Frenk}  \&
  {White}}{{Davis} et~al.}{1985}]{DavisEtal1985}
{Davis} M.,  {Efstathiou} G.,  {Frenk} C.~S.,   {White} S.~D.~M.,  1985,
  \mn@doi [\apj] {10.1086/163168}, \href
  {https://ui.adsabs.harvard.edu/abs/1985ApJ...292..371D} {292, 371}

\bibitem[\protect\citeauthoryear{{Dolag}, {Meneghetti}, {Moscardini}, {Rasia}
  \& {Bonaldi}}{{Dolag} et~al.}{2006}]{DolagEtal2006}
{Dolag} K.,  {Meneghetti} M.,  {Moscardini} L.,  {Rasia} E.,   {Bonaldi} A.,
  2006, \mn@doi [\mnras] {10.1111/j.1365-2966.2006.10511.x}, \href
  {https://ui.adsabs.harvard.edu/abs/2006MNRAS.370..656D} {370, 656}

\bibitem[\protect\citeauthoryear{Dorigo \& St{\"{u}}tzle}{Dorigo \&
  St{\"{u}}tzle}{2004}]{ACObook}
Dorigo M.,  St{\"{u}}tzle T.,  2004, Ant colony optimization.
{MIT} Press

\bibitem[\protect\citeauthoryear{{Doroshkevich}, {Kotok}, {Poliudov},
  {Shandarin}, {Sigov}  \& {Novikov}}{{Doroshkevich}
  et~al.}{1980}]{DoroshkevichEtal1980}
{Doroshkevich} A.~G.,  {Kotok} E.~V.,  {Poliudov} A.~N.,  {Shandarin} S.~F.,
  {Sigov} I.~S.,   {Novikov} I.~D.,  1980, \mn@doi [\mnras]
  {10.1093/mnras/192.2.321}, \href
  {https://ui.adsabs.harvard.edu/abs/1980MNRAS.192..321D} {192, 321}

\bibitem[\protect\citeauthoryear{Edelsbrunner, Letscher  \&
  Zomorodian}{Edelsbrunner et~al.}{2002}]{EdelsbrunnerEtal2002}
Edelsbrunner Letscher  Zomorodian 2002, \mn@doi [Discrete Comput. Geom.]
  {10.1007/s00454-002-2885-2}, 28, 511–533

\bibitem[\protect\citeauthoryear{{Falck}}{{Falck}}{2013}]{Origami2}
{Falck} B.,  2013, {ORIGAMI: Structure-finding routine in N-body simulation},
  Astrophysics Source Code Library, record ascl:1304.012 (\mn@eprint {ascl}
  {1304.012})

\bibitem[\protect\citeauthoryear{{Falck}, {Neyrinck}  \& {Szalay}}{{Falck}
  et~al.}{2012}]{Origami1}
{Falck} B.~L.,  {Neyrinck} M.~C.,   {Szalay} A.~S.,  2012, \mn@doi [\apj]
  {10.1088/0004-637X/754/2/126}, \href
  {https://ui.adsabs.harvard.edu/abs/2012ApJ...754..126F} {754, 126}

\bibitem[\protect\citeauthoryear{Forman}{Forman}{1998}]{Forman1998}
Forman R.,  1998, \mn@doi [Advances in Mathematics]
  {https://doi.org/10.1006/aima.1997.1650}, 134, 90

\bibitem[\protect\citeauthoryear{{Genovese}, {Perone-Pacifico}, {Verdinelli}
  \& {Wasserman}}{{Genovese} et~al.}{2010}]{GenoveseEtal2010}
{Genovese} C.~R.,  {Perone-Pacifico} M.,  {Verdinelli} I.,   {Wasserman} L.,
  2010, arXiv e-prints, \href
  {https://ui.adsabs.harvard.edu/abs/2010arXiv1003.5536G} {p. arXiv:1003.5536}

\bibitem[\protect\citeauthoryear{{Gonz{\'a}lez} \& {Padilla}}{{Gonz{\'a}lez} \&
  {Padilla}}{2010}]{Gonzalez&Padilla2010}
{Gonz{\'a}lez} R.~E.,  {Padilla} N.~D.,  2010, \mn@doi [\mnras]
  {10.1111/j.1365-2966.2010.17015.x}, \href
  {https://ui.adsabs.harvard.edu/abs/2010MNRAS.407.1449G} {407, 1449}

\bibitem[\protect\citeauthoryear{{Graham}, {Clowes}  \& {Campusano}}{{Graham}
  et~al.}{1995}]{GrahamEtal1995}
{Graham} M.~J.,  {Clowes} R.~G.,   {Campusano} L.~E.,  1995, \mn@doi [\mnras]
  {10.1093/mnras/275.3.790}, \href
  {https://ui.adsabs.harvard.edu/abs/1995MNRAS.275..790G} {275, 790}

\bibitem[\protect\citeauthoryear{Gyulassy}{Gyulassy}{2008}]{Gyulassy2008}
Gyulassy A.~G.,  2008, PhD thesis, University of California at Davis, USA

\bibitem[\protect\citeauthoryear{{Hahn} \& {Abel}}{{Hahn} \&
  {Abel}}{2011}]{Hahn&Abel2011}
{Hahn} O.,  {Abel} T.,  2011, \mn@doi [Monthly Notices of the Royal
  Astronomical Society] {10.1111/j.1365-2966.2011.18820.x}, \href
  {https://ui.adsabs.harvard.edu/abs/2011MNRAS.415.2101H} {415, 2101}

\bibitem[\protect\citeauthoryear{{Hahn}, {Porciani}, {Carollo}  \&
  {Dekel}}{{Hahn} et~al.}{2007}]{HahnEtal2007}
{Hahn} O.,  {Porciani} C.,  {Carollo} C.~M.,   {Dekel} A.,  2007, \mn@doi
  [\mnras] {10.1111/j.1365-2966.2006.11318.x}, \href
  {https://ui.adsabs.harvard.edu/abs/2007MNRAS.375..489H} {375, 489}

\bibitem[\protect\citeauthoryear{{Hoffman}, {Metuki}, {Yepes}, {Gottl{\"o}ber},
  {Forero-Romero}, {Libeskind}  \& {Knebe}}{{Hoffman}
  et~al.}{2012a}]{HoffmanEtal2012}
{Hoffman} Y.,  {Metuki} O.,  {Yepes} G.,  {Gottl{\"o}ber} S.,  {Forero-Romero}
  J.~E.,  {Libeskind} N.~I.,   {Knebe} A.,  2012a, \mn@doi [\mnras]
  {10.1111/j.1365-2966.2012.21553.x}, \href
  {https://ui.adsabs.harvard.edu/abs/2012MNRAS.425.2049H} {425, 2049}

\bibitem[\protect\citeauthoryear{{Hoffman}, {Metuki}, {Yepes}, {Gottl{\"o}ber},
  {Forero-Romero}, {Libeskind}  \& {Knebe}}{{Hoffman}
  et~al.}{2012b}]{HoffmanEtala2012}
{Hoffman} Y.,  {Metuki} O.,  {Yepes} G.,  {Gottl{\"o}ber} S.,  {Forero-Romero}
  J.~E.,  {Libeskind} N.~I.,   {Knebe} A.,  2012b, \mn@doi [\mnras]
  {10.1111/j.1365-2966.2012.21553.x}, \href
  {https://ui.adsabs.harvard.edu/abs/2012MNRAS.425.2049H} {425, 2049}

\bibitem[\protect\citeauthoryear{{Jenkins} et~al.,}{{Jenkins}
  et~al.}{1998}]{Jenkins1998}
{Jenkins} A.,  et~al., 1998, \mn@doi [\apj] {10.1086/305615}, \href
  {https://ui.adsabs.harvard.edu/abs/1998ApJ...499...20J} {499, 20}

\bibitem[\protect\citeauthoryear{{Jhee}, {Song}, {Smith}, {Shin}, {Park}  \&
  {Laigle}}{{Jhee} et~al.}{2022}]{JheeEtal2022}
{Jhee} H.,  {Song} H.,  {Smith} R.,  {Shin} J.,  {Park} I.,   {Laigle} C.,
  2022, arXiv e-prints, \href
  {https://ui.adsabs.harvard.edu/abs/2022arXiv220109540J} {p. arXiv:2201.09540}

\bibitem[\protect\citeauthoryear{{Jones} et~al.,}{{Jones}
  et~al.}{2004}]{6DFGSOrg}
{Jones} D.~H.,  et~al., 2004, \mn@doi [\mnras]
  {10.1111/j.1365-2966.2004.08353.x}, \href
  {https://ui.adsabs.harvard.edu/abs/2004MNRAS.355..747J} {355, 747}

\bibitem[\protect\citeauthoryear{{Jones} et~al.,}{{Jones} et~al.}{2009}]{6DFGS}
{Jones} D.~H.,  et~al., 2009, \mn@doi [\mnras]
  {10.1111/j.1365-2966.2009.15338.x}, \href
  {https://ui.adsabs.harvard.edu/abs/2009MNRAS.399..683J} {399, 683}

\bibitem[\protect\citeauthoryear{{Kim}, {Smith}  \& {Shin}}{{Kim}
  et~al.}{2022}]{KimEtal2022}
{Kim} Y.,  {Smith} R.,   {Shin} J.,  2022, \mn@doi [\apj]
  {10.3847/1538-4357/ac7e45}, \href
  {https://ui.adsabs.harvard.edu/abs/2022ApJ...935...71K} {935, 71}

\bibitem[\protect\citeauthoryear{{Kitaura} \& {Angulo}}{{Kitaura} \&
  {Angulo}}{2012}]{Kitaura&Angulo2012}
{Kitaura} F.-S.,  {Angulo} R.~E.,  2012, \mn@doi [\mnras]
  {10.1111/j.1365-2966.2012.21614.x}, \href
  {https://ui.adsabs.harvard.edu/abs/2012MNRAS.425.2443K} {425, 2443}

\bibitem[\protect\citeauthoryear{{Kleiner}, {Pimbblet}, {Jones}, {Koribalski}
  \& {Serra}}{{Kleiner} et~al.}{2017}]{KleinerEtal2017}
{Kleiner} D.,  {Pimbblet} K.~A.,  {Jones} D.~H.,  {Koribalski} B.~S.,   {Serra}
  P.,  2017, \mn@doi [\mnras] {10.1093/mnras/stw3328}, \href
  {https://ui.adsabs.harvard.edu/abs/2017MNRAS.466.4692K} {466, 4692}

\bibitem[\protect\citeauthoryear{{Klypin} \& {Shandarin}}{{Klypin} \&
  {Shandarin}}{1983}]{KlypinEtal1983}
{Klypin} A.~A.,  {Shandarin} S.~F.,  1983, \mn@doi [\mnras]
  {10.1093/mnras/204.3.891}, \href
  {https://ui.adsabs.harvard.edu/abs/1983MNRAS.204..891K} {204, 891}

\bibitem[\protect\citeauthoryear{{Kraljic} et~al.,}{{Kraljic}
  et~al.}{2018}]{KraljicEtal2018}
{Kraljic} K.,  et~al., 2018, \mn@doi [\mnras] {10.1093/mnras/stx2638}, \href
  {https://ui.adsabs.harvard.edu/abs/2018MNRAS.474..547K} {474, 547}

\bibitem[\protect\citeauthoryear{{Kraljic} et~al.,}{{Kraljic}
  et~al.}{2019}]{KraljicEtal2019}
{Kraljic} K.,  et~al., 2019, \mn@doi [\mnras] {10.1093/mnras/sty3216}, \href
  {https://ui.adsabs.harvard.edu/abs/2019MNRAS.483.3227K} {483, 3227}

\bibitem[\protect\citeauthoryear{{Laigle} et~al.,}{{Laigle}
  et~al.}{2015}]{LaigleEtal2015}
{Laigle} C.,  et~al., 2015, \mn@doi [\mnras] {10.1093/mnras/stu2289}, \href
  {https://ui.adsabs.harvard.edu/abs/2015MNRAS.446.2744L} {446, 2744}

\bibitem[\protect\citeauthoryear{{Laigle} et~al.,}{{Laigle}
  et~al.}{2018}]{LaigleEtal2018}
{Laigle} C.,  et~al., 2018, \mn@doi [\mnras] {10.1093/mnras/stx3055}, \href
  {https://ui.adsabs.harvard.edu/abs/2018MNRAS.474.5437L} {474, 5437}

\bibitem[\protect\citeauthoryear{{Lambert}, {Kraan-Korteweg}, {Jarrett}  \&
  {Macri}}{{Lambert} et~al.}{2020}]{2MRS}
{Lambert} T.~S.,  {Kraan-Korteweg} R.~C.,  {Jarrett} T.~H.,   {Macri} L.~M.,
  2020, \mn@doi [\mnras] {10.1093/mnras/staa1946}, \href
  {https://ui.adsabs.harvard.edu/abs/2020MNRAS.497.2954L} {497, 2954}

\bibitem[\protect\citeauthoryear{{Lewis} \& {Challinor}}{{Lewis} \&
  {Challinor}}{2011}]{LewisEtal2000}
{Lewis} A.,  {Challinor} A.,  2011, {CAMB: Code for Anisotropies in the
  Microwave Background} (\mn@eprint {ascl} {1102.026})

\bibitem[\protect\citeauthoryear{{Libeskind} et~al.,}{{Libeskind}
  et~al.}{2018}]{LibeskindEtal2018}
{Libeskind} N.~I.,  et~al., 2018, \mn@doi [\mnras] {10.1093/mnras/stx1976},
  \href {https://ui.adsabs.harvard.edu/abs/2018MNRAS.473.1195L} {473, 1195}

\bibitem[\protect\citeauthoryear{Little, Maggioni  \& Murphy}{Little
  et~al.}{2020}]{LLPD}
Little A.,  Maggioni M.,   Murphy J.~M.,  2020, Journal of Machine Learning
  Research, 21, 1

\bibitem[\protect\citeauthoryear{{Luber}, {van Gorkom}, {Hess}, {Pisano},
  {Fern{\'a}ndez}  \& {Momjian}}{{Luber} et~al.}{2019}]{LuberEtal2019}
{Luber} N.,  {van Gorkom} J.~H.,  {Hess} K.~M.,  {Pisano} D.~J.,
  {Fern{\'a}ndez} X.,   {Momjian} E.,  2019, \mn@doi [\aj]
  {10.3847/1538-3881/ab1b6e}, \href
  {https://ui.adsabs.harvard.edu/abs/2019AJ....157..254L} {157, 254}

\bibitem[\protect\citeauthoryear{{Macri} et~al.,}{{Macri}
  et~al.}{2019}]{2MRSOrg}
{Macri} L.~M.,  et~al., 2019, \mn@doi [\apjs] {10.3847/1538-4365/ab465a}, \href
  {https://ui.adsabs.harvard.edu/abs/2019ApJS..245....6M} {245, 6}

\bibitem[\protect\citeauthoryear{{Metuki}, {Libeskind}, {Hoffman}, {Crain}  \&
  {Theuns}}{{Metuki} et~al.}{2015}]{MetukiEtal2015}
{Metuki} O.,  {Libeskind} N.~I.,  {Hoffman} Y.,  {Crain} R.~A.,   {Theuns} T.,
  2015, \mn@doi [\mnras] {10.1093/mnras/stu2166}, \href
  {https://ui.adsabs.harvard.edu/abs/2015MNRAS.446.1458M} {446, 1458}

\bibitem[\protect\citeauthoryear{Mohammadi, Tino  \& Bunte}{Mohammadi
  et~al.}{2022}]{MohammadiEtal2022}
Mohammadi M.,  Tino P.,   Bunte K.,  2022, \mn@doi [Neural Computation]
  {10.1162/neco_a_01478}, 34, 595

\bibitem[\protect\citeauthoryear{{Pauls} \& {Melott}}{{Pauls} \&
  {Melott}}{1995}]{PaulsEtal1995}
{Pauls} J.~L.,  {Melott} A.~L.,  1995, \mn@doi [\mnras]
  {10.1093/mnras/274.1.99}, \href
  {https://ui.adsabs.harvard.edu/abs/1995MNRAS.274...99P} {274, 99}

\bibitem[\protect\citeauthoryear{{Peebles}}{{Peebles}}{1980}]{Peebles1980}
{Peebles} P.~J.~E.,  1980, {The large-scale structure of the universe}.
Princeton University Press

\bibitem[\protect\citeauthoryear{{Ramachandra} \& {Shandarin}}{{Ramachandra} \&
  {Shandarin}}{2015}]{Ramachandra&Shandarin2015}
{Ramachandra} N.~S.,  {Shandarin} S.~F.,  2015, \mn@doi [\mnras]
  {10.1093/mnras/stv1389}, \href
  {https://ui.adsabs.harvard.edu/abs/2015MNRAS.452.1643R} {452, 1643}

\bibitem[\protect\citeauthoryear{{Sathyaprakash}, {Sahni}  \&
  {Shandarin}}{{Sathyaprakash} et~al.}{1996}]{SathyaprakashEtal1996}
{Sathyaprakash} B.~S.,  {Sahni} V.,   {Shandarin} S.~F.,  1996, \mn@doi [\apjl]
  {10.1086/310024}, \href
  {https://ui.adsabs.harvard.edu/abs/1996ApJ...462L...5S} {462, L5}

\bibitem[\protect\citeauthoryear{{Schaap} \& {van de Weygaert}}{{Schaap} \&
  {van de Weygaert}}{2000}]{Schaap&Weygaert2000}
{Schaap} W.~E.,  {van de Weygaert} R.,  2000, \aap, \href
  {https://ui.adsabs.harvard.edu/abs/2000A&A...363L..29S} {363, L29}

\bibitem[\protect\citeauthoryear{{Shandarin}}{{Shandarin}}{2011}]{Shandarin2011}
{Shandarin} S.~F.,  2011, \mn@doi [\jcap] {10.1088/1475-7516/2011/05/015},
  \href {https://ui.adsabs.harvard.edu/abs/2011JCAP...05..015S} {2011, 015}

\bibitem[\protect\citeauthoryear{{Shen}, {Abel}, {Mo}  \& {Sheth}}{{Shen}
  et~al.}{2006}]{ShenEtal2006}
{Shen} J.,  {Abel} T.,  {Mo} H.~J.,   {Sheth} R.~K.,  2006, \mn@doi [\apj]
  {10.1086/504513}, \href
  {https://ui.adsabs.harvard.edu/abs/2006ApJ...645..783S} {645, 783}

\bibitem[\protect\citeauthoryear{{Sheth}}{{Sheth}}{2004}]{Sheth2004}
{Sheth} J.~V.,  2004, \mn@doi [\mnras] {10.1111/j.1365-2966.2004.08191.x},
  \href {https://ui.adsabs.harvard.edu/abs/2004MNRAS.354..332S} {354, 332}

\bibitem[\protect\citeauthoryear{{Sheth} \& {van de Weygaert}}{{Sheth} \& {van
  de Weygaert}}{2004}]{ShethWeygaert2004}
{Sheth} R.~K.,  {van de Weygaert} R.,  2004, \mn@doi [\mnras]
  {10.1111/j.1365-2966.2004.07661.x}, \href
  {https://ui.adsabs.harvard.edu/abs/2004MNRAS.350..517S} {350, 517}

\bibitem[\protect\citeauthoryear{{Smith} et~al.,}{{Smith}
  et~al.}{2021}]{SmithEtal2021}
{Smith} R.,  et~al., 2021, \mn@doi [\apj] {10.3847/1538-4357/abe1b1}, \href
  {https://ui.adsabs.harvard.edu/abs/2021ApJ...912..149S} {912, 149}

\bibitem[\protect\citeauthoryear{{Smith}, {Calder{\'o}n-Castillo}, {Shin},
  {Raouf}  \& {Ko}}{{Smith} et~al.}{2022a}]{Smith2Etal2022}
{Smith} R.,  {Calder{\'o}n-Castillo} P.,  {Shin} J.,  {Raouf} M.,   {Ko} J.,
  2022a, \mn@doi [\aj] {10.3847/1538-3881/ac8053}, \href
  {https://ui.adsabs.harvard.edu/abs/2022AJ....164...95S} {164, 95}

\bibitem[\protect\citeauthoryear{{Smith} et~al.,}{{Smith}
  et~al.}{2022b}]{SmithEtal2022}
{Smith} R.,  et~al., 2022b, \mn@doi [\apj] {10.3847/1538-4357/ac7ab5}, \href
  {https://ui.adsabs.harvard.edu/abs/2022ApJ...934...86S} {934, 86}

\bibitem[\protect\citeauthoryear{{Sousbie}}{{Sousbie}}{2011}]{Disperse1}
{Sousbie} T.,  2011, \mn@doi [\mnras] {10.1111/j.1365-2966.2011.18394.x}, \href
  {https://ui.adsabs.harvard.edu/abs/2011MNRAS.414..350S} {414, 350}

\bibitem[\protect\citeauthoryear{{Sousbie}, {Pichon}  \& {Kawahara}}{{Sousbie}
  et~al.}{2011}]{Disperse2}
{Sousbie} T.,  {Pichon} C.,   {Kawahara} H.,  2011, \mn@doi [\mnras]
  {10.1111/j.1365-2966.2011.18395.x}, \href
  {https://ui.adsabs.harvard.edu/abs/2011MNRAS.414..384S} {414, 384}

\bibitem[\protect\citeauthoryear{{Springel}}{{Springel}}{2005}]{Gadget2}
{Springel} V.,  2005, \mn@doi [MNRAS] {10.1111/j.1365-2966.2005.09655.x}, \href
  {http://adsabs.harvard.edu/abs/2005MNRAS.364.1105S} {364, 1105}

\bibitem[\protect\citeauthoryear{Taghribi, Bunte, Smith, Shin, Mastropietro,
  Peletier  \& Tiňo}{Taghribi et~al.}{2022a}]{TaghribiEtal2022}
Taghribi A.,  Bunte K.,  Smith R.,  Shin J.,  Mastropietro M.,  Peletier R.~F.,
    Tiňo P.,  2022a, \mn@doi [IEEE Transactions on Knowledge and Data
  Engineering] {10.1109/TKDE.2022.3177368}, pp~1--1

\bibitem[\protect\citeauthoryear{Taghribi, Canducci, Mastropietro, {De Rijcke},
  Bunte  \& Tiňo}{Taghribi et~al.}{2022b}]{TAGHRIBI2022376}
Taghribi A.,  Canducci M.,  Mastropietro M.,  {De Rijcke} S.,  Bunte K.,
  Tiňo P.,  2022b, \mn@doi [Neurocomputing]
  {https://doi.org/10.1016/j.neucom.2021.05.108}, 470, 376

\bibitem[\protect\citeauthoryear{{Tempel}, {Stoica}, {Mart{\'\i}nez},
  {Liivam{\"a}gi}, {Castellan}  \& {Saar}}{{Tempel}
  et~al.}{2014}]{TempelEtal2014}
{Tempel} E.,  {Stoica} R.~S.,  {Mart{\'\i}nez} V.~J.,  {Liivam{\"a}gi} L.~J.,
  {Castellan} G.,   {Saar} E.,  2014, \mn@doi [\mnras] {10.1093/mnras/stt2454},
  \href {https://ui.adsabs.harvard.edu/abs/2014MNRAS.438.3465T} {438, 3465}

\bibitem[\protect\citeauthoryear{{Tempel}, {Stoica}, {Kipper}  \&
  {Saar}}{{Tempel} et~al.}{2016}]{TempelEtal2016}
{Tempel} E.,  {Stoica} R.~S.,  {Kipper} R.,   {Saar} E.,  2016, \mn@doi
  [Astronomy and Computing] {10.1016/j.ascom.2016.03.004}, \href
  {https://ui.adsabs.harvard.edu/abs/2016A&C....16...17T} {16, 17}

\bibitem[\protect\citeauthoryear{\VAN{Weygaert}{Van}{van}~de Weygaert \&
  {Schaap}}{\VAN{Weygaert}{Van}{van}~de Weygaert \&
  {Schaap}}{2009}]{Weygaert&Schaap2009}
\VAN{Weygaert}{Van}{van}~de Weygaert R.,  {Schaap} W.,  2009, in
  {Mart{\'\i}nez} V.~J.,  {Saar} E.,  {Mart{\'\i}nez-Gonz{\'a}lez} E.,
  {Pons-Border{\'\i}a} M.~J.,  eds, , Vol.~665, Data Analysis in Cosmology.
Springer Berlin Heidelberg, Berlin, Heidelberg, pp 291--413,
  \mn@doi{10.1007/978-3-540-44767-2_11}

\bibitem[\protect\citeauthoryear{Wang \& Carreira-Perpi{\~n}{\'a}n}{Wang \&
  Carreira-Perpi{\~n}{\'a}n}{2010}]{MBMS}
Wang W.,  Carreira-Perpi{\~n}{\'a}n M.~{\'A}.,  2010, 2010 IEEE Computer
  Society Conference on Computer Vision and Pattern Recognition, pp 1759--1766

\bibitem[\protect\citeauthoryear{{Wang}, {Szalay}, {Aragon-Calvo}, {Neyrinck}
  \& {Eyink}}{{Wang} et~al.}{2014}]{WangEtal2014}
{Wang} X.,  {Szalay} A.~S.,  {Aragon-Calvo} M.~A.,  {Neyrinck} M.~C.,   {Eyink}
  G.~L.,  2014, in American Astronomical Society Meeting Abstracts \#223. p.
  457.16

\bibitem[\protect\citeauthoryear{{Wu}, {Bertholet}, {Huang}, {Cohen-Or}, {Gong}
   \& {Zwicker}}{{Wu} et~al.}{2018}]{WuEtal2018}
{Wu} S.,  {Bertholet} P.,  {Huang} H.,  {Cohen-Or} D.,  {Gong} M.,   {Zwicker}
  M.,  2018, \mn@doi [IEEE Transactions on Pattern Analysis and Machine
  Intelligence] {10.1109/TPAMI.2017.2754254}, 40, 2529

\bibitem[\protect\citeauthoryear{{York} et~al.,}{{York} et~al.}{2000}]{SDSS}
{York} D.~G.,  et~al., 2000, \mn@doi [\aj] {10.1086/301513}, \href
  {https://ui.adsabs.harvard.edu/abs/2000AJ....120.1579Y} {120, 1579}

\makeatother
\end{thebibliography}

%\begin{thebibliography}{99}
%\end{thebibliography}

%%%%%%%%%%%%%%%%%%%%%%%%%%%%%%%%%%%%%%%%%%%%%%%%%%

%%%%%%%%%%%%%%%%% APPENDICES %%%%%%%%%%%%%%%%%%%%%

\appendix

\section{Further Method Analysis}
\label{sec:MethodAnalysis}
%------------------------------%

In this section, we assess the effect of two prominent factors on the functioning of the 1-DREAM toolbox: the effect of parameter changes, and the effect of stochasticity. As presented in Section~\ref{sec:GeneralFormalism}, 1-DREAM makes use of many hyper-parameters which allow the user to adjust to the data and the structures expected within the data. This is especially apparent given that LAAT and EM3A+ are based on random walks within the data which requires several parameters for the construction of the walks. As a reference to these parameters, we provide in Table~\ref{table:parameters} a list of all parameters used by each algorithm along with their recommended values and a brief description. We note that all values listed in bold can be changed at will to adapt to the given structures and data, while the remaining parameters are fixed as the default settings and do not need to be changed. As for the stochastic element, it is clear that stochasticity is present in some of the algorithms which use random initializations namely LAAT and MMCrawling. 
For LAAT, this is apparent given the random distribution of the agents in the data set and the random walk that follows afterwards, while for MMCrawling this can be seen during the initialization phase when choosing a random point on the manifolds for the construction of their skeletal representations. 
We therefore explore how sensitive the extraction of structures and their subsequent probabilistic modelling are to changes in free parameters and initializations.

\subsection{Effect of Parameter Change}
\label{sec:ParameterChange}

The effect of the parameters on changes in the random walks have been studied in the initial propositions of LAAT and EM3A, namely the work presented in \citet{TaghribiEtal2022} and \citet{MohammadiEtal2022}. 
We therefore recount briefly the results of the parameter analysis presented in those works. With regards to LAAT, we first note the trade-off effect between the influence of alignment with the manifold and the influence of pheromone concentration when specifying the parameter $\kappa$ (refer to equation~\ref{vformula}). 
$\kappa$ has an effect on the choice of the particles visited by the agents, as does the parameter $\omega$ which tunes the transition probabilities in the random walk (refer to equation~\eqref{pformula}). 
\citet{TaghribiEtal2022} have demonstrated that results show little difference in response to small changes of $\kappa$ and $\omega$, but certain trends could be observed in the behaviour of the agents for large changes of these parameters. 
Increasing $\kappa$ to larger values regardless of the choice of $\omega$ showed little changes in results, which highlights the fact that the alignment with the manifold is of greater importance to the choice of jumps than is the concentration of the pheromone when it comes to highly-curved manifolds. 
As for $\omega$, taking larger values enforces the agents to exclusively choose the jumps with the highest preference. 
This produces larger deviations when runs are repeated with different initial distributions for the agents. 
On the other hand, smaller values for $\omega$ enforce a more random choice for the jumps.
This reduces the denoising effect by allowing more particles away from the manifold to be selected in the random walk. In the case of cosmological data sets of the cosmic web, we provide the recommended values for $\kappa$ and $\omega$ in Table~\ref{table:parameters}.

\begin{center}
  \begin{table}
  \caption{Full list of 1-DREAM parameters: recommended values and definitions for cosmological Cosmic Web data. Parameters in bold can be adapted to other values while the rest are advised to keep fixed.} 
  \renewcommand{\arraystretch}{1.2}%{1.5} % Default value: 1
  \centering
  \begin{tabular}{@{\extracolsep{\fill}} p{0.18\linewidth}p{0.16\linewidth}p{0.09\linewidth} l%p{0.4\linewidth}
  }
  %{p{1cm} p{1.5cm} p{0.5cm} p{3cm}} %\hline
   Alg. & Param. & Value & Description\\ 
   \toprule%\hline\hline
    LAAT & $r \in \RR$    & \textbf{0.5}   &  \textbf{Neighborhood radius} \\
      & $\zeta \in \RR$  &  0.05  & Evaporation rate\\
      & $\kappa \in \RR$ & 0.8  & Shape v. Pheromone\\
      & $\omega \in \RR$  &  10  & Inverse temperature\\
      & $\gamma \in \RR$ &  0.05  & Deposited pheromone\\
      & $ F \in \RR$ &  3  & Pheromone threshold\\
      & $N_\mathrm{epoch} \in \NN$ &  100  & Number of Epochs\\
      & $N_\mathrm{steps} \in \NN$ &  12000  & Steps per epoch\\
      & $N_\mathrm{ants} \in \NN$ &  20 & Number of Agents\\
   \hline
    EM3A+ & $r \in \RR$ &  \textbf{1}  &  \textbf{Neighborhood radius} \\
     %& $b \in \RR$  &  0.25  & Jump weight \\
     & $\eta \in \RR$ &  0.1  & Learning rate \\
     & $N_s \in \NN$  &  100  & Number of steps\\
     & $N_g \in \NN$  &  10  & Number of generations\\ 
  \hline
    DimIndex &  $r \in \RR$  &  \textbf{1}  & \textbf{Neighborhood radius}\\
    & $\tau \in \NN$ &  5  & Filtering threshold \\
   \hline
    MMCrawling & $r \in \RR$ &  \textbf{1}  & \textbf{Neighborhood radius}\\
    & $\beta \in \RR$ &  \textbf{0.6}  & \textbf{Jump tolerance} \\
  \hline
    SGTM & $r \in \RR$ &  \textbf{0.5}  &  \textbf{Neighborhood radius} \\
    & $\varepsilon \in \NN$ & \textbf{1} & \textbf{RBF centers min. distance}\\
     %& $\upsilon \in \RR$ &  1000  &  Regularization Cov. matrix\\
   \bottomrule%\hline\hline
  \end{tabular}
  \label{table:parameters}
  \end{table}
\end{center}

Similarly, we recount the effect on the results due to changing the amount of deposited pheromone, and its specified evaporation rate $\zeta$. 
\citet{TaghribiEtal2022} have shown that the results are very robust against changes of these two parameters especially when modifying the amount of pheromone deposited when an agent visits a given particle. 
When increasing that parameter, the pheromone value on all particles at the end of the run increases in a proportional way, therefore, the ratio of this amount between the different particles in the data set remains the same. 
As for $\zeta$ (refer to equation~\eqref{evaporationformula}) taking larger values for this parameter leads to faster evaporation of the pheromone on all the particles visited. 
This enforces more random jump choices for the agents and so larger deviations between the runs of the algorithm. It follows that in this case, fainter structures would be less likely to be visited by the agents, and hence more likely to be discarded as noise. Similarly, recommended values for these parameters are provided in Table~\ref{table:parameters}.

A similar robustness of results is observed when changing the parameters pertaining to EM3A+.
Aside from the choice of neighborhood radius centered at the different points, the effect of modifying the learning rate $\eta$ (refer to equation~\eqref{EM3Aupdate}) should also be specified. 
\citet{MohammadiEtal2022} have shown that selecting a smaller value for $\eta$ leads to less variation in the results across different runs but requires a larger number of iterations for one run. 
This highlights the trade-off effect between the number of iterations specified for the random walk and between the quality of the results. The recommended value for $\eta$ is given in Table~\ref{table:parameters}.

\begin{figure}
\centering
%\begin{subfigure}[b]{0.95\columnwidth}
   \includegraphics[width=0.99\linewidth]{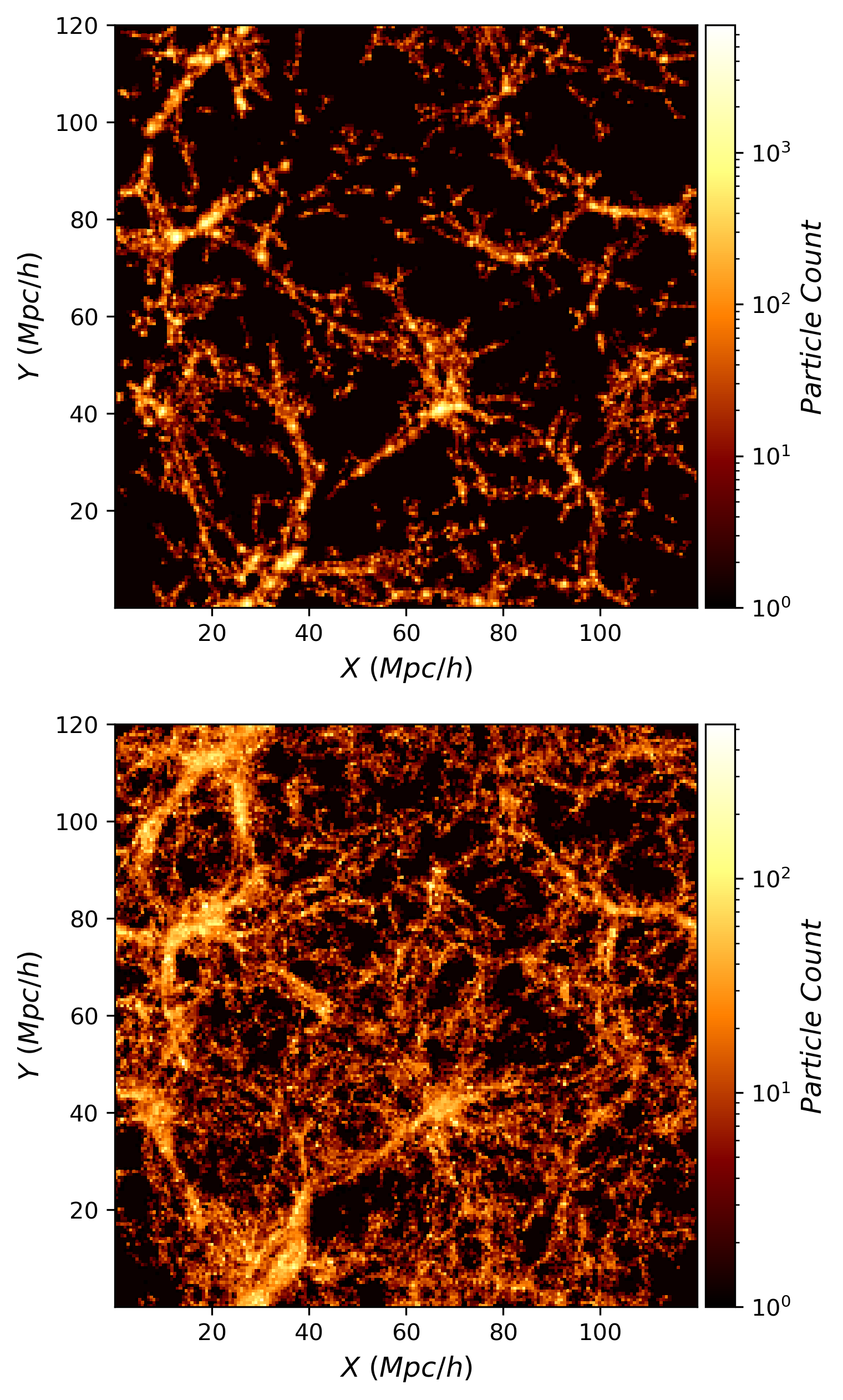}\\

\caption{The top panel %part of \kb{the upper} panel %(a)
demonstrates a slice of thickness $20$~Mpc/h after extraction of the structures retrieved with one full run of LAAT on the entire data cube.
The bottom panel shows the result of re-running LAAT on the filtered-out particles and the structures extracted in the second run. The results are demonstrated on the same slice as the top panel.
%The results are portrayed on the same slice as Panel (a).  % part of Panel (b).
}
\label{fig:LAAT2Runs}
\end{figure}

\begin{figure}
\centering
%\begin{subfigure}[b]{0.95\columnwidth}
   \includegraphics[width=0.99\linewidth]{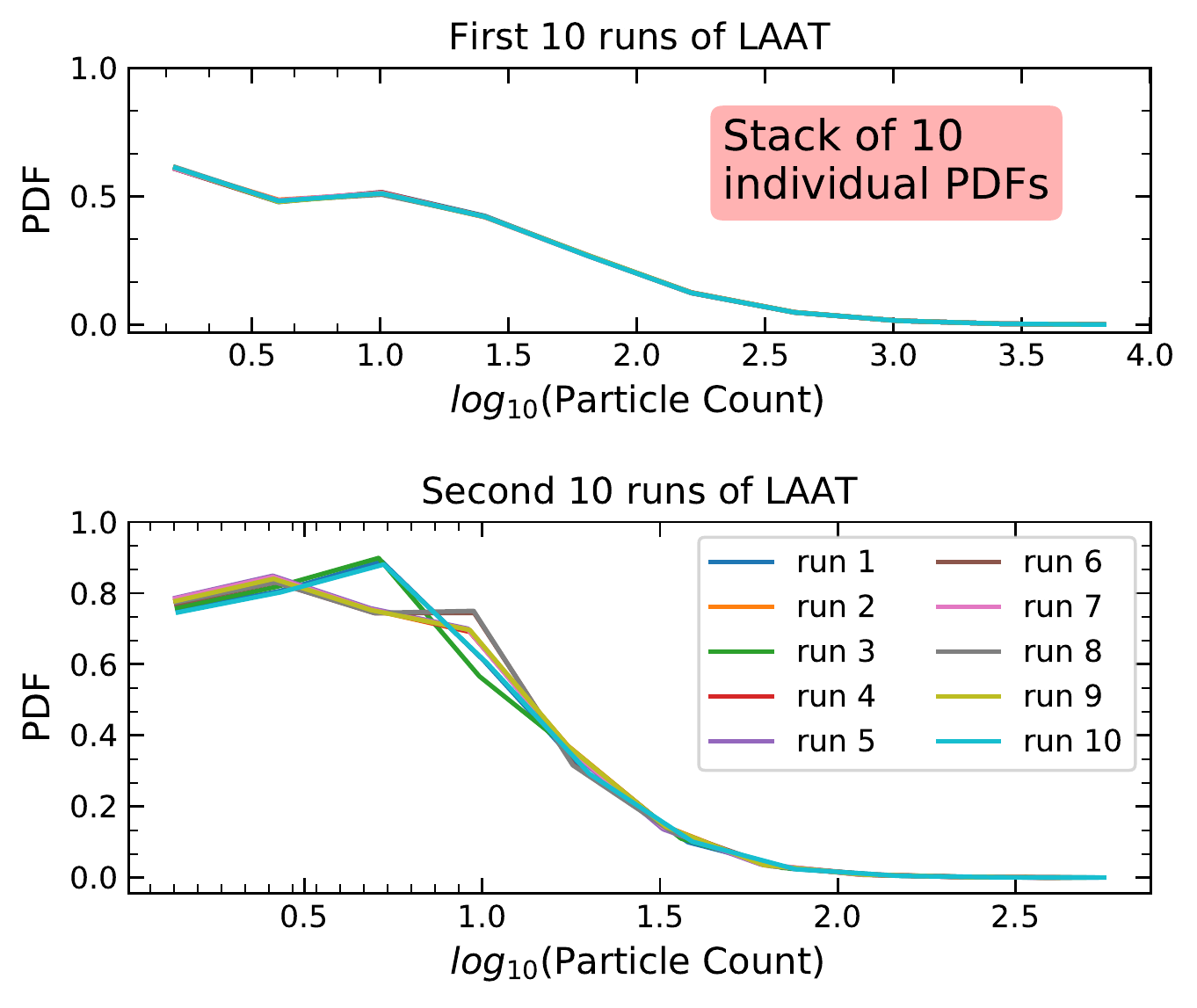}
   %\caption{}
   %\label{fig:Ng2}
%\end{subfigure}
\caption{Top panel: The result of applying LAAT $10$~times, each time changing the initial distribution of agents.
For each run, we bin the remaining particles of the data cube in a $100^3$ grid and calculate the particle count of each cell. 
The corresponding PDF of all grid cell number densities is shown as a graph. The panel therefore shows 10 superimposed PDFs each corresponding to an individual run of LAAT.
Bottom panel: The result of applying LAAT $10$~times on the filtered out particles of the first $10$~runs. Similar to the upper panel, we plot the PDFs obtained from the $10$ runs.}
\label{fig:LAATPDFs}
\end{figure}

\begin{figure*}
\centering
%\begin{subfigure}[b]{\textwidth}
   \includegraphics[width=0.8\linewidth]{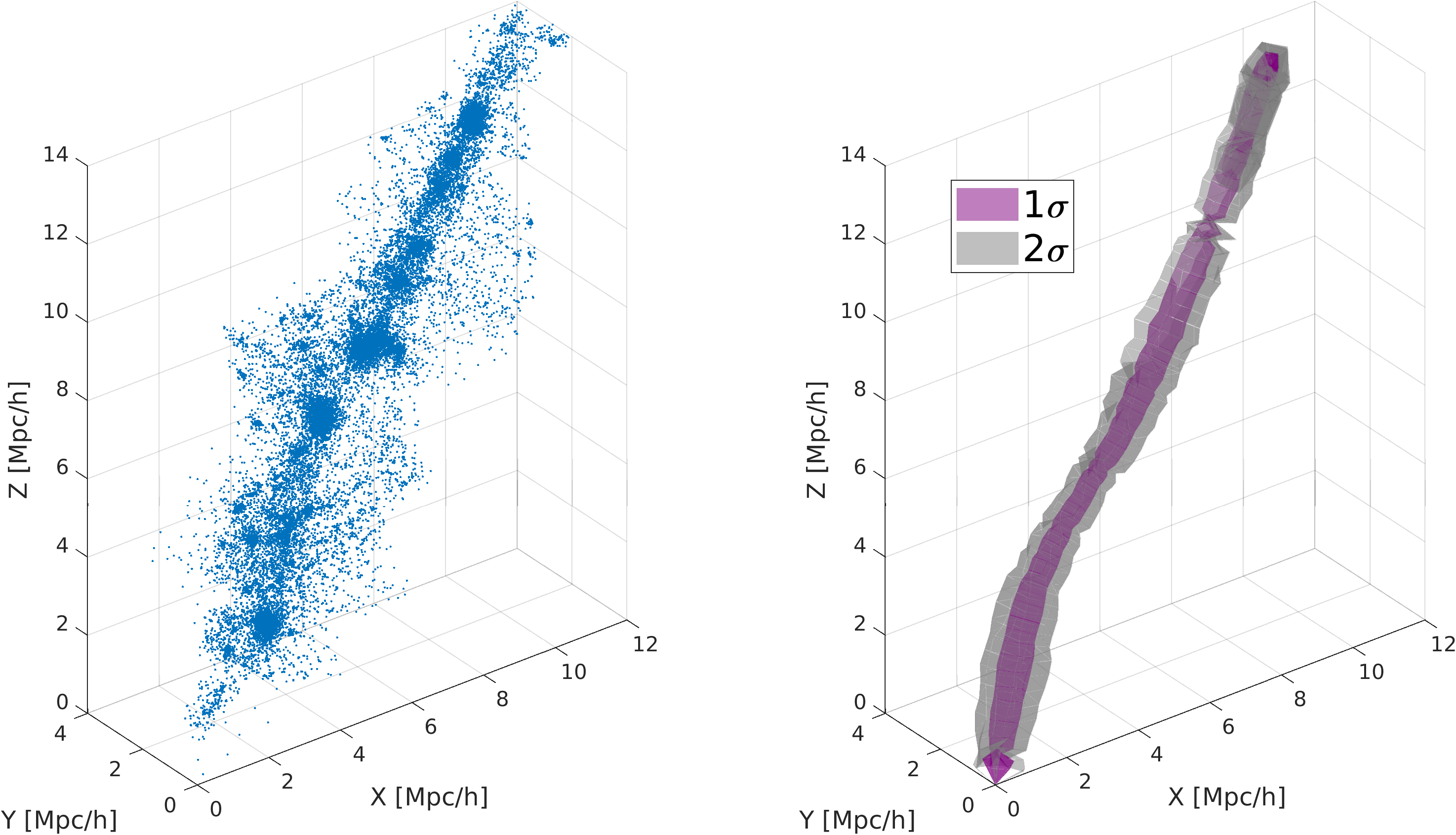}\\
   %\label{fig:Tube}
   %\caption{100\% of the Filament.}
%\end{subfigure}
%\begin{subfigure}[b]{\textwidth}
   \includegraphics[width=1\linewidth]{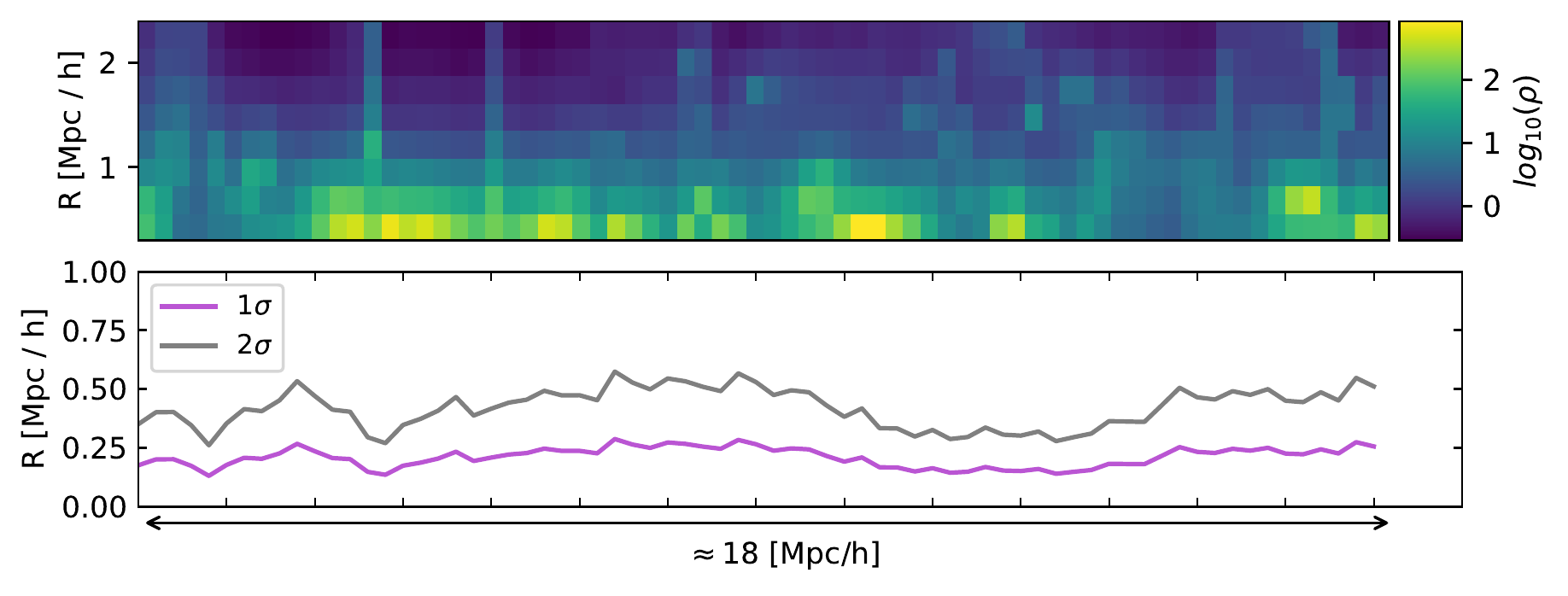}
   %\caption{80\% of the Filament.}
   %\label{fig:Sigma}
%\end{subfigure}
\caption{The top left-panel shows the original filament extracted from the N-cluster simulation data. 
The top-right panel shows two tubes corresponding to the position changes in the trained axes at each level on the filament such that  
the tubes are centered around the mean axis of the 100 trained axes, and the radius of the tubes at each position on the filament corresponds to these variations within a distance of $\sigma$ (purple) and $2\sigma$ (grey) from the mean. 
In the middle panel we show a density biplot of the same filament with an estimated length of $\approx 18$~Mpc/h along the $x$-axis. 
The bottom panel shows the variations in $\sigma$ and $2\sigma$ along the length of the filament.}
\label{fig:Tube}
\end{figure*}

%of applying a threshold for the filtration of Dark Matter particles.
%The threshold corresponds to the minimum amount of pheromone concentrated on the particles during one full run of LAAT.}
In regards to N-body simulation data of the Cosmic Web, since the structures span a very wide range of density, we present the effect of running LAAT twice. 
In the first run the agents are initiated on the entire data set, detecting the most prominent and dense clusters and filaments, while for the second run, the agents are initiated on the particles filtered out by the first run (i.e. the particles that accumulated the least amount of pheromone). 
The threshold applied for filtering the particles in this case is chosen to be the minimum value of pheromone concentration achieved in the first run. 
The results are provided in Figure~\ref{fig:LAAT2Runs} where we demonstrate the remaining particles within a slice of $20$~Mpc/h of the whole data cube. 
We observe that after the first run of LAAT (upper panel of Figure~\ref{fig:LAAT2Runs}), the concentration of the pheromone accumulates on the most dense regions which are the clusters and the most dense filaments. 
Meanwhile, after the second run of LAAT on the particles filtered out by the first run (lower panel of Figure~\ref{fig:LAAT2Runs}), we see that the more tenuous structures are getting highlighted. 
This demonstrates that it might be necessary to perform several filtrations using LAAT in order to capture the hierarchical structures of the Cosmic Web. 
However, if the most dense structures in a data cube is what is sought after, then it is sufficient to run LAAT once. 
In the next section, we discuss further the stability of these different runs.    

\subsection{Effect of Stochasticity}
\label{sec:Stochasticity}

We begin the discussion of the effect of the stochastic nature of the algorithms on the results they achieve. 
With respect to LAAT, under fixed parameters, we run the algorithm on the entire data cube and filter out the particles with the same criteria as presented in Section~\ref{sec:ParameterChange}. 
The remainder of the particles are binned within a $100 \times 100 \times 100$ grid, and the particle count is calculated within each element of the volume grid. 
We then plot the probability density function (PDF) of the obtained density distribution.
This procedure is repeated for each of the 10 runs of LAAT on the same data and using the same initialization parameters. 
The upper PDF of Figure~\ref{fig:LAATPDFs} shows the result of the 10 runs. 
We clarify that the function observed in the figure is in fact a stack of 10 lines each corresponding to a run of LAAT under a random initial distribution of agents. 
The %perfect 
compatibility between the 10 functions across all densities shows that there is no observed effect from the randomization of the agents' initial positions on finding the distinction between regions of low/high density.

We then remove the densest structures through thresholding as performed in section~\ref{sec:ParameterChange} to allow the ants to better trace out the fainter structures. We perform a second run on the particles filtered out by the 10 first runs and repeat the described analysis. 
Thus, we obtain 10 new PDFs displayed in the bottom panel of Figure~\ref{fig:LAATPDFs}. 
We see that there is great agreement between runs for regions of high density ($\log_{10}$(Particle Count) $> 1.5$), and that this agreement is slowly disrupted for regions of smaller density until $\log_{10}$(Particle Count) $\leq 0.5$, where the agreement is restored again. 
In the range of densities where we observe disagreement between the runs, we can see the PDF branches into 2-3 solutions.
We suspect that these are local minima that could confuse LAAT with regards to their structure.
Such regions could be faint filaments or faint walls. 
On the other hand, achieving agreement between the runs again for smaller densities indicates that LAAT does not have difficulties in discerning extremely faint regions. 
In total, we see that LAAT works well in separating very dense and very faint regions, while we attribute the slight difficulty in discerning the regions in between to the multi-scale nature of the Cosmic Web.

With respect to MMCrawling, the initial seed for the construction of graph representations of structures is chosen at random. 
This could then affect the positions of the Gaussian centers obtained after using SGTM to ``learn" the shape of the structure and model it (see Section~\ref{sec:AGTM} for more details). 
In other words, the graph representation constructed by MMCrawling has the potential to affect the axis trained through SGTM. 
For testing this effect with respect to a change in the initial seed, we first extract a random filament from our data set with varying density along its length as a demonstration.
The filament is shown in the top-left panel of Figure~\ref{fig:Tube}. 
We then run EM3A+ on the filament to recover its central axis. 
This allows us to run MMCrawling 100 times on the same filament and recovered central axis in an attempt to compare the differences between these runs. 
To quantify this difference, we first choose one of the trained axes at random and move along it, considering the orthogonal planes
%.While moving, we consider the orthogonal plane to the axis %and 
centered on it, and calculate the intersection points with the trained axes.
%between the trained axes and this plane. 
Since we have 100 axes, we obtain 100 intersection points with the plane at each moving position %Given a moving position, we 
and calculate the ``spread" of results. % \kb{as follows:} 
%in the following way: 
For all intersection points with coordinates $(x_i, y_i, z_i)$, $i=1,2\dots n$ and $n=100$, the mean in 3-dimensions $(x,y,z)$ and respective standard deviation 
$\sigma = \sqrt(\sigma^2)$ are calculated: % as:
\begin{align}
%\begin{equation}
\label{eq:median}
    (x, y, z) &= \frac{1}{n} \sum^{n}_{i=1} (x_i, y_i, z_i)\enspace, \\
%\end{equation}
%\begin{equation}
\label{eq:sigma2}
    \sigma^2 &= \frac{1}{n} \sum^{n}_{i=1}\left((x-x_i)^2 + (y-y_i)^2 + (z-z_i)^2\right) \enspace.
%\end{equation}
\end{align}
The calculated spread in the trained axes at each position along the filament is demonstrated in the top-right panel of Figure~\ref{fig:Tube}. 
The result is visualized as two tubes showing a change in $1\sigma$ (purple tube) and $2\sigma$ (grey tube) around the mean trained axis of the filament.
The locations where the tube is more constricted represent smaller $\sigma$ at those locations, and so better agreement between the 100 trained axes. %Vice versa 
The opposite is true for regions with a wider tube thickness. 
In the middle panel of Figure~\ref{fig:Tube} we show the density variations along the longitudinal and radial directions to the considered filament (see Section~\ref{sec:Application} for more details). 
This allows us to trace connections between the local density of the filament and variations in $\sigma$, as similarly plotted in the bottom panel of Figure~\ref{fig:Tube}. 
We can see that the highest variations occur close to the middle of the filament where we find several local regions of high density. 
We see therefore that it is easier to train an axis to fit regions of very high or very low density, while some variation may occur in regions that stand in between. 
However, despite variations in $\sigma$ we see that its value does not exceed $\approx 0.25$~Mpc/h which is still predominantly close to the center of the filament especially compared to the radial density profiles presented in the upper panel of Figure~\ref{fig:DensityProfiles}.

%After this step we would therefore have 100 models of the same structure each giving a value for the likelihood of every particle to belong to the filament. The results of the 100 models are then averaged using the Probabilistic Hough Transform (PHT): the volume containing the filament is discretized into a 100x100x100 bin grid where the likelihood of the points belonging to the same cell are averaged and then normalized so that the total likelihood of the cube is 1. In this way the likelihoods calculated by each model for the particles are averaged.

%We present here the methods used for plotting the biplots and orthogonal planes to the recovered filaments as presented for figures ? and ?. 

%%%%%%%%%%%%%%%%%%%%%%%%%%%%%%%%%%%%%%%%%%%%%%%%%%

% Don't change these lines
\bsp	% typesetting comment
\label{lastpage}
\end{document}